\documentclass[twocolumns,reprint,noshowpacs]{revtex4-1}
\usepackage{graphicx}
 \usepackage{amsmath}
 \usepackage{amssymb}
\usepackage{epsfig}

 \usepackage{color,colordvi}
 \definecolor{Red}{rgb}{0.9,0.0,0.1}
 \newcommand{\Cr}[1]{#1}

 \usepackage[
         pdftitle={Controlling adsorption of semiflexible polymers on planar 
             and curved substrates},
          pdfauthor={Tobias Kampmann, Horst-Holger Boltz, Jan Kierfeld}
          ]{hyperref}

 \newcommand*{\br}{{\bf r}}
 \newcommand*{\bt}{{\bf t}}

 \newcommand*{\abs}[1]{\left|#1\right|}
 \newcommand{\tavg}[1]{\left\langle #1 \right\rangle} 
 
 \newcommand*{\Kma}[0]{\text{ ,}}
 \newcommand*{\Airy}[1]{\mathrm{Ai}\left(#1\right) }
 \def\D#1{\mathclose{\!\mathrm{d}}#1\,} 
 \let \Beta B

\begin{document}

 \title{Controlling adsorption of semiflexible polymers on planar 
 and curved substrates}

 \author{Tobias A. Kampmann}
 \email{tobias.kampmann@udo.edu}
 \author{Horst-Holger Boltz}
 \author{Jan Kierfeld}
 \email{jan.kierfeld@tu-dortmund.de}
 \affiliation{Physics Department, TU Dortmund University, 
 44221 Dortmund, Germany}

 \date{\today}

 \begin{abstract}
 We study  the  adsorption  of semiflexible polymers  
 such as  polyelectrolytes or DNA 
 on planar and  curved substrates, e.g.,
 spheres or washboard substrates via
 short-range potentials using extensive Monte-Carlo simulations,
 scaling arguments, and analytical transfer matrix techniques. 
 We show that the adsorption threshold of stiff or semiflexible polymers 
 on a planar substrate can be controlled by polymer stiffness: 
 adsorption requires the highest potential strength if the 
  persistence length of the polymer matches the range of the adsorption 
 potential.
  On  curved substrates, i.e., an adsorbing sphere 
 or an adsorbing washboard surface, the adsorption 
 can  be additionally controlled 
 by the curvature of the surface structure. 
 The additional 
 bending energy in the adsorbed state leads to an increase of  the  
 critical adsorption strength, which depends on the 
 curvature radii of the substrate structure. 
 For an adsorbing sphere, this gives rise to an  optimal  
 polymer stiffness  for adsorption, i.e., a local minimum in 
  the critical potential strength 
 for adsorption, which can be controlled by curvature. 
 For two- and three-dimensional washboard substrates,  we 
  identify the range of persistence lengths and the mechanisms
 for an effective control of 
  the adsorption threshold by the substrate curvature. 
 \end{abstract}

 \pacs{???}

 \maketitle 

 \section{Introduction}

 Typical synthetic polymers are flexible and effects from their bending
 rigidity can be neglected on  length scales comparable to 
 their contour length. 
 For semiflexible polymers, on the other hand, 
 the bending rigidity is relevant for large scale fluctuations. 
  Many biopolymers such
 as DNA, filamentous (F-) actin, or microtubules belong  to the class of
 semiflexible polymers. The biological function of these polymers requires
 considerable mechanical rigidity; for example, actin filaments are the main
 structural elements of the cytoskeleton, in which they form a
 network rigid enough to maintain the shape of the cell and to 
 transmit forces. Prominent examples for synthetic semiflexible 
 polymers are polyelectrolytes at sufficiently low salt concentration 
 \cite{Skolnick1977} or  dendronized polymers \cite{foerster1999}, 
 where the electrostatic repulsion of charges along the
 backbone or the steric interaction of side groups give rise to considerable
 bending rigidity. 

 The bending rigidity of semiflexible polymers is 
 characterized by their  bending rigidity $\kappa$. The ratio
 of bending rigidity and thermal energy determines the 
 persistence length $L_p\sim \kappa/k_BT$, which is  the
 characteristic length scale for the decay of tangent correlations
 \cite{Gutjahr2006}.
   The physics of semiflexible polymers becomes qualitatively
 different from the physics of flexible synthetic polymers on length scales
 smaller than the persistence length where bending energy dominates over
 conformational entropy. Typical biopolymer persistence lengths range 
 from $50\,{\rm nm}$ for DNA \cite{bednar1995} (at high salt concentrations)
 to the $10\,{\rm \mu m}$-range for F-actin \cite{ott1993}
 or even up to the mm-range for
 microtubules \cite{venier1994}
 and are, thus, comparable to typical contour lengths.

 Polymer adsorption is the most important  
 phase transition for single polymer chains with numerous applications 
 \cite{degennes,eisenriegler,Netz2003}.
 Here, we consider the 
 adsorption of a single semiflexible polymer to a planar surface and 
 investigate, in particular, how the bending rigidity allows to 
 control the adsorption transition, i.e., to control the 
 critical potential strength necessary for adsorption. 
 Relevant adsorption interactions 
  are  van der Waals interactions or depletion interactions 
 for uncharged semiflexible polymers and  screened 
 electrostatic interactions or 
 counterion-induced interactions for charged polymers such as DNA or 
 F-actin \cite{angelini2003}. 
 F-actin can also be bound via  crosslinking molecules.
 \cite{dos2003,winder2005,Fletcher2010}

  From a theoretical point of view, the adsorption of semiflexible 
 polymers exhibits a rich behavior because several relevant 
 length scales compete. For a free semiflexible polymer the 
 persistence length $L_p$ and its contour length $L$ are the relevant length 
 scales, and for $L<L_p$ the behavior is dominated by bending energy.
 In the adsorption problem, both length scales 
  compete with the correlation length $\xi$ of the adsorption transition
 and the range $\ell$ of the adsorption potential. 
 \Cr{
 The correlation length $\xi$ is given by the characteristic length 
 of desorbed segments (loops) and diverges at the adsorption transition. 
 }
 If the contour length is small compared to the correlation length,
  $L <\xi$, finite size effects are relevant and affect the critical 
 behavior close to the adsorption transition. 
 If the persistence length is small compared to the correlation length, 
 $L_p < \xi$, there is a crossover in the   critical properties 
 of the adsorption transition to 
 those  of a flexible polymer \cite{Maggs1989,Kierfeld2006},
which can only be observed close to the adsorption 
 transition, and corrections to the critical potential strength
 are small.  If the persistence length becomes even smaller than the potential 
 range,  $L_p \lesssim \ell$, we expect that  the critical potential strength 
 for adsorption itself will cross over to the flexible polymer result. 
 This crossover is the central subject of this paper.

 Various aspects of the 
  adsorption transition of semiflexible polymers to planar substrates
 by short-range potentials have been studied theoretically. 
 An early study of semiflexible polymer adsorption 
 by Birshtein \cite{Birshtein1979} was based on an 
 analytical transfer matrix calculation for lattice polymers.
 The main finding was that the  critical potential 
 strength for adsorption decreases with stiffness, i.e., stiff polymers adsorb
 more easily, and that the transition sharpens with increasing stiffness
 but remains continuous.
 These findings were confirmed by numerical calculations 
 \cite{VanderLinden1996}.
 A decreasing critical potential strength for adsorption has also
 been found in  off-lattice
  molecular dynamics simulations \cite{kramarenko96} and 
 Monte-Carlo  simulations \cite{sintes01}.
 Scaling arguments for the critical potential strength
 for the adsorption of semiflexible polyelectrolytes \cite{Netz1999}
 indicate the same trend, whereas  
Monte-Carlo simulations on polyelectrolytes found a 
critical potential strength  increasing with stiffness \cite{Kong1998}.
There has also been some rigorous
 mathematical analysis of the binding transition onto a hard wall 
 \cite{Caravenna2008}.
 \Cr{
 Adsorption of semiflexible polymers can be studied 
 both for lattice polymers \cite{Birshtein1979,Hsu2013} and off-lattice as in 
 the present approach. 
 In this paper, we focus on the dependence of the critical potential 
 strength for adsorption on the polymer stiffness, i.e., 
 as a function of the dimensionless ratio $L_p/\ell$ 
 of persistence length (stiffness)
 and potential range, which are the two basic length scales 
 for semiflexible polymer adsorption. We will address both 
 the flexible limit $L_p \ll \ell$ and the stiff limit $L_p\gg \ell$. 
 For a lattice polymer, the lattice constant $a$ provides a 
 third intrinsic length scale of the problem, which introduces 
 lattice effects on small scales. The 
 flexible limit is unaffected from lattice effects 
 only if   $a \ll L_p \ll \ell$,
 which is difficult to achieve in simulations and
  motivates the use of an off-lattice model. 
 }

 Frequently, transfer matrix methods \cite{freed}
 have been applied  to the adsorption of continuous  off-lattice 
 semiflexible polymers to planar surfaces
   \cite{Maggs1989, Gompper1989, Gompper1990, Kuznetsov1997,bundschuh00,Stepanow2001,Semenov2002,kierfeld03,Benetatos2003,Kierfeld2005a, deng10}.
 Many transfer matrix approaches have employed
  a weak bending approximation (or {\it Monge parametrization})
  \cite{Maggs1989, Gompper1989, Gompper1990,bundschuh00,
  Semenov2002,kierfeld03,Kierfeld2005a}, 
where it is assumed that 
 deviations from the orientation parallel to the 
adsorbing surface are small. 
 The main findings of the  transfer matrix studies in 
 Refs.\ \citenum{kierfeld03,Kierfeld2005a} were as follows:
  The adsorption transition is continuous for an 
 orientation-independent adsorption potential, whereas it can 
 become discontinuous if an additional orientation-dependence is present. 
  All critical exponents governing the transition  were determined, 
and  analytical results  for the scaling function 
 governing the segment distribution were derived. 
 In the present paper, we will focus on the  critical potential strength  for
 adsorption and give an analytical derivation  of 
 its value for   weakly bent semiflexible polymers
 adsorbing on  a planar surface with a  short-range adsorption potential 
 using a transfer matrix approach.
 We confirm our analytical results  quantitatively  
 by extensive  Monte-Carlo  simulations, which are 
 {\em not} limited by  a weak bending approximation.

 There are two limitations of  the 
  weak bending approximation:
 (i) If the persistence length $L_p$ becomes 
  smaller than the potential range $\ell$,  the  polymer can 
  change orientation in the adsorbed state, and we have to use 
 a flexible polymer model with a Kuhn length $b_K=2L_p$ 
 to treat the adsorption transition.
 (ii) The correlation
 length $\xi$, which is  closely related to the 
 length of unbound polymer segments (so-called loops),
 diverges upon approaching the transition. 
 Sufficiently close to the transition,  $\xi$ will exceed the 
 persistence length $L_p$, and we have to use a flexible 
 polymer model with a Kuhn length $b_k = 2L_{p}$ to obtain the correct
 critical properties \cite{Kierfeld2006}. 
 The corrections for (ii) $L_p <\xi$
 mainly affect the critical behavior in the vicinity of 
  the adsorption transition and  have 
 already been addressed in the Supporting Information of Ref.\ 
 \citenum{Kierfeld2006}.
 In the present paper, we will address the consequences of 
 a crossover to  a flexible  limit  with (i) $L_p \lesssim \ell$
 in detail. 
 The corrections for  $L_p\lesssim \ell$ are more serious and strongly 
 affect the value for the critical potential strength itself. 
 This dependence can be exploited to control the adsorption 
 of semiflexible polymers.

 We argue that not only stiff polymers 
 adsorb more easily to a planar 
 surface but also flexible polymers adsorb more easily.
 This has also been observed in Ref.\ \citenum{Stepanow2001}
 in a transfer matrix treatment in an expansion around the 
 flexible limit. 
 We find that the combination of 
 easy adsorption in both the stiff and flexible limit 
  results in a
 {\rm  maximum} in the critical potential strength  or 
 a minimum in the critical temperature for adsorption in 
 the intermediate range.
 This maximum is a result of the  
 competition between $L_p$ and the potential range $\ell$
 and  is attained for $L_p \sim \ell$.
 It is, therefore,  closely connected to the problem (i) of the 
 weak bending approximation  explained above. 
 We confirm the existence of a maximum in the 
 critical adsorption potential strength as a function of the 
 polymer rigidity by performing extensive Monte-Carlo simulations
 without any  weak bending approximation. 
 An equivalent crossover in adsorption behavior
 has also been observed in numerical transfer matrix
 studies in Ref.\ \citenum{deng10}. 
 \Cr{
The numerical transfer matrix results of 
 Ref.\ \citenum{deng10} are in agreement with our 
 Monte-Carlo simulations  (because 
 lengths are measured in units of $L_p$  in Ref.\ \citenum{deng10} the 
critical potential strength  for adsorption 
  does not show the existence of a maximum as a function 
 of polymer stiffness).}
 We  derive two different interpolation functions, which 
 describe this crossover behavior of the critical potential 
 strength for adsorption  accurately 
 over the whole range of polymer rigidities.
 \Cr{
 Our interpolation functions
  are based on  exact analytical results that we obtain 
 in the stiff limit and differ from the interpolation 
 function proposed in Ref.\ \citenum{deng10}.
 }

\begin{figure}
  \includegraphics[width=0.45\textwidth]{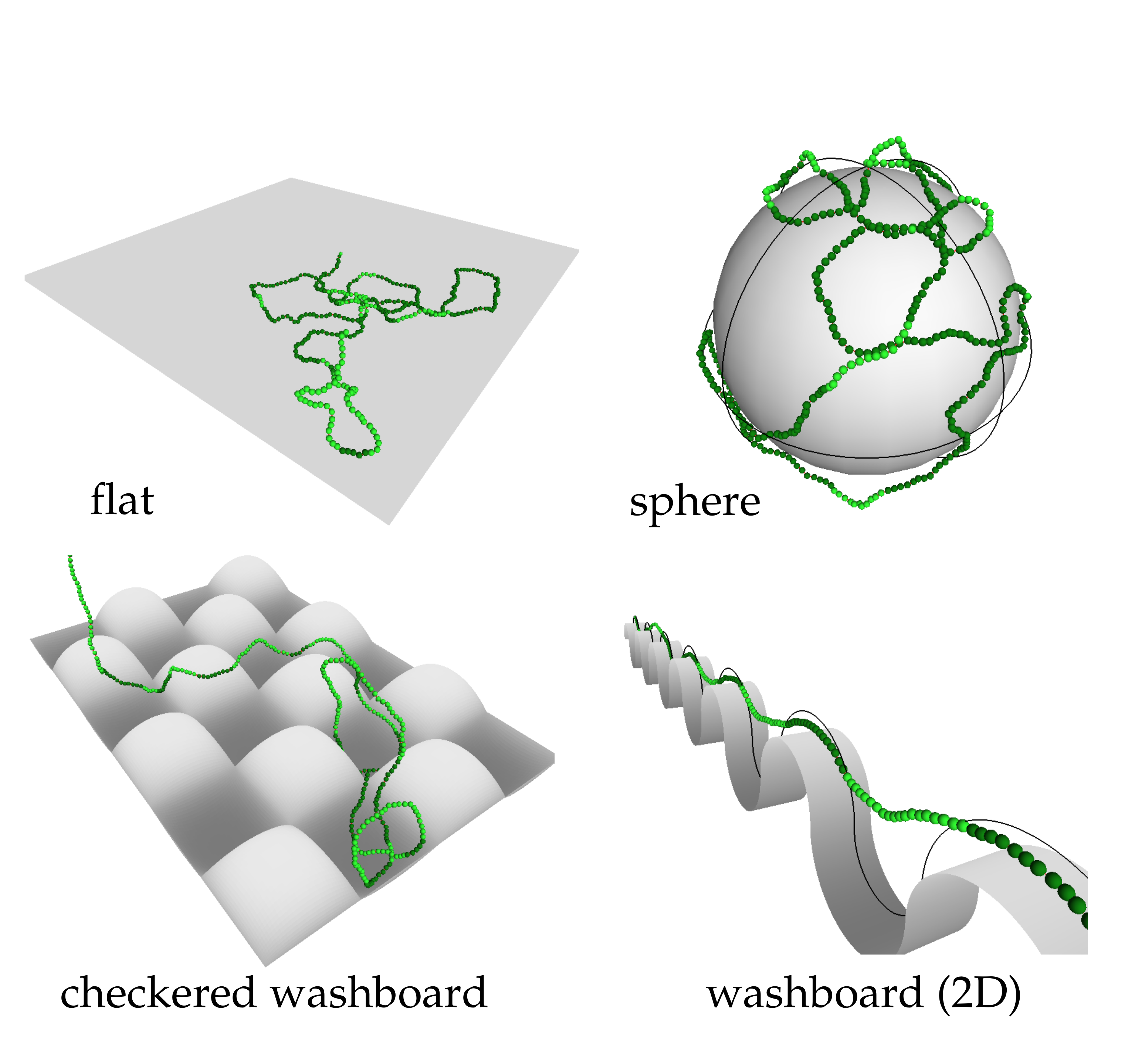}
  \caption{
 Schematic figure of adsorption  geometries 
 studied in this paper. The darker beads are adsorbed onto the surface.
 }
  \label{fig:geom}
 \end{figure}

 The existence of a maximum in the critical potential strength 
 for adsorption  has interesting and potentially useful consequences 
 for applications because 
  tuning the polymer rigidity versus the  potential range 
 allows to control 
 the adsorption threshold. For example, 
  tuning $L_p$ or $\ell$ to match each other 
 can prevent adsorption. 
 Another attractive  option to control the adsorption transition 
 are modifications of the adsorbing surface geometry. 
 \Cr{
 Adsorption of polyelectrolytes has been studied for various 
 different geometries including spherical 
 \cite{Wallin1996,Kong1998,Netz1999b,Schiessel2000,netz00,Cherstvy2011},
 cylindrical \cite{Cherstvy2011} or pore \cite{Cherstvy2012} geometries. 
 }
 Because semiflexible polymers
 have a bending rigidity, we will explore adsorption control by 
 additional {\em curvature} of the adsorbing surface. Then, the 
 adsorption free energy gain will compete with the additional 
 bending energy imposed by the substrate curvature such that systematic
 variation of the persistence length will also allow to control 
 the adsorption. 
 We will study three different types of curved substrate geometries,
 which are an adsorbing sphere, an adsorbing  washboard and a 
  checkered washboard surface as schematically shown in 
 Fig.\  \ref{fig:geom}.
 For such curved  substrates the radii of curvature 
 will set additional length scales competing with the persistence 
 length $L_p$ and allowing to control the adsorption threshold.

 The spherical adsorption geometry is relevant for 
 complexation of DNA or other polyelectrolytes 
 with oppositely charged colloids or histone proteins 
 \cite{Wallin1996,Kong1998,Netz1999b,Schiessel2000,netz00,Cherstvy2011}.
 In Refs.\ \citenum{Netz1999b} and \citenum{netz00}, a minimum of the critical
 charge for a wrapping transition  
 has been found as a function of the electrostatic screening length
 using numerical minimization of bending and electrostatic energies.
 Here, we include thermal fluctuations in the adsorption problem. 
 Based on the results for the planar geometry we can make 
 analytical estimates for a spherical adsorber, 
 which also  exhibit a local minimum in the critical potential 
 strength for adsorption as a function of the potential range
 due to a crossover from thermally driven  to bending energy driven 
 desorption. 
 These results are in quantitative agreement with 
 our Monte-Carlo simulations for this geometry.

 Structured adsorbing substrates can give rise to interesting 
 shape transitions of semiflexible polymers 
 in the adsorbed state \cite{Hochrein2007,Gutjahr2010} 
 and also give rise to activated dynamics of polymers
 \cite{Kraikivski2004,Kraikivski2005a}.
 Here, we focus on the influence of the surface structure 
 on the adsorption transition itself for a washboard and 
 checkered washboard geometry.
 Adsorption of semiflexible polymers on 
 washboard structures has been considered 
 analytically in Ref.\ \citenum{pierrelouis2011}. 
 Using scaling arguments and  Monte-Carlo simulations,
  we find a single adsorption transition for a washboard surface structure 
  with a critical adsorption strength, which exhibits a rich behavior
 as a function of polymer stiffness:
 Whereas we have a single local maximum in the 
 critical adsorption strength  as a function of polymer stiffness
 for a planar substrate, we find  two 
 maxima and one  local minimum for a washboard surface structure.
 For  a checkered washboard surface structure the local minimum 
 is suppressed again and we find a single maximum, which is 
 broadened as compared to a planar substrate.
 We identify the range of persistence lengths, for which  
 the adsorption threshold can be effectively changed and, thus, 
 controlled by the substrate curvature.

 The paper is organized as follows.  In the next section,
  we introduce the theoretical and simulation model for 
 semiflexible polymers, and the  Monte-Carlo simulation technique.
 The paper is then divided into two parts.
  First, we study the adsorption onto a
 flat surface.  We focus on the critical potential strength and 
 its dependence on polymer rigidity,
 which is  determined analytically by scaling arguments and transfer
 matrix calculations and  numerically from Monte-Carlo simulations. 
 In the second part, we investigate
 adsorption onto curved substrates
  for a sphere, a washboard  and a  checkered washboard surface.
 Using scaling arguments and  Monte-Carlo simulations, 
 we determine 
  the critical potential strength for  these surfaces as a function 
 of polymer rigidity.
 This allows us to predict how adsorption 
 of semiflexible polymers can be controlled by the curvatures of 
 the surface structure. 
\Cr{
Finally, we relate our results to experiments on polyelectrolyte 
adsorption. 
}

 \section{Model and Simulation}

 \subsection{Model} 

 The fundamental model to describe freely fluctuating 
 inextensible semiflexible polymers on all length scales
 is the Kratky-Porod model, 
 also known as {\it worm-like chain model} \cite{kratkyporod,harris66}.
 The Hamiltonian for a polymer with contour $\br(s)$, which is 
 parametrized by the arc length $s$, is given by the 
 bending energy
   \begin{align}
     \mathcal{H}_b[{\br}(s)] = \frac{\kappa}{2} \int_0^{L} \mathrm{d}s
      \left(\partial_s^2 \br(s)\right)^2 ,
   \label{eq:Ham_WLC}
   \end{align}
   where $\kappa$ is the bending rigidity and $L$ the contour length
  of the polymer. The
   persistence length $L_{p,D}$ of the worm-like chain is defined as the 
  decay length  of tangent correlations in $D$ 
   spatial dimensions  and is given by \cite{kleinert,Gutjahr2006}
   \begin{align}
    L_{p,D} = \frac{2\kappa}{(D-1)k_BT}.
   \label{LpD}
 \end{align}
  In the following, we  will use use $L_p$ to denote the three-dimensional 
 persistence length,
$L_p \equiv L_{p,3} = \kappa/k_BT$.
   The  worm-like chain model contains the two
   limiting cases  of flexible polymers for  $L_p \ll L$ and 
  rod-like polymers for $L_p \gg L$.  

  \begin{figure}
      \includegraphics[width = 0.45\textwidth]{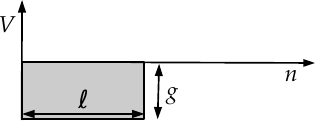}
       \caption{Square-well adsorption potential $V(n)$ as a function of 
 the coordinate $n$ perpendicular to the wall (for  a flat
 substrate $n=z$).
 }
       \label{fig:potential}
   \end{figure}

 Adsorption by an attractive planar or curved 
 surface is modeled by a
 potential $V(n)$ per polymer length, which only 
 depends on the coordinate $n$ perpendicular to the surface.
 The surface is at $n=0$; for a planar surface in the xy-plane, 
 the coordinate $n$  is the Cartesian coordinate $z$. 
 The adsorption potential  consists of a 
  short-ranged  attractive 
 square-well potential $V_a(n)$  of potential range $\ell$
 and with an energy gain  $g>0$  per unit length of an adsorbed polymer
 and a hard wall potential $V_{\rm wall}(n)$, see  Fig.\ \ref{fig:potential},
 \begin{equation}
   V(n) = V_a(n)  + V_{\rm wall}(n) 
        =  \begin{cases}
                 \infty & \mbox{ for }  n < 0 \\
                   -g         & \mbox{ for } 0<n \le  \ell \\
                   0         & \mbox{ for } n >  \ell.
         \end{cases}
    \label{eq:Vz}
 \end{equation}  
  The total adsorption energy is 
 $\mathcal{H}_{ad}[{\br}(s)]  = 
    \int_0^{L}\mathrm{d}s \,V(n(s))$.
 Short-range attractive  potentials 
  can arise from van der Waals forces or   screened electrostatic
 interactions. In these cases, the potential range $\ell$  is comparable
 to the polymer thickness or  the Debye-H{\"u}ckel screening length, 
 respectively.
 For polyelectrolytes,  we have $g/k_BT\sim \sigma \tau l_B/\tilde \kappa$ 
 and $\ell \sim \tilde \kappa^{-1}$, 
 where $l_B = e^2/4\pi \varepsilon_0\varepsilon k_BT$ is the 
 Bjerrum length, $1/\sigma$ is the area per  unit charge on 
 the surface, $1/\tau$ is the  length per unit charge on the polymer
 and $\tilde \kappa = (8\pi l_Bc)^{1/2}$ is the  inverse Debye screening length 
 depending on the total concentration $c$ of (monovalent) counter-ions
 \cite{Netz1999}. The polyelectrolyte persistence length is given by
 the sum of the bare mechanical persistence length $L_{p,\text{mech}}$
  and an electrostatic contribution,  
 $L_p = L_{p,\text{mech}}+ l_B\tau^2/4\tilde \kappa^2 $ 
\cite{Odijk1977,Skolnick1977}.

 Whereas our findings for the critical potential strength 
$g_c$ for adsorption  will depend 
 on microscopic features  of the attractive potential such as 
 the potential range, 
  results for critical exponents are expected
 to apply to all short-ranged interaction potentials, i.e., 
 to all potentials which decay
 faster than $z^{-2/3}$ for large separations $z$.
 \cite{lipowsky89}.
\Cr{We also expect the parameter dependence of the critical potential 
strength for all short-range adhesion potentials, which have only 
one characteristic length scale $\ell$ for the potential range 
and one energy scale $g$ for the potential strength, to be 
identical to our generic square-well potential. Numerical 
prefactors can vary. Therefore,  
our results for the parameter dependence 
of the critical potential strength  
for adsorption should also
 apply, for example, to  polyelectrolytes
using the above identifications $\ell \sim \tilde\kappa$ and $g
\sim \sigma \tau l_B/\tilde \kappa$.
}

 A weakly bent semiflexible polymer without overhangs 
 has a preferred orientation parallel 
 to the adsorbing surface. 
 For a planar surface this allows for a particularly simple 
   {\it Monge parametrization} by 
 choosing the $x$-coordinate along the 
 the preferred orientation and 
   $\br(x) =  (x,y(x),z(x))$, which leads to 
 \begin{align}
     &\mathcal{H}[z(x)] =\mathcal{H}_b[z(x)] +\mathcal{H}_{ad}[z(x)] 
    \nonumber\\
     &\approx \int_0^{L} \mathrm{d}x\, \frac{\kappa}{2} (\partial^2_x z)^2+
       \int_0^{L} \mathrm{d}x\,  V_a(z(x))
    \label{eq:Ham_monge}
 \end{align}
 where we used $ds =  dx (1+  (\partial_x y)^2 + (\partial_x z)^2)^{1/2}
  \approx dx$ assuming weakly bent configurations.
 Then, fluctuations in $y(x)$ 
 decouple and can be neglected for the planar 
 adsorption problem, which can be 
 treated as a two-dimensional problem  of configurations $(x,z(x))$ 
 in a plane as done in eq.\ (\ref{eq:Ham_monge}). 
 We use the Hamiltonian (\ref{eq:Ham_monge}) for 
 analytical transfer matrix calculations for adsorption on planar 
 substrates.

 For a free polymer, 
 the assumption of weak bending  is valid 
 as long as $\langle (\partial_x z)^2 \rangle, 
  \langle (\partial_x y)^2  \rangle \sim L/L_p \ll 1$,
 i.e., for contour lengths  below the persistence length.
 At the adsorption transition, the correlation length 
 $\xi$ of the transition  is an additional relevant length 
 scale, which  is  comparable to the typical 
 length of  unbound polymer segments (so-called loops).
 For an adsorbed polymer, the weak bending approximation remains valid 
 as long as the contour length of these unbound segments is smaller
 than the persistence length,  $\xi \ll L_p$, even if 
 $L \gg L_p$.

 We neglect effects from  self-avoidance. 
 Generally, we expect this to be a good approximation as long as 
 typical polymer configurations are elongated and 
 contain only few loops, as it is the case for sufficiently stiff polymers. 
 We will discuss possible effects of self-avoidance on 
 our results in the end of the paper.

 \subsection{Simulation model}

 In order to perform Monte-Carlo simulations we use a discrete 
 and extensible representation  of the semiflexible polymer 
 in terms of the {\it semiflexible harmonic
   chain (SHC) model} \cite{kierfeld04}.
 The SHC model represents a discretization of the original 
 worm-like chain model 
 (\ref{eq:Ham_WLC}) with additional bond extensibility
 and is, therefore, not limited to the weak bending regime. 
 The SHC model  consists of a fixed number $N+1$ of
   beads $\br_i$ ($i=0,...,N$) 
 connected by $N$ bonds $\bt_i \equiv \br_{i+1}-\br_i$ ($i=0,...,N-1$)
 with equilibrium length $b_0$, such that $N=L/b_0$.  
 Each bond is extensible with  a harmonic stretching energy 
 $(k/2) \left( |\bt_i| - b_0 \right)^2$. 
 The SHC Hamiltonian containing the 
  discretized version of the bending energy (\ref{eq:Ham_WLC})
  and the harmonic  stretching energies is given by 
 \begin{equation}
    \mathcal{H}_{\rm SHC}
       =  \frac{\kappa}{b_0}
      \sum_{i=1}^{N-2} \left(1-\hat{\bt}_i \cdot \hat{\bt}_{i+1} \right)
 + \frac{kb_0^2}{2} \sum_{i=0}^{N-1}
     \left( \frac{|\bt_i|}{b_0} - 1\right)^2,
 \label{eq:Ham_SHC}
 \end{equation}
 where $\hat{\bt}_i = {\bt_i}/{\abs{\bt_i}}$ are the 
 unit tangent  vectors. 
 Extensible bonds allow for a more effective Monte-Carlo simulation 
 using displacement moves of the beads. The use of bead displacement
 moves is preferred over bond rotation moves because the adsorption energy 
  is naturally given in terms of bead positions. 
 The discretized adsorption energy used in the 
  SHC simulations is
$\mathcal{H}_{ad} [\{{\bf r_i}\}] = \sum_{i=0}^{N-1} b_0 V(n_i(s))$.

 In order to approximate an inextensible worm-like chain we have to  
 choose the bond extensibility $k$ as large as possible. 
 Real polymers always have a certain mechanical extensibility. 
 Moreover, 
  the SHC model is a coarse-grained model of a worm-like chain,
 which does not contain configurational fluctuations on length scales 
 smaller than $b_0$. Therefore, an upper bound for $k$ 
 is given by the entropic elasticity associated with the stored filament 
 length within each segment of contour length $b_0$. 
  If we use 
 $N=100$ segments in simulations to discretize a polymer of 
 micrometer length, for example an actin filament 
 of contour length $L=  10~{\rm \mu m}$ the bond length 
 $b=0.1~{\rm \mu m}$ is much larger than the actin monomer size and 
 each bond can perform configurational  fluctuations reducing its 
 length to an apparent bond length $b$.
 In  response to a  tensile force $f$ fluctuations are pulled out,
 and  we find \cite{MacKintosh1995}
 \begin{equation}
   b_0-  b 
   =  
          \frac{b}{\pi}\int_{\pi/b_0}^\infty \D{q}  \frac{k_B T}{\kappa q^2+ f}
\approx
\frac{b_0^2}{\pi^2L_p} +\frac{b_0^4}{3\pi^4k_BT L_p^2}f,
 \label{Deltab}
 \end{equation}
 where we only integrate over small wavelength 
 fluctuations with $q>\pi/b_0$, which correspond to 
 shape fluctuations of  a single segment.
 If $f\ll \kappa/b^2$ the stretching is weak and the expansion
 in eq.\ (\ref{Deltab}) justified.
 From the  last term we obtain the  entropic stiffness 
\begin{equation}
   k = 3\pi^4 k_BT L_p^2/b_0^4.
\label{k}
\end{equation}
 For actin filaments, this entropic stiffness is much smaller than 
 the mechanical stiffness\cite{Kojima1994} and, therefore, 
 dominates the elasticity (entropic and mechanical
  springs have to be considered
 loaded in series rather than parallel).

 \subsection{Adsorption geometries}

 Apart from adsorption to a planar substrate, we also 
 investigate  adsorption to three different 
 curved substrates (shown in  Fig.\ \ref{fig:geom})
  in $D=2$ or $D=3$ spatial dimensions:
 (i) adsorption to a sphere  of radius $R_s$ in $D=3$ dimensions, 
 (ii)  adsorption to a washboard surface consisting of a sequence of 
  alternating concave and convex 
     half-circles of radius $R_w$ in $D=2$ dimensions. 
 (iii)  adsorption to a checkered washboard potential consisting of 
    a square lattice of alternating concave and convex 
    spherical  pieces   
     of radius $R_c$ in $D=3$ dimensions. 

 \Cr{
 For all three geometries we focus on the 
 critical potential strength $g_c$ as a function of 
 polymer stiffness, which is captured by the dimensionless 
  ratio $L_p/\ell$  of persistence length 
 and potential range. }

 \subsection{Simulation}

\begin{figure}
       \includegraphics[width = 0.45\textwidth]{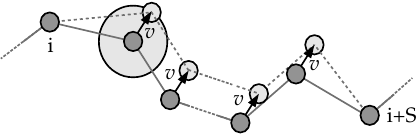}
       \caption{
 MC moves used in the simulation. 
  Each move displaces segments of $S$ successive beads, 
 but  the bending and stretching 
 energy for only {\em two} tangents at the ends of the 
  segment have to be updated. The choice of the 
  distribution of $S$ depends on the simulated 
   geometry; large  values of $S$ are only suitable for free polymers.
 A single  monomer displacement corresponds to  the special case $S=1$. 
 }
       \label{fig:moves}
     \end{figure}

 We perform  extensive Monte-Carlo (MC) simulations of the adsorption 
 process for all four geometries. We use the Metropolis 
 algorithm with bead displacement moves 
 of single beads or segments of beads as shown in  Fig.\ \ref{fig:moves}. 
 We always attach one end of the polymer to the boundary 
 of the attractive potential, i.e., at $n=\ell$, in order to 
 suppress  diffusive motion of the polymer center of mass   in the 
 desorbed phase. 

  In the simulation we measure  lengths in units of the bond length $b_0$ 
 and energies in units of $k_BT$.
  Typical simulated polymers consist of several
   hundreds of beads. The number of beads $N$ is at least
  $200$ to minimize finite size effects. Each MC  sweep consists of $N$
  MC moves, where segments of $S$ successive  beads are moved by a random
  vector of length $v$. The MC displacement  $v$ is
  determined  before each simulation to realize  an 
     acceptance rate of  about $50\%$ (typical values are $v\simeq 0.05$).
 A typical MC simulation consists of $10^7$ sweeps. The
    entropic 
    spring constant $k \sim 300\, k_BT L_p^2/b_0^4$ 
  (see eq.\ \eqref{k}) is very high due to the large
     prefactor and the quadratic dependence on the persistence length.
  We use smaller values
   such as  $k=100\,k_BT/b_0^2$ or
     $k=1000\,k_BT/b_0^2$,
     which are also independent of $L_p$, to speed up the
     simulation, because the displacement length $v$ is determined by
     the dominant energy and has to be chosen very small for 
    stiff springs $k$.  

    It turns out that the  most important control 
     parameter for the  adsorption transitions is the ratio
     $L_p/\ell$. 
   The parameter ranges that we explore are $0.5 \le \ell/b_0 \le 10$ and 
    $1 \le L_p/b_0 \le 10000$, i.e., $10^{-1} \le L_p/\ell \le 10^4$. 
     We only consider  persistence lengths larger than  one bond length,
 $L_p > b_0$.
   For smaller persistence lengths  $L_p<b_0$,
   effects from the stiffness are always negligible 
  as compared to discretization effects and the effective persistence length 
   will be $L_p \sim b_0$ such that lowering $L_p$ below $b_0$ 
  would not further decrease  the effective stiffness.

  For simulations with curved surfaces, the curvature can be
  characterized by a curvature radius $R$. For  typical applications, 
   curvature radii larger than the potential range $\ell$ 
 should be most interesting. Therefore we focus on this
  parameter regime. We also choose $R$ larger than  $b_0$, otherwise 
   discretization effects  dominate 
   curvature effects.

 \section{Adsorption to a planar substrate}

 \subsection{Analytical results}

 The parameter dependence of the critical potential strength for 
 adsorption to a planar substrate 
 can be obtained from a simple scaling argument,
 which takes into account the competition between the 
 entropic free energy cost of 
 confining an adsorbed polymer to the attractive 
  region  $0 <z<\ell$ of the square-well potential and the 
 the adsorption energy gain. 
 The entropic free energy cost  can be estimated by the 
 deflection length $\lambda$, which is the typical 
 length scale between contacts with the confining boundaries \cite{Odijk1983}.
 The resulting free energy cost per length 
  is approximately $1k_BT$ per deflection length \cite{Maggs1989,Gompper1990}.
 The adsorption energy gain per length is $-g$. Therefore the 
 total free energy change per length upon adsorption is 
 $\Delta f = k_BT/\lambda -g$. Adsorption requires $\Delta f<0$ or 
 $g>g_c$ with  a critical potential strength of the form 
 $g_c\sim {k_BT}/{\lambda}$.

 For a semiflexible polymer with $L_p\gg \ell$, the 
 collision condition $\langle z^2 \rangle(\lambda) \sim
 \lambda^3/L_p = \ell^2$ for a thermally fluctuating 
 segment of contour length $\lambda$ gives
 \cite{Odijk1983,Koster2007,Koster2008}
$\lambda_{SF} \sim L_p^{1/3}\ell^{2/3} \gg L_p$.
 The resulting  critical potential strength for adsorption is
 \cite{Gompper1990,kierfeld03}
 \begin{equation}
   g_c\sim \frac{k_BT}{\lambda_{SF}}  = c_{SF} \frac{k_BT}{\ell}
      \left(\frac{L_p}{\ell}\right)^{-1/3}.
 \label{eq:gc_stiff}
 \end{equation} 
 \Cr{This parameter dependence of $g_c$ has been obtained previously  in 
 Refs.\ \citenum{Gompper1990,Semenov2002,kierfeld03} and in the context of 
 polyelectrolytes in Ref.\ \citenum{Netz1999}.
 In Ref.\ \citenum{Benetatos2003}, adsorption by discrete linker molecules 
 instead of a continuous adsorption potential
 has been considered by transfer matrix methods
 taking  the thermodynamic limit of an 
 infinite linker number per length.
 The  results of Ref.\ \citenum{Benetatos2003} 
 can only be compared  to the other approaches 
 if the polymer length $L_m$ between linkers 
 becomes small; they  lead to the same 
  parameter dependence  (\ref{eq:gc_stiff}) only if the 
 linker distance  $L_m$  is identified with the 
 deflection length of the adsorption potential 
 (i.e., setting $J\sim g L_m$ and $L_m \sim \lambda_{SF}$ in 
 Ref.\ \citenum{Benetatos2003}).
 }

 \Cr{
 Here, we use transfer matrix methods to go beyond the 
 parameter dependence  (\ref{eq:gc_stiff}) and also derive 
 the analytical result
 }
 \begin{equation}
  c_{SF} = 3^{-1/3}\Gamma(1/3)/2\simeq 0.929
 \label{eq:cSF}
 \end{equation}
 \Cr{
  for the numerical prefactor in Supplement\cite{suppl} (IC) for a 
 square-well potential in the limit $\ell\ll L_p$. 
 }
 In this limit,
 the adsorbed semiflexible polymer is sufficiently stiff that 
 coiling {\em within} the potential range $\ell$ is 
 suppressed and $\langle z^2 \rangle(\lambda) \sim
 \lambda^3/L_p$ holds.

 For $L_p \ll \ell$, on the other hand, such coiling occurs, and 
 we have to employ a flexible polymer model with a Kuhn length 
 $b_K =2L_{p,D}$ in order to describe the adsorption transition. 
 For such a flexible polymer segment of  length $\lambda$,  
  we have $\langle z^2 \rangle(\lambda) \sim \lambda b_k \sim \lambda L_p$  and 
 collisions with the confining boundary happen on 
  a deflection length  scale
$\lambda_{F} \sim \ell^2/L_p  \ll L_p$
  as obtained from the 
 collision condition $\langle z^2 \rangle(\lambda_F) \sim \ell^2$.
 Adsorption then requires  a critical potential strength
 \begin{equation}
   g_c\sim \frac{k_BT}{\lambda_F} = c_F \frac{k_BT}{\ell}
     \frac{L_p}{\ell}.
 \label{eq:gc_flexible}
 \end{equation} 
 Using standard  transfer matrix methods for flexible polymers
 and solving the equivalent  problem of a quantum mechanical particle 
 in a square-well potential
 \cite{degennes}, one finds 
 $c_F = 2\pi^2/4D(D-1) $ for the numerical prefactor in $D$ dimensions.

 Remarkably, the results (\ref{eq:gc_stiff}) for  a semiflexible polymer with 
 $L_p\gg \ell$ and  (\ref{eq:gc_flexible}) for a flexible polymer 
 with $L_p\ll \ell$  show a rather different behavior as a function of 
 polymer stiffness $L_p$: whereas the critical adsorption 
 potential strength $g_c$ is {\em increasing} with stiffness
 in the flexible regime, it {\em decreases} with stiffness 
 in the stiff regime. 
 \Cr{
 Both in the stiff and in the flexible limit the critical potential 
 strength for adsorption becomes small. 
 The driving force behind this behavior is, however, different. 
 In the stiff limit, the entropic  free energy cost for adsorption 
 becomes small because for a stiff polymer shape fluctuations are 
 small, and the stiff polymer does not lose much configurational entropy
 upon adsorption. 
 In the flexible limit, on the other hand, the effective monomer 
 length decreases (and the effective monomer number increases) 
 with stiffness.  For a small monomer size, positional fluctuations 
  and, thus, the entropy cost of confinement  also decrease.
 Therefore, also in the flexible limit the configurational entropy 
 cost during adsorption becomes small. 
 }
 As a result, we expect a {\em maximum} of the critical potential 
 strength for adsorption in the intermediate stiffness regime,
 where $L_p \sim \ell$. Hence, adsorption is most difficult 
 if polymer persistence length  and adsorption potential range are tuned 
 to match each other. 
 This   has interesting consequences 
 for applications because 
  tuning the polymer rigidity such that the persistence length 
 $L_p$ matches   the  potential range $\ell$ 
 could be a measure to prevent adsorption.

 In order to quantify the location of the 
 maximum, we will use  an   interpolating 
     function  $I(x)$ describing the 
 critical adsorption potential as a function of the 
 dimensionless stiffness parameter $L_p/\ell$, 
 \begin{equation}
   \frac{g_c\ell}{k_BT} =  I\left(\frac{L_p}{\ell}\right),
 \label{eq:I}
 \end{equation}
 which captures 
  the correct asymptotics in the semiflexible and flexible limit, 
 \begin{align}
    I(x) &\approx c_F x ~~\mbox{for}~~ x\ll 1
    \label{eq:Ismall}\\
    I(x) &\approx c_{SF} x^{-1/3} ~~\mbox{for}~~ x\gg 1
    \label{eq:Ilarge}
 \end{align}
 according to (\ref{eq:gc_stiff}) and (\ref{eq:gc_flexible}).
 \Cr{
 Furthermore, in the stiff limit, the leading 
 corrections from flexibility are of the 
 order 
 \begin{equation}
    I(x) \approx c_{SF} x^{-1/3} +  {\mathcal O}(x^{-1}),
   \label{eq:Ilarge2}
 \end{equation}
 see Ref.\ \citenum{Kierfeld2006} and 
 the   discussion in Supplement\cite{suppl} (IC3). 
 An interpolation function $I(x)$ has to fulfill the three constraints
 (\ref{eq:Ismall}), (\ref{eq:Ilarge}), and (\ref{eq:Ilarge2}). 
 We will obtain a fourth constraint below. 
 }

 An  interpolating scaling function $I(x)$ can be  motivated 
 by the behavior in the stiff limit (\ref{eq:Ilarge}). 
 Additional compatibility with the flexible limit (\ref{eq:Ismall}) 
  suggests $I(x) =c_1 x/(1+ c_2x^{4/3})^{-1}$ with 
 $c_1 = c_F$ and $c_2 = c_F/c_{SF}$. 
 The constraint (\ref{eq:Ilarge2}) on the next to leading order term 
 in the stiff limit 
 then suggests the presence of another term, 
 \begin{equation}
 I(x) = c_1x (1+c_2 x^{4/3}+c_3 x^{2/3} )^{-1}.
  \label{eq:fit1}
 \end{equation}
 This scaling function 
  contains three free fit parameters $c_1$, $c_2$, and $c_3$.
 The choices  $c_1=c_F$  
 and $c_2  =  c_F/c_{SF}$  
 will reproduce the known flexible and semiflexible
 limits (\ref{eq:Ismall}) and (\ref{eq:Ilarge}), respectively, 
 and the remaining  parameter $c_3$ allows to vary the position 
 of the maximum. 

 \Cr{
 A scaling function similar to eq.\ (\ref{eq:fit1}) 
 has also  been used to describe numerical transfer matrix 
 calculations in Ref.\ \citenum{deng10}. The interpolation function
 of Deng {\it et al.}  differs in two aspects:
(i)   the numerical prefactor  $c_{SF}$, see eq.\ (\ref{eq:Ilarge}),
   has been treated as a completely
 free fit parameter
  because an analytical result was not available, and (ii) the interpolation 
 function does not obey the constraint  (\ref{eq:Ilarge2})
   because of a   next to leading term in the stiff limit
 $I_{\text{Deng}}(x)\approx c_{SF}
 x^{-1/3}+\mathcal{O}(x^{-4/3})$.
 A more detailed discussion  of this scaling function is given 
  in Supplement\cite{suppl} (IA).
 }

 Alternatively, we can use the above  scaling argument for the 
 deflection length $\lambda$ and  $g_c \sim k_BT/\lambda$ to 
 motivate an alternative  functional form of the  interpolation 
 function $I(x)$ with only two free fit parameters, which 
  is described in Supplement\cite{suppl} (IB).
 The derivation in Supplement\cite{suppl} (IB) also suggests as a fourth constraint, 
 that the interpolation 
 function $I(x)$ should obey the functional form 
 \begin{equation}
   I(x) \sim \frac{x^{-1/3}}{\tilde{g}({\rm const}\, x^{-2/3})}~~
  \mbox{for}~x\gg 1
 \label{eq:constraint}
 \end{equation}
 in the stiff limit, where $\tilde{g}(x)$ is an 
  analytical function $\tilde{g}(x)$ with $\tilde{g}(0)\neq 0$.
 The interpolation function  (\ref{eq:fit1}) as well as the alternative 
 interpolation function given in the Supplement\cite{suppl}
  fulfill  this constraint, whereas the scaling function used in 
 Ref.\ \citenum{deng10} does not obey this constraint.

 \subsection{Numerical results}

We determine the critical potential strength from the 
MC simulations using two different methods: (i) by order parameter 
cumulants and (ii) finite size scaling. Finite size scaling (ii) 
also allows us to determine the free energy exponent $\nu$. 
The simulation 
results for the critical potential strength are summarized 
  in the phase diagrams 
   Fig.\ \ref{fig:kritpot_flach}(A)  for $D=3$ and Fig.\ 
  \ref{fig:kritpot_flach}(B) for $D=2$.  The resulting  
    fit parameters $c_1$, $c_2$, and $c_3$  for the interpolation
  function $I(x)$ from eq.\  (\ref{eq:fit1}) 
   are shown in Table  \ref{tab:fits}.

 \subsubsection{Critical potential strength via third order parameter 
  cumulant}

 An effective method to determine the critical potential strength
 uses the fact that the derivative of the free energy density
 with respect to the potential strength $g$ gives the 
mean  fraction of polymer length  in the square-well potential, 
which provides an  order parameter for the adsorption transition. 
Derivatives of the free energy with respect to $g$ generate 
cumulants of the mean fraction of adsorbed polymer length.
In the Supplement\cite{suppl} (IIA) we discuss in detail  that 
the second cumulant is expected to have a maximum at the transition and, 
thus, the third cumulant a zero. 
\Cr{We use this criterion both for the planar and for curved 
geometries to locate the adsorption transition.}

\begin{figure}
  \begin{center}
   \includegraphics[width=0.45\textwidth]{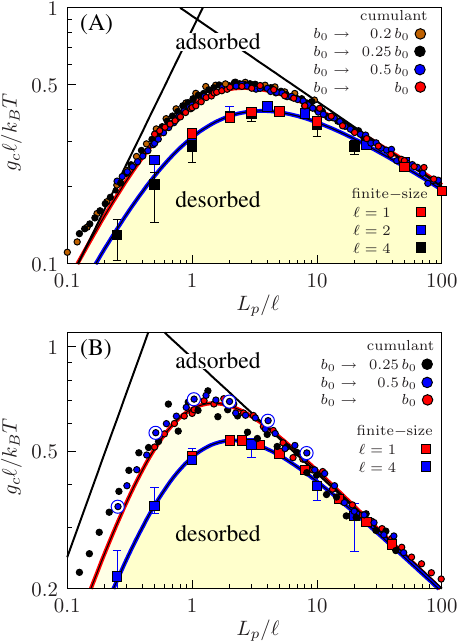}
   \caption{    
   Phase diagram for a flat substrate in
     (A) $D=3$ and (B) $D=2$ as obtained from MC simulations. 
   The  double logarithmic plot  shows
   the dimensionless critical potential strength $g_c\ell/k_BT$ as 
  a function of the dimensionless stiffness parameter $L_p/\ell$
  with  increasing bending stiffness from left to right. 
The yellow region
     marks the desorbed state. 
  The analytical results (\ref{eq:gc_stiff}) in the stiff limit 
and (\ref{eq:gc_flexible}) in the flexible limit are shown as 
  straight black lines.
     Circles show  results for the critical
     potential strength as determined from the zero 
   of  the third order parameter cumulant using 
simulation parameters  $N=200$, $\ell=2b_0$, and $k=1000\,k_BT/b_0^2$.
    By changing $b_0$  we check that the discretization length
     $b_0$ has no influence on our results. Squares 
  show results  from finite size scaling for  $k=1000\,k_BT/b_0^2$. 
  The colored curves show  
  interpolation function \eqref{eq:fit1} with fit parameters 
  $c_1$, $c_2$, and $c_3$ 
    as given in   Table \ref{tab:fits} for  
   the  cumulant method  (red curves)
   and  the finite size method   (blue curves).
 Large blue  circles in (B) correspond to the simulation snapshots 
  shown in Fig.\ \ref{fig:kritpot_flach_2d_snapshots}.
    }
      \label{fig:kritpot_flach}
\end{center}
\end{figure}

The simulation results for the critical potential strength $g_c$ 
as determined by the third cumulant are summarized 
  in  the phase diagrams 
  Fig.\ \ref{fig:kritpot_flach}(A)  for $D=3$ and Fig.\ 
  \ref{fig:kritpot_flach}(B) for $D=2$ (circles).  The resulting  
    fit parameters of the  interpolation
  functions from eqs.\  (\ref{eq:fit1}) and (S3) from Supplement\cite{suppl} (IB)
   are shown in Table  \ref{tab:fits}.

In the stiff limit,  we find good agreement with the 
 analytical result   \eqref{eq:gc_stiff} 
 as can also be seen from the values for the 
     fit  parameter combination 
 $c_2c_{SF}/c_1$  of the interpolating function 
 $I(x)$ from eq.\  (\ref{eq:fit1})  in  Table  
 \ref{tab:fits}, which are close to our analytial result 
  $c_2c_{SF}/c_1=1$    in the stiff limit.
 In the flexible limit, 
   simulation results for  $g_c$ are slightly  larger than 
  the analytical result (\ref{eq:gc_flexible}) in $D=3$ 
and  smaller in $D=2$ as can be seen from the 
   values for the   fit  parameter $c_1$ of the interpolating function 
 in  Table   \ref{tab:fits}: 
 we expect  $c_1/c_{F}=1$  from the analytical results
  in the flexible  limit and  find 
  $c_1/c_F\gtrsim 1$  for $D=3$  and  $c_1/c_F\lesssim 1$ 
for $D=2$.  
 One  reason for this deviation is the 
   finite size of the polymer;  simulation results
    tend to the analytical 
   result \eqref{eq:gc_flexible} with increasing length of the
    polymer (using the same discretization $b_0$).
   We also examine discretization effects by changing the bond length 
   $b_0 \rightarrow a b_0$  by a factor $a$ and accordingly 
   $N  \rightarrow {N}/{a}$ to keep the polymer length $L=Na$ constant. 
 This allows further to explore smaller values for $L_p /\ell$,
   without violating the condition $L_p >b_0$, which ensures that 
  the persistence length is not cut off by the discretization $b_0$.
 The simulation data in 
  Fig.\  \ref{fig:kritpot_flach} shows that  
 changing the discretization $b_0$ has only little effect on the critical 
potential strength.

\begin{figure}
 \begin{center}
    \includegraphics[width=0.45\textwidth]{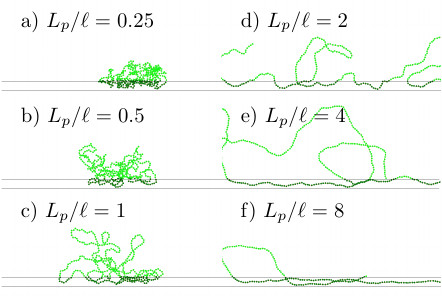} 
      \caption{ 
          Typical simulation configurations  for adsorption on 
      a flat substrate in $D=2$ for increasing stiffness parameters
      $L_p/\ell$. Snapshots are taken  in the adsorbed phase
  close to the critical potential strength. 
 The  simulation parameters are $N=400$, $\ell=4b_0$, 
   and $k=250\,k_BT/b_0^2$.
      }
      \label{fig:kritpot_flach_2d_snapshots}
\end{center}
\end{figure}

For  $D=2$ we also show simulation snapshots of typical 
polymer configurations in Fig.\ \ref{fig:kritpot_flach_2d_snapshots}. 
These snapshots are taken  in the adsorbed phase
  close to the critical potential strength. 
   We can clearly distinguish  the different characteristics of configurations 
in the the stiff and flexible limit, which give rise 
to the different adsorption behavior:
in the flexible limit $L_p/\ell < 1$, 
the  adsorbed polymer  exhibits turns within the 
attractive potential layer of width $\ell$ giving rise to 
compact adsorbed configurations. In the stiff limit $L_p/\ell > 1$, 
on the other hand, the configurations are elongated without turns 
within the attractive potential layer.

A comparison of our numerical results for $D=2$ and $D=3$ (see Fig. \ref{fig:kritpot_flach} (A) and (B)) shows that the critical adsorption strength is indeed independent
 of the number of transversal dimensions in the stiff limit in agreement with eq. \eqref{eq:gc_stiff} and \eqref{eq:cSF}. This justifies our treatment with only one transversal dimension within the Monge approximation see eq. \eqref{eq:Ham_monge}.
 Also the asymptote in the flexible limit is independent of dimensionality except for the prefactor in agreement with eq. \eqref{eq:gc_flexible}. Therefore,  the general shape of the critical potential strength as a function of stiffness with maximum for $L_p \sim \ell$ is valid for all dimensions.

\begin{table}
    \begin{tabular}{c||cccc}
\hline
data set  & $c_1 /c_{F}$    & $c_2 c_{SF} /c_1$   & $c_3$ & $\text{max}(I)$ \\ 
\hline
theory (D=3) &  $1 $       & $1$  &  free & \\
cumulant      & $1.13{\pm} 0.05$  & $1.03 {\pm} 0.01$     & $0.26 {\pm} 0.06$& $2.61$
\\
finite size   & $0.9{\pm} 0.1$  & $1.01 {\pm} 0.03$     & $0.6 {\pm} 0.2$& $3.60$  \\
theory (D=2) &  1          &$1$  &  free &  \\
cumulant  &  $0.51 {\pm} 0.03$  & $0.97 {\pm} 0.01$ & ${-}0.48 {\pm} 0.05$ &  $1.49$   \\
finite size   &  $0.40 {\pm} 0.01$  & $0.98 {\pm} 0.01$ & ${-}0.03 {\pm} 0.04$ &  $2.13$
  \\
\hline
\end{tabular}
\caption{Simulation results for the fit 
  parameters  $c_1$, $c_2$, and $c_3$  for the interpolation
  function $I(x)$ from eq.\  (\ref{eq:fit1}) 
in comparison 
  with  theoretical expectations. 
  The maximum value of the resulting interpolation function
  is given for comparison. 
  All fits for the cumulant method are performed for 
MC data from simulations with 
 $N=200$, $\ell=2b_0$, $k=1000\,k_BT/b_0^2$.
  For the analysis of simulation data we use the cumulant method 
  or finite size scaling as explained in the text. 
}
 \label{tab:fits}
\end{table}

 \subsubsection{Critical exponent and potential strength 
  via finite size scaling}
\label{sec:finite-size}

We also use finite size scaling of the specific heat 
to corroborate our results for the critical 
potential strength and to calculate the 
 critical exponent $\nu$ for the correlation length and the 
free energy. 
The (extensive) specific heat
 $C=\beta^2\left(\tavg{\mathcal{H}^2}-\tavg{\mathcal{H}}^2\right)$ 
 exhibits finite size scaling according to 
  \begin{equation}
   L^{-{2}/{\nu}}\beta^2\left(\tavg{\mathcal{H}^2}-\tavg{\mathcal{H}}^2\right) = 
   f\left( (g-g_c) L^{{1}/{\nu}} \right)  
 \label{eq:finitesize}
  \end{equation}
with a scaling function $f(x)$. 
This expression is rather natural for a continuous transition, but it is
important to note that it also applies if the adsorption 
transition is of first order, where \cite{Binder1987}
\begin{equation}
  L^{-2}\beta^2\left(\tavg{\mathcal{H}^2}-\tavg{\mathcal{H}}^2\right) = 
    f_1\left( (g-g_c) L^{d} \right)  
  \label{eq:finitesize1}
\end{equation}
with a scaling function $f_1(x)$ and $d$ as the internal dimension  
of the system. 
This expression  is actually 
of the same form as expression  (\ref{eq:finitesize}) for a continuous 
transition because a polymer has internal dimension
$d=1$, and we have $\nu=1$ for a first order transition. 
Therefore, there is no
systematic bias regarding the order of the transition if eq.\
(\ref{eq:finitesize})  is used for finite size scaling.
In Supplement\cite{suppl} (IIB), we explain in detail how the best parameter 
set $(\nu,\,g_c)$ for the finite size 
scaling (\ref{eq:finitesize}) is determined 
using a systematic error minimization procedure.

Data for the critical potential strength $g_c$ as obtained from the 
 finite size scaling procedure is  presented in the phase diagrams 
\ref{fig:kritpot_flach} (A) and \ref{fig:kritpot_flach} (B)
(squares). 
We find good  agreement  with our analytical result. 
 The resulting    fit parameters of the  interpolation
  functions from eqs.\  (\ref{eq:fit1}) and (S3) from Supplement\cite{suppl} (IB)
   are shown in Table  \ref{tab:fits}.

The finite size scaling procedure  also allows us to 
determine the critical exponent $\nu$. 
As discussed in Supplement\cite{suppl} (III), we find an exponent
$\nu$ around $\nu=2$ for
small bending rigidity, which lowers towards $\nu=1$ 
 with increasing stiffness. This is in agreement with the theoretical 
expectation that a semiflexible polymer  should 
exhibit a critical behavior corresponding to  $\nu_{SF}=1$
with a  crossover to a flexible behavior with $\nu=\nu_F=2$ 
in the small regime $|g-g_{c,SF}| < k_BT/L_p$ around the transition,
where the correlation length $\xi$ exceeds  $L_p$.

\section{Adsorption to curved substrates}

Because semiflexible polymers
have a bending rigidity, adsorption can be controlled by an 
additional curvature of the adsorbing surface.
We investigate the influence of surface curvature for 
 three different geometries, an  adsorbing sphere, 
an adsorbing  washboard  and a 
 checkered washboard surface as shown in 
Fig.\  \ref{fig:geom} both in $D=2$ and $D=3$ spatial 
dimensions. 

Depending on the polymer stiffness several additional effects can 
occur for adsorption on curved substrates:
(i) Flexible polymers with $L_p <\ell$ 
adsorb in a compact conformation on a flat substrate, 
as can be seen 
in Fig.\ \ref{fig:kritpot_flach}a).
 Concave curvatures with radii $R>\ell$ can give rise to 
an {\em increased} adhesion energy gain because 
the polymer  can realize  a larger 
 contact area with the adhesive potential, 
see  Fig.\ \ref{fig:kritpot_washboard}a).
This effect favors adsorption on a curved substrate and is relevant
for washboard potentials.  
(ii) For stiff polymers with $L_p >\ell$ 
an additional bending energy cost occurs
during  adsorption on a curved substrate.
This effect favors desorption and 
is the most relevant effect to effectively control 
the adsorption for all geometries 
we consider by tuning the substrate curvature radius $R$.

 \subsection{Adsorption to a sphere}

An additional bending energy cost arises for adsorption of 
a stiff semiflexible 
polymer with $L_p>\ell$ to a sphere with radius $R_s$. 
For a polymer of length $L$ firmly adsorbed with curvature 
$1/R_s$ the additional total bending energy 
is $E_{R} \sim \frac{1}{2} L \kappa /R_s^2$. We can include this energy 
into the simple scaling argument for the critical potential 
strength. The  total free energy change per length upon adsorption becomes 
$\Delta f = k_BT/\lambda -g+ k_BT L_p/2R_s^2$ with the deflection length 
$\lambda \sim L_p^{1/3}\ell^{2/3}$, which we assume to be unchanged 
by curvature effects (which should be justified for $R_s\gg \lambda$
  \cite{Koster2008}). 
The adsorption condition $\Delta f<0$ leads to the following  estimate 
for the critical potential strength,
\begin{equation}
  \frac{g_c \ell}{k_BT} = g_c(R_s=\infty)+ \frac{L_p\ell}{2R_s^2} =
 c_{SF} \left(\frac{L_p}{\ell}\right)^{-1/3}
     +  \frac{L_p\ell}{2R_s^2} 
\label{eq:gc_stiff_R}
\end{equation} 
with the critical potential strength $g_c(R_s=\infty)$ 
for a planar substrate from eq.\ (\ref{eq:gc_stiff}).

The result (\ref{eq:gc_stiff_R}) can also be interpreted in terms
of the contact curvature radius $R_{co} \sim (\kappa/|\Delta f|)^{1/2}$
for polymer adsorption by an effective  contact potential of 
strength\cite{Kierfeld2006} $\Delta f =  k_BT/\lambda -g$, 
which is given by the 
free energy of adsorption to a planar substrate. 
The adsorption condition $g>g_c$ with $g_c$ as given by (\ref{eq:gc_stiff_R})
corresponds to the condition $R_{co} < R_s$ that the
contact curvature is smaller than the sphere radius.

\begin{figure*}
 \includegraphics[width = \textwidth]{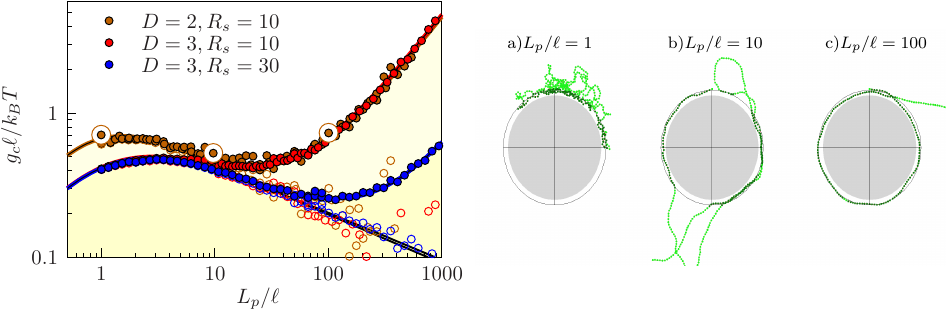}
      \caption{ 
  Left:  Phase diagram for an adhesive sphere in  $D=3$ and $D=2$ as obtained 
   from MC simulations:  double logarithmic plot  of
   the dimensionless critical potential strength $g_c\ell/k_BT$ as 
  a function of the dimensionless stiffness parameter $L_p/\ell$ 
 as obtained with the cumulant method.
 Simulation parameters are $N=200$, $\ell=b_0$, and $k=1000\,k_BT/b_0^2$.
  Equation \eqref{eq:gc_stiff_R} 
is used to fit the data via an effective sphere radius $R_s$;
   the results are $R_{s,1}/\ell =10.46 {\pm}  0.06$, 
   $R_{s,2}/\ell  = 10.27 {\pm} 0.02$ and $R_{s,3}/\ell  =30.6 {\pm}  0.2$.
  Filled circles are original MC results, 
   hollow circles show the reduced potential strength 
  $g_c \ell/k_BT - {L_p\ell}/{2R_s^2}$, which agrees well with our simulation 
  results for a planar surface. 
\Cr{
  Right:  Typical simulation configurations for increasing stiffness parameters
      $L_p/\ell$ and $R_s=10\ell$. 
Snapshots are taken  in the adsorbed phase
  close to the critical potential strength. 
 The  simulation parameters are $N=400$, $\ell=2b_0$, 
   and $k=100\,k_BT/b_0^2$. 
   The simulation snapshots correspond to the 
    large brown circles in the phase diagram on the left. 
}
}
 \label{fig:kritpot_sphere}
\end{figure*}

MC simulation data agree well with the analytical result 
(\ref{eq:gc_stiff_R}) for the adsorption threshold. 
We describe our MC simulation results for the critical potential strength 
(as obtained by the cumulant method)
by eq.\  (\ref{eq:gc_stiff_R}) with the radius $R_s$ as fit parameter. 
We expect the resulting effective adsorption radius 
 to be of the order $R_{s}+\ell/2$. The simulation 
results are shown in the phase diagram in Fig.\ \ref{fig:kritpot_sphere}
and exhibit  good agreement with eq.\ (\ref{eq:gc_stiff_R}) 
with effective curvature radii within the interval
$[R_s,R_s+\ell]$.
In Fig.\ \ref{fig:kritpot_sphere} we also show the reduced 
critical potential strength $g_c \ell/k_BT - {L_p\ell}/{2R_s^2}$, 
which agrees very well with our results for a planar substrate. 

Remarkably, the critical potential strength 
  (\ref{eq:gc_stiff_R})
has a {\em local minimum} at 
\begin{equation}
   \frac{L_p}{\ell} \sim  \left(\frac{R_s}{\ell}\right)^{3/2}
\label{eq:min_sphere}
\end{equation}
 with  $g_{c,\text{min}} \sim   ({k_BT}/{\ell})
\left({\ell}/{R_s}\right)^{1/2}$
 because of the bending energy correction. 
This can be used to design an ``optimally sticky'' sphere 
for adsorption of a semiflexible polymer by choosing a radius 
$R_{s,opt} \sim \ell (L_p/\ell)^{2/3}$ for given 
persistence length and potential range or by choosing an 
optimal potential range $\ell_{opt} \sim R_s^3/L_p^2$ for given 
sphere radius and persistence length. The latter can be realized for 
polyelectrolytes by adjusting the salt concentration. 
In the absence of thermal fluctuations, such a minimum 
in the critical potential strength for adsorption has also been 
found for the complexation of polyelectrolytes with oppositely 
charges spheres \cite{Netz1999b,netz00}.

\subsection{Washboard  surface}

In $D=3$ dimensions 
the washboard surface is translationally invariant in one direction 
and is composed of
alternating half-cylinders of radius $R_w$, see 
Fig.\ \ref{fig:geom}.  Then the  polymer
orients {\em parallel} to the  half-cylinders during adsorption 
in order to avoid additional 
bending energies, and the critical potential strength for adsorption 
 is very similar to the planar surface result.
Consequently, the substrate structure radius $R_w$ gives no control 
on the adsorption threshold for a cylindrical washboard in $D=3$.
  
Therefore, we focus first on washboard surfaces in $D=2$  dimensions, which 
are composed of half-circles of radius $R_w$.
This two-dimensional adsorption geometry is equivalent to the 
situation where 
the polymer is confined to a two-dimensional plane {\em perpendicular} 
to the half-cylinders of a three-dimensional washboard surface. 
We will show  that under this confinement, pronounced effects 
from the surface structure occur, and the substrate curvature radius 
can be used to control adsorption. 
Afterwards, 
we will discuss the checkered washboard surface as an 
alternative substrate structure  to effectively 
control the adsorption threshold in $D=3$ spatial dimensions
without applying additional constraints.

\begin{figure*}
\begin{center}
  \includegraphics[width=0.98\textwidth]{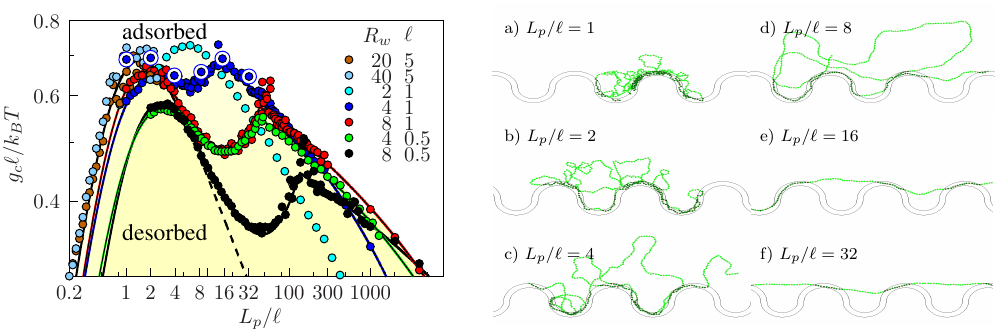}
   \caption{   
   Left: 
  Phase diagram for an adhesive washboard surface in  $D=2$ as obtained 
   from MC simulations:  double logarithmic plot  of
   the dimensionless critical potential strength $g_c\ell/k_BT$ as 
  a function of the dimensionless stiffness parameter $L_p/\ell$ 
 as obtained with the cumulant method for different values 
  $R_w/\ell = 2,4,8,16$.
 The remaining  
  simulation parameters are $N=200$, and $k=100\,k_BT/b_0^2$. 
    The solid lines show 
    fits using  eq.\ (\ref{eq:gc_stiff}) (using $R_s$ as fit parameter) 
   for an adsorbing sphere 
      for small $L_p/\ell$  and fits using  eq.\  (\ref{eq:gc_incomplete1}) 
     (using $w_1$ and $w_2$ as fit parameters)
    for larger stiffnesses $L_p/\ell$. 
 The dashed line is the fit for the flat substrate. 
  Right: 
   Typical simulation configurations for increasing stiffness parameters
      $L_p/\ell$ and $R_w=4\ell$. Snapshots are taken  in the adsorbed phase
  close to the critical potential strength. 
 The  simulation parameters are $N=400$, $\ell=2b_0$, 
   and $k=100\,k_BT/b_0^2$. 
   The simulation snapshots correspond to the 
    large blue circles in the phase diagram on the left. 
}
      \label{fig:kritpot_washboard}
\end{center}
 \end{figure*}

Fig.\ 
\ref{fig:kritpot_washboard} shows typical simulation snapshots 
 and the phase diagram for the adsorption transition on 
 a washboard surface in $D=2$.
The simulation snapshots 
illustrate the following four distinct 
regimes of characteristic adsorption 
behavior:
\begin{itemize}
\item[a)]  $R_w\gg  L_p\approx \ell$: 
    The concavely curved valleys of the washboard surface support adsorption
    of a flexible polymer in a compact shape.
\item[b),c)]  $R_w> L_p > \ell$: The curvature is negligible on the scale 
  of the persistence length. We find  adsorption to an 
  effectively planar substrate in an elongated shape. 
\item[d)]  $R_w\approx L_p\gg \ell$: The scale of the 
persistence length corresponds to  a single half-sphere.
  The adsorption behavior is similar to 
          adsorption on a single sphere.
\item[e),f)] $L_p > R_w \gg \ell$: The persistence length is larger
 than a half-sphere radius. This results in ``incomplete'' adsorption  
   on the tips of the washboard substrate. 
\end{itemize}
For $L_p\lesssim R_w$ (snapshots a)-d)), we have a ``complete'' adsorption 
into the concavely curved valleys of the washboard structure. 
Because there is only one such valley on the scale of the 
persistence length, the 
adsorption threshold for ``complete'' adsorption 
is well described 
by the previous result (\ref{eq:gc_stiff_R}) for adsorption 
to a single sphere, where we use $R_s=R_w$.

This result is only modified in the regime of 
``incomplete'' adsorption on top of the 
washboard surface  for stiff polymers with $L_p > R_w$. 
In this regime,  bound configurations consist of an alternating sequence of 
short adhered segments 
on top of the half-spheres  and free segments of (projected)
length $4R_w$ between adhered segments. 
To estimate the critical potential strength we calculate the 
free energy difference 
of such an incompletely adsorbed  configuration to the completely 
unbound state.

The free segments between the adsorption points have a partition 
sum $ Z(4R_w) \sim  Z_0(4R_w)  \ell \alpha L_p/(4R_w)^2$.
where $Z_0(4R_w)$ is the unconstrained partition sum of the free 
polymer and $Z(4R_w)$ the partition sum constrained to hit the 
top of the half-circle within a distance $\ell$ and with a 
tangent\cite{Gompper1989,Benetatos2003} $v \le \alpha$. 
The scaling $z^2 \sim L^3/L_p$ and 
$v^2 \sim L/L_p = (z/L_p)^{2/3}$ implies 
$\alpha \sim (\ell/L_p)^{1/3}$.
This results in an entropic free energy loss  per length of 
\begin{align}
   \Delta f_{\rm free} &=   -w_1
        \frac{k_BT}{4R_w} \ln( Z(4R_w)/Z_0(4R_w) ) \nonumber\\
     &= w_1 \frac{k_BT}{4R_w} \ln\left[w_2 \frac{16R_w^2}{L_p\ell} 
    \left(\frac{L_p}{\ell}\right)^{1/3} \right]
\label{Deltaf1}
\end{align} 
with two numerical constants $w_1$ and $w_2$.
Each  short adhered segment of length $\Delta L$ 
on  top of the half-circle within the attractive layer of 
thickness $\ell$ contributes a free energy difference 
$\Delta f \Delta L= (-g+ k_BT/\lambda_{SF})\Delta L$ 
comparable to a segment adsorbed on planar substrate. 
Because there is one segment per length $4R_w$ between
adhered segments
the resulting free energy difference per length
is given by 
\begin{align}
  \Delta f_{\rm ad} &= \frac{\Delta L}{4R_w}
     \left( -g + c_{SF} \frac{k_BT}{\ell^{2/3}L_p^{1/3}}\right)
\end{align}
The condition $\Delta f_{\rm free}+ \Delta f_{\rm ad} <0$ for adsorption 
results in a critical potential strength 
\begin{equation}
   g_c =  c_{SF}\frac{k_BT}{\ell^{2/3}L_p^{1/3}}+
    w_1 \frac{k_BT}{\Delta L}
       \ln\left[ 16w_2 \left(\frac{R_w}{\ell} \right)^2
    \left(\frac{\ell}{L_p}\right)^{2/3} \right]
\label{eq:gc_incomplete}
\end{equation}

For {\it straight} polymers ($T=0$, $L_p \gg R_w$), the length $\Delta L$ 
is calculated from the geometrical relation $R_w^2+\Delta L^2 = (R_w+\ell)^2$,
which gives
$\Delta L = 2\sqrt{R_w\ell} \left(1+{\cal O}(\ell/R_w)\right)$.
In the presence of thermal fluctuations the polymer can use 
an energy $k_BT$ to adapt further to the potential and increase 
the adsorbed length $\Delta L$. 
Equating the thermal angular fluctuations $\Delta \alpha =(\Delta L/L_p)^{1/2}$ 
over a length $\Delta L$ with the  curvature angle 
$\alpha = \Delta L/ R_w$ of the half-circle, we obtain 
$\Delta L = {R_w^2}/{L_p}$.
Both effects should add up to give 
$\Delta L = 2\sqrt{R_w\ell} + {R_w^2}/{L_p}$.
The first contribution dominates for 
larger stiffnesses 
\begin{equation}
 \frac{L_p}{\ell} \gg \left(\frac{R_w}{\ell}\right)^{3/2}.
\label{largeLp}
\end{equation}
For these stiffnesses we can also neglect the first 
 entropic contribution in the result 
(\ref{eq:gc_incomplete}) for the critical adsorption  strength
in eq.\ (\ref{eq:gc_incomplete}) and find 
\begin{align}
  \frac{g_c\ell}{k_BT}  &=  
 w_1 \frac{1}{2\sqrt{R_w/\ell}} 
        \ln\left[ 16w_2 \left(\frac{R_w}{\ell} \right)^2
    \left(\frac{\ell}{L_p}\right)^{2/3} \right]
\label{eq:gc_incomplete1}
\end{align}
which is only logarithmically $L_p$-dependent.

For smaller stiffness $(L_p/\ell) \ll (R/\ell)^{3/2}$ we find 
\begin{align} 
 \frac{g_c\ell}{k_BT}  &=  c_{SF}\left(\frac{L_p}{\ell}\right)^{-1/3}
\nonumber\\
&~~+ w_1 \frac{L_p\ell}{R_w^2} 
      \ln\left[ 16w_2 \left(\frac{R_w}{\ell} \right)^2
    \left(\frac{\ell}{L_p}\right)^{2/3} \right]
\label{weak2}
\end{align}
which is very similar to the complete adsorption result 
as given by the single sphere result 
(\ref{eq:gc_stiff_R}) with  $R_s=R_w$. 
The crossover from complete to incomplete adsorption should 
happen if the  condition (\ref{largeLp}) 
is fulfilled. 
Therefore, incomplete adsorption on top of the washboard only
involves  short straight segments $\Delta L \sim 2\sqrt{R\ell}$
without much curvature in agreement with   the simulation snapshots 
e) and f) in Fig.\ \ref{fig:kritpot_washboard}.

Incomplete adsorption is  different from the adsorption transitions 
discussed in Refs.\ 
\citenum{pierrelouis2008,pierrelouis2011}, where it has been 
proposed that adsorption 
proceeds via the shortening 
of desorbed bridges between segments strongly adsorbed in the 
 concave valleys of the washboard surface.
\Cr{
The main differences  in Refs.\ \citenum{pierrelouis2008,pierrelouis2011}
are (i) the use of a contact adsorption potential in  Refs.\ 
\citenum{pierrelouis2008,pierrelouis2011}, which corresponds to the 
limit  $\ell \approx 0$, (ii) the presence of a tension, which is 
absent in our system, as we do not consider external stretching forces and
(iii) a surface undulation amplitude, which is  small compared 
to the wavelength, where we  consider  a washboard 
structure consisting of half-circles, i.e., the undulation amplitude
equals the wavelength. 
For a contact potential, 
incomplete adsorption on top of the substrate has for zero temperature only been found 
in the presence of tension in Ref.\ \citenum{pierrelouis2008}.
A finite potential range $\ell >0$  as used in the present work 
favors incomplete adsorption because it gives rise to a 
 length $\Delta L>0$ of 
adsorbed segments an, thus, an extensive adhesion energy 
for a  straight rod, i.e., in the limit of infinite stiffness.
For a contact potential, on the other hand, 
a straight rod touches the adhesive 
structure only at single points an the adhesion energy is zero.
Furthermore a
finite potential range allows for thermal fluctuations within the
potential at
non-zero temperatures, also favoring a (incompletely) adsorbed phase.
}

The MC simulation results in the phase diagram in 
Fig.\ \ref{fig:kritpot_washboard}
show good agreement with the analytical 
result (\ref{eq:gc_incomplete1}) for the adsorption threshold 
in the regime of  incomplete adsorption for stiff 
polymers  fulfilling  (\ref{largeLp}).
We can successfully fit the MC data for the  critical potential strength 
(as obtained using the cumulant method) using eq.\ (\ref{eq:gc_incomplete1}) 
with the numerical constants $w_1$ and $w_2$ as   fit parameters.
The resulting values for the 
\Cr{leading order fit parameter $w_1$}
(see  Table \ref{tab:washboard_2d}) are indeed independent of the 
substrate curvature radius $R_w/\ell$. 

\begin{table}[ht]
\begin{tabular}{r|l||c|c|c}
\hline
 $R_w/b_0$&$\ell/b_0$&$R_s/ \ell$&$w_1$&$w_2$\\
\hline
$4 $ & $1$& $4.6{{\pm}}0.1$& $0.532{{\pm}}0.005$& $4.5{{\pm}}0.2$\\
$8 $ & $1$& $8.29{{\pm}}0.09$ &$0.509{{\pm}}0.008$ & $7.5{{\pm}}0.7$\\
$4 $ & $0.5$& $8.78{{\pm}}0.07$& $0.512{{\pm}}0.006$&$ 5.9{{\pm}}0.4$ \\
$8 $ & $0.5$& $15.7{{\pm}}0.1$& $0.47{{\pm}}0.02$& $11{{\pm}}3$\\
\hline
\end{tabular}
\caption{Simulation results for the fit parameters 
$R_s$, $w_1$ and $w_2$ for the interpolation functions
 \eqref{eq:gc_stiff_R} for smaller and \eqref{eq:gc_incomplete1} 
for larger stiffnesses.}
\label{tab:washboard_2d}
\end{table}

For more flexible polymers violating (\ref{largeLp}), fits with the 
 result  (\ref{eq:gc_stiff}) for an adsorbing sphere with 
$R_s$ as fit parameter  work well. All resulting values 
for the effective curvature radii lie in the interval
$[R_w,R_w+\ell]$ as  expected.

MC simulations and our  analytical results  for the adsorption threshold 
show  that adsorption control by the substrate curvature radius $R_w$
is most efficient in the regime 
\begin{equation}
   \frac{L_p}{\ell} \sim  \left(\frac{R_w}{\ell}\right)^{3/2}
\label{eq:min_wash}
\end{equation}
where the critical potential strength exhibits a {\em local minimum} 
as for adsorption on a single sphere. 
If the stiffness is increased such that (\ref{largeLp}) holds, 
we find incomplete adsorption, where the dependence of $g_c$ on 
the curvature radius $R_w$ is much weaker according to eq.\ 
(\ref{eq:gc_incomplete1}). This is also  clearly supported by the 
MC simulation results in the phase diagram 
Fig.\ \ref{fig:kritpot_washboard}.
If the stiffness is much smaller than the minimum 
value (\ref{eq:min_wash}) curvature effects become negligible on the 
scale of the persistence length, and we find the crossover to 
effectively planar adsorption. 
The MC phase diagram  in Fig.\ \ref{fig:kritpot_washboard} clearly 
indicates a {\em window of stiffnesses} for adsorption control
around the local minimum and in 
between two local maxima of the critical potential strength. 
The maximum at small stiffness is located at $L_p/\ell \sim 1$ 
as for planar adsorption, the maximum for large stiffnesses at 
$L_p/\ell \sim  ({R_w}/{\ell})^{3/2}$ as given by the condition 
(\ref{largeLp}) for the crossover between complete and incomplete 
adsorption.
The stiffness window for  adsorption control vanishes if the 
substrate curvature radius $R_w$ approaches the potential range $\ell$, 
as can be seen in the phase diagram  Fig.\ \ref{fig:kritpot_washboard}  
(cyan data points for ${R_w}/{\ell} =2$). 

Our results 
not only apply to the control of the adsorption of polymers on washboard 
surfaces, for example, to control polyelectrolyte adsorption by 
tuning the salt concentration and, thus, the range $\ell$ of the 
adsorption potential. 
Another technologically important application is the 
control of adsorption of graphene sheets to 
adhesive washboard potentials consisting of a sequence of 
  alternating concave and convex 
    half-cylinders. Our results  apply to this 
problem as well, and a transition from incomplete to complete adsorption 
has also been discussed for graphene  sheets \cite{Li2010}.

\subsection{Checkered washboard surface}

Now we want to discuss  washboard substrates in $D=3$ 
spatial dimensions, which is the relevant case for 
applications. We propose the checkered washboard (see Fig.\  \ref{fig:geom})
as a substrate, which allows 
to effectively control adsorption by 
substrate curvature for  $D=3$.
As pointed out above, a washboard substrate consisting  of 
cylinders (see Fig.\  \ref{fig:geom})
does not allow for an adsorption control as the 
polymer can reorient parallel to the cylinders. 
To construct a  washboard structure in three spatial 
dimensions where the polymer cannot simply reorient 
to avoid curvature during adsorption,
we consider  a checkered washboard consisting of 
rectangular subunits described by 
a Cartesian product of two washboard half-circles with a 
height function 
$z_{R_c}(x,y) = {\pm} R_c \sqrt{1-(x/R_c)^2} \sqrt{1-(y/R_c)^2}$ ($x,y \in
[{-}R_c,R_c]$)
with a radius $R_c$. 
These rectangular subunits differ from half-spheres but exhibit the 
same curvature at the tips of the surface. 
We expect adsorption on the checkered washboard for  $L_p\lesssim R_c$ 
to be similar to adsorption to the $D=2$ 
washboard surface for $L_p\lesssim R_w$ because there is only 
a single valley or top on the scale of the persistence length.
We expect adsorption to be well described by the 
result (\ref{eq:gc_stiff_R}) for adsorption 
to a single sphere, where we use $R_s=R_c$.
For $L_p > R_c$, however, the alternating checkered structure 
of valleys and top  will modify the adsorption behavior.

\begin{figure}[th]  
      \includegraphics[width=0.45\textwidth]{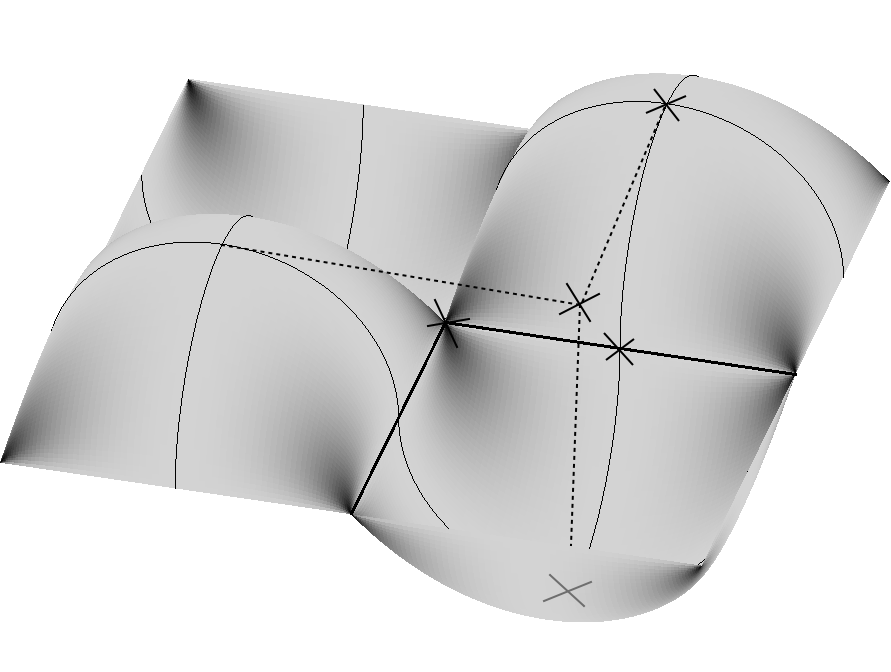}
      \caption{
    The checkered washboard substrate. Brightness
        codes for the error in the potential range implementation 
   (see text). Dark color indicates larger errors.
    Crosses indicate different attachment points 
  of the polymer used in simulations. 
  Thick black lines indicate the square lattice of lines of 
  preferential adsorption.  
  }
      \label{fig:wobble}
\end{figure}

In the simulation, we need an effective approximate method 
to implement  the attractive range: 
For the checkered washboard structure we do not determine 
the normal distances $n$ from eq.\ (\ref{eq:Vz})  exactly, but 
approximate the attractive region via $z_{R_c} < z < z_{R_c+\ell}$.
The relative error is worst at the corners
 and can be estimated as $\frac{1}{\ell}(z_{R_c +
  \ell}-z_{R_c})-1 \approx 0.155$ for $\ell \ll R_c$. 
The checkered substrate and the numerically determined  error 
for the attractive range 
is illustrated in \ref{fig:wobble}.

\begin{figure*}[th]  
\includegraphics[width=0.95\textwidth]{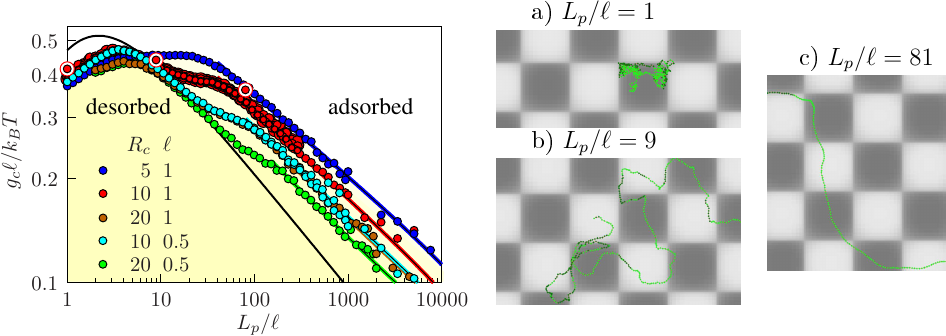} 
      \caption{ 
 Left:
    Phase diagram for a checkered substrate as obtained from MC
        simulations: double logarithmic plot of the dimensionless critical
        potential strength $g_c\ell/k_BT$ as a function of the dimensionless
        stiffness parameter $L_p/\ell$ as obtained with the cumulant
        method.    The solid lines show 
     fits using  eq.\  (\ref{eq:gc_cheq}) 
     (using $a_1$ and $a_2$ as fit parameters)
    for larger stiffnesses $L_p/\ell$. 
        Simulation parameters are $N=200$ and $k=100\,k_BT/b_0^2$.  
  The red data points for $R_c/\ell=10$ are results for four 
  different attachment points as shown in Fig.\ \ref{fig:wobble}.
  Right:  Typical simulation configurations  for adsorption on 
      a checkered washboard for increasing stiffness parameters
      $L_p/\ell$. Snapshots are taken  in the adsorbed phase
  close to the critical potential strength. 
The simulation snapshots correspond to the 
    large red circles in the phase diagram on the left. 
}
      \label{fig:kritpot_cheqwash}\label{fig:kritpot_cheqwash_snapshots}
    \end{figure*}

The MC simulation results in the phase diagram 
 Fig.\ \ref{fig:kritpot_cheqwash} 
show that the  adsorption threshold 
can be  controlled by the substrate curvature radius $R_c$ 
in a similar fashion as for the washboard substrate in $D=2$ 
dimensions for $L_p \lesssim R_c$. 
However, we find two characteristic differences for stiffer polymers
$L_p > R_c$:
(i) As opposed to the $D=2$ washboard substrate we do not 
find a local minimum of the critical potential strength but 
 we find a single broad maximum or shoulder in the regime 
${L_p}/{\ell} \sim  \left({R_c}/{\ell}\right)^{3/2}$
for $R_c/\ell \lesssim 10 b_0$. 
(ii)  The critical potential strength
exhibits a remarkably stronger dependence on the substrate 
curvature radius $R_c$ for larger polymer stiffness as 
compared to the $D=2$ washboard. For large stiffnesses
$L_p \gg R_c$ the critical potential decreases with increasing
$R_c$ as can be seen in the phase diagram Fig.\
\ref{fig:kritpot_cheqwash}.

In order to illustrate the different adsorption mechanism 
underlying these characteristic differences we also present 
 typical simulation snapshots in Fig.\ \ref{fig:kritpot_cheqwash_snapshots}.
Whereas we find ``incomplete'' adsorption on the 
 substrate  tips for  the washboard substrate in $D=2$  (see simulation 
snapshots \ref{fig:kritpot_washboard} e),f)) for large stiffnesses, 
the polymer 
preferentially adsorbs ``between'' the tops and valleys, 
i.e., along the straight boundaries of  the square subunits
for  a checkered washboard in the regime of  large stiffnesses $L_p \gg R_c$
as illustrated by the simulation snapshot 
Fig.\ \ref{fig:kritpot_cheqwash_snapshots} c). 
These boundaries are  the equal height lines 
$z(x,y)=0$ and  form a plane-filling square lattice of lines for preferential
adsorption with
in-plane lattice constant $2R_c$,
see thick black lines in Fig.\ \ref{fig:wobble}).
No out-of-plane curvature is required for adsorption
onto this lattice of straight lines. 
The  regions around each straight line segment is almost vertically 
tilted but locally flat such that for each adsorbing segment 
our results for adsorption on a flat substrate will apply.
In order to connect between 
neighboring straight lines of the square lattice of adsorption sites, 
 a stiff
polymer has to  run through the discrete square lattice
of intersection points where four subunits meet.

This restricts the thermal  fluctuations of the adsorbed polymer
parallel to the surface and 
 implies an additional entropy cost during adsorption
because in-plane configurations are restricted. 
This additional entropy cost can be used to control the 
adsorption via the distance $2R_c$ between the 
lattice points. 
This entropy cost  can be  estimated in an analogous manner as 
the entropy cost for  adsorption on the discrete array of tips 
for the $D=2$ washboard. Adapting the corresponding 
estimate (\ref{Deltaf1}) for the entropic free energy cost 
appropriately, we obtain 
   $\Delta f_{\rm ch}
     = a_1 ({k_BT}/{2R_c}) \ln\left[a_2 ({4R_c^2}/{L_p\ell} )
    \left({L_p}/{\ell}\right)^{1/3} \right]$.
The additional free energy cost leads to a corresponding 
shift of the critical potential strength for adsorption as 
compared to a flat substrate, $ g_c  =  g_c(R_s=\infty) + \Delta f_{\rm ch}$,
or 
\begin{align}
  \frac{g_c \ell}{k_BT} &=  c_{SF} \left(\frac{L_p}{\ell}\right)^{-1/3}
  \nonumber\\
   &~~~  +  a_1 \frac{\ell}{2R_c} \ln\left[4a_2 \left(\frac{R_c}{\ell} \right)^2
    \left(\frac{\ell}{L_p}\right)^{2/3}  \right].
\label{eq:gc_cheq}
\end{align} 
This result predicts an offset in the  critical potential strength, 
which increases for smaller substrate curvature radii  $R_c$. It  
depends only  logarithmically  on the  
polymer stiffness $L_p$. The result (\ref{eq:gc_cheq}) exhibits 
a   different  dependence of  the critical potential strength
 on the substrate 
curvature radius $R_c$ for larger polymer stiffness as 
compared to the result (\ref{eq:gc_incomplete1}) for the 
$D=2$ washboard.

The MC simulation results in the phase diagram in 
Fig.\ \ref{fig:kritpot_cheqwash}
confirm  both of these predictions qualitatively. 
We can successfully fit the MC data for the  critical potential strength 
(as obtained using the cumulant method) in the 
stiff regime $L_p \gg R_c$ using eq.\ (\ref{eq:gc_cheq}) 
with the numerical constants $a_1$ and $a_2$ as   fit parameters,
see Table  \ref{tab:wobbel}. The fit parameters are indeed roughly 
independent of the 
substrate curvature radius $R_c/\ell$.

\begin{table}[ht]
\begin{tabular}{r|l||c|c}
\hline
 $R_c/b_0$&$\ell/b_0$&$a_1$&$a_2$\\
\hline
$5 $ & $1$&  $0.25{\pm}0.01$& $73{\pm}15$\\
$10 $ & $1$& $0.38{\pm}0.04$ & $16{\pm}10$\\
$20 $ & $1$&  $0.42{\pm}0.04 $&$ 12{\pm}7$ \\
$10 $ & $0.5$&  $0.46{\pm}0.05 $& $10{\pm}7$\\
$20 $ & $0.5$&  $0.42{\pm}0.10$& $24{\pm}43$\\
\hline
\end{tabular}
\caption{Simulation results for the
fit parameters $a_1$ and $a_2$ for larger stiffness.
We use eq.\ \eqref{eq:gc_cheq} as fit function.
}
\label{tab:wobbel}
\end{table}

The MC simulations and the scaling argument for the adsorption 
threshold show that, using a checkered washboard structure,
  adsorption can be controlled rather effectively by the substrate 
curvature radius $R_c$ in the entire stiff limit, where 
$L_p \gg R_c$. As opposed to the washboard substrate in $D=2$ 
discussed in the previous section,  where a window of 
stiffnesses for effective adsorption control emerged, adsorption 
control by a checkered washboard in $D=3$ 
is still effective  for large polymer stiffnesses.

\section{Discussion of experimental results}

\Cr{
Many experimental results are available for polyelectrolyte 
adsorption or complexation. For polyelectrolyte adsorption 
the potential strength $g/k_BT \sim  \sigma \tau l_B/\tilde \kappa$ 
can be controlled by the surface charge $\sigma$. 
Experimentally, the  critical surface charge $\sigma_c$ for polyelectrolyte 
adsorption can be  measured as a function of the  inverse 
 Debye screening length 
$\tilde \kappa = (8\pi l_Bc)^{1/2}$, which is controlled by the salt
concentration, resulting in a relation\cite{Zhang2000,Cooper2006}
$\sigma_c\propto \kappa^a$ 
with a characteristic exponent $a$. 
Using our results for adsorption to a planar surface, 
$g_c\ell/k_BT\sim (L_p/\ell)^{-1/3}$ in the stiff regime, see 
eq.\ (\ref{eq:gc_stiff})  and 
$g_c\ell/k_BT\sim L_p/\ell$ in the flexible regime, 
see eq.\ (\ref{eq:gc_flexible}), and   using 
a potential range $\ell \sim 1/\tilde\kappa$ given by the Debye screening
 length, we find 
\begin{align}
   \sigma_c &\propto {\tilde \kappa}^2 (L_p/\ell)^{-1/3} ~(\mbox{stiff}),
\nonumber\\
  \sigma_c &\propto {\tilde \kappa}^2 (L_p/\ell) ~(\mbox{flexible}).
\label{eq:sigmac}
\end{align}
 According to Odijk \cite{Odijk1977}, 
 Fixman and Skolnick \cite{Skolnick1977}
 the polyelectrolyte persistence length is given by
 the sum of the bare mechanical persistence length $L_{p,\text{mech}}$
  and an electrostatic contribution due to the electrostatic 
self-repulsion of the polymer,  
 $L_p = L_{p,\text{mech}}+ l_B\tau^2/4\tilde \kappa^2 $.
For a mechanically dominated  persistence length we have 
$L_p/\ell \propto \tilde\kappa$, whereas we have 
$L_p/\ell \propto \tilde\kappa^{-1}$ for an electrostatically dominated 
persistence length. 
Combining this with (\ref{eq:sigmac}), we obtain 
four possible scaling behaviors $\sigma_c \propto \kappa^a$ with 
exponents
\begin{align*}
    a&= 5/3 ~~\mbox{mechanical stiffness, stiff limit}\\
    a&= 3 ~~\mbox{mechanical stiffness, flexible limit}\\
    a&= 7/3 ~~\mbox{electrostatic stiffness, stiff limit}\\
    a&= 1 ~~\mbox{electrostatic stiffness, flexible limit},
\end{align*}
which characterize polyelectrolyte adsorption
onto planar surfaces. For curved surfaces such as a sphere, there 
are additional corrections in the stiff limit according to 
eq.\ (\ref{eq:gc_stiff_R}), such that 
$\sigma_c \propto \tilde\kappa^{0}(L_p/\ell)$, which leads to 
$\sigma_c \propto \tilde\kappa$ ($a=1$) for mechanical stiffness
and $\sigma_c \propto \tilde\kappa^{-1}$ ($a=-1$) for 
electrostatic stiffness.
}

\Cr{
The experimental results on polyelectrolyte adsorption onto 
proteins and micelles in Ref.\ \citenum{Cooper2006} agree best with 
an exponent $a=1$ corresponding an electrostatic stiffness and the flexible 
limit for a planar substrate, which is reasonable in view of the  
short mechanical persistence
lengths of the studied polyelectrolytes and with protein radii larger 
than these persistence lengths.  
}

\section{Conclusion}

We studied adsorption of semiflexible polymers on 
planar and curved substrates. Using extensive Monte-Carlo 
simulations and analytical arguments we showed that the 
interplay between three characteristic length scales --
(i) the persistence length  $L_p$ characterizing polymer stiffness, 
(ii) the range $\ell$ of the 
attractive adsorption potential, and 
(iii) a characteristic curvature radius $R$ of the surface structure
-- allows to control the adsorption threshold for 
semiflexible polymers  effectively.

For a planar adsorbing surface we find a maximum of the critical 
potential strength for adsorption, i.e., a ``minimally 
sticky'' surface if the persistence length matches the potential range,
 $L_p\sim \ell$. 
We presented two scaling functions which can quantitatively describe 
the crossover between flexible and stiff limit and the location 
of the maximum in the critical potential strength in agreement with 
MC simulations, see Fig.\ \ref{fig:kritpot_flach}. 
We also quantified the exact asymptotic  value of the 
critical potential strength for adsorption in the stiff limit, 
see eqs.\ (\ref{eq:gc_stiff}) and (\ref{eq:cSF}). 
\Cr{
Our results can also resolve  contradictory 
statements in the literature. Simulations of 
adsorbing semiflexible polymers in Refs.\ 
\citenum{kramarenko96,sintes01} found 
a critical potential strength decreasing with stiffness:
these simulations probed the stiff limit with a persistence length 
exceeding the potential range. 
On the other hand, simulations of adsorbing polyelectrolytes found a 
critical potential strength  increasing with electrostatic 
stiffness \cite{Kong1998}:
these simulations probed the flexible limit with a 
small electrostatic persistence length. 
}

For an  adsorbing sphere of radius $R_s$ the critical 
potential strength is increased for large persistence lengths
by the additional bending energy involved in adsorption to a 
curved object. 
This results in an ``optimally sticky'' 
adsorbing sphere if  the condition (\ref{eq:min_sphere}),
$L_p/\ell \sim (R_s/\ell)^{3/2}$ holds. 
MC simulation results in Fig.\ \ref{fig:kritpot_sphere}
confirm this result. 

For a washboard surface consisting of cylinders, adsorption control 
is not possible because the polymer can orient parallel to the 
cylinders and avoid additional bending during adsorption. 
The situation is different if we restrict the polymer to 
a two-dimensional plane perpendicular to the cylinders. 
 For such a washboard surface in two spatial dimension,
we find an additional crossover 
from complete adsorption to an incomplete adsorption on the 
tips of the surface structure for large persistence lengths. 
The condition $L_p/\ell \gg  (R_w/\ell)^{3/2}$, see eq.\ (\ref{largeLp}),
 characterizes the regime of incomplete adsorption. 
The adsorption threshold can be controlled by the substrate 
curvature in a 
polymer stiffness window given by 
$1 \lesssim  L_p/\ell \lesssim  ({R_w}/{\ell})^{3/2}$,
as also shown by MC simulations, see Fig.\ \ref{fig:kritpot_washboard}
with an ``optimally  sticky'' curvature radius for
$L_p/\ell \sim  (R_w/\ell)^{3/2}$, see eq.\ (\ref{eq:min_wash}).

Checkered washboard structures offer a  possibility
to control  adsorption also in three spatial dimensions. 
On the one hand, the checkered washboard suppresses 
polymer reorientation as for a cylindrical three-dimensional washboard. 
On the other hand, it suppresses  an incomplete adsorption 
on the tips of the substrate as it occurs for
the two-dimensional washboard. For a checkered washboard,
stiff polymers rather adsorb 
in the locally flat straight boundaries between tops and valleys 
of the structure, which form a square lattice. 
The driving force for 
adsorption control on this type of substrate 
is the restriction to the square array of straight adsorption 
lines within the adsorbing plane  rather than the 
control of the out-of-plane curvature. 
As a result, there is 
no polymer stiffness window for adsorption control, 
 adsorption control always effective for large 
stiffnesses $L_p\gg R_c$.

We expect similar results for other, eventually more irregularly 
curved substrates, where  the substrate curvatures can be 
adjusted to the polymer persistence length and the potential 
range to create sticky or non-sticky regions. 
Our results for the washboard surfaces demonstrate that 
 not only the out-of-plane curvature 
will be important but also the shape and curvature of 
locally flat preferred lines of adsorption sites, which are 
given by the lines of equal substrate height.

In this work, we neglected all effects from self-avoidance. 
Generally, we expect this to be a good approximation as long as 
typical polymer configurations are elongated and 
contain only few loops, as it is the case for sufficiently stiff polymers. 
For the adsorption of semiflexible polymers this is 
typically the case in the stiff regime $L_p/\ell >1$. 
However, we expect pronounced corrections in the flexible limit
 $L_p/\ell <1$. On the other hand, it is well-known \cite{degennes}
that the critical potential strength 
 adsorption of a self-avoiding chain on a planar substrate  
is $g_c \sim k_BT (L_p/\ell)^{5/3}$  rather than 
$g_c \sim k_BT (L_p/\ell)$ in the absence of self-avoidance 
(see eq.\ (\ref{eq:gc_flexible})). Therefore, 
the critical potential remains an  {\em increasing} function of 
polymer stiffness.  Consequently, we  expect the most 
 important features of the phase diagrams,
such as the maximum of the critical potential strength 
as a function of polymer stiffness for adsorption on a planar 
substrates, to be similar also in the presence of 
self-avoidance. This remains to be investigated quantitatively
in future work.

%

\setcounter{section}{0}
\setcounter{equation}{0}
\setcounter{figure}{0}
\renewcommand{\theequation}{{S\arabic{equation}}}
\renewcommand{\thefigure}{{S\arabic{figure}}}

\section*{Supplemental Material for ``Controlling adsorption of semiflexible polymers on planar 
and curved substrates''}

The supplemental material is structured as follows:
   (i) Discussion of different interpolation functions $I(x)$. 
  (ii) Analytical transfer matrix calculation for planar substrate
  in the stiff limit, which gives the critical potential strength 
   for a square-well potential 
   in the limit of small potential range. 
 (iii) Details of the cumulant method and 
  finite size scaling procedure used to obtain the 
   critical potential strength in Monte-Carlo simulations. 
  (iv) Results for the critical correlation length exponent $\nu$ 
   of the adsorption transition for a planar substrate. 
 (v) Additional simulation snapshots in the desorbed state. 

\section{Adsorption to a planar substrate}

\subsection{Interpolation function of Deng {\it et al.}}
\label{App_Ideng}

\Cr{
In Ref.\ \citenum{sdeng10}, Deng {\it et al.} also 
use  an  interpolation function 
 to describe the crossover of the critical potential strength 
$g_c$ for adsorption  between the stiff and flexible 
limit. 
They measure the 
 critical potential strength  in units of $k_BT/2L_p$ 
rather than $k_BT/\ell$ as in (11); 
 the critical potential strength does not exhibit a maximum if 
measured  in these units. 
In Ref.\ \citenum{sdeng10},
an interpolation 
$g_c (2L_p/k_BT) = \tilde{I}(2L_p/\ell)$
with a scaling function 
\begin{equation*}
\tilde{I}_{\rm Deng}(x) = \frac{(c_F/2) x^2}{[C_2 x^2+ C_1 x+1]^{2/3}}
\end{equation*}
is used with two fit parameters $C_1\simeq 0.94$ and $C_2\simeq 0.38$,
which are determined from numerical transfer matrix calculations. 
}

\Cr{
Comparing the two scaling forms, we find 
the relation
\begin{equation*}
   \tilde{I}(x) = x I(x/2)~,~~I(x) = \tilde{I}(2x)/2x
\end{equation*}
Consequently the scaling function proposed in Ref.\ \citenum{sdeng10}
corresponds to 
\begin{align}
     I_{\rm Deng}(x) &= \frac{c_F  x}{[4C_2x^2+2C_1x+1]^{2/3}}
\label{eq:Ideng}  
\end{align}
}

\Cr{
This scaling function does not  
obey the constraints (13) and (14) 
listed in the main text:
\begin{itemize}
\item[(i)]
 The numerical prefactor  $c_{SF}$ has been treated as a fit parameter
 because an analytical result was not available.
 Using their fit results,  
we find $c_{SF} = 2^{-4/3} c_F C_2^{-2/3} \simeq 0.619$,
 which is  close but smaller than  our
analytical result $c_{SF} \simeq 0.929$. 
The reason is a different determination of the critical potential 
strength from simulations. Deng {\it et al.\ } use an 
 extrapolation of adsorbed fraction, which is the 
first cumulant of the adsorption energy, to zero, whereas we mainly use 
the third cumulant. 
\item[(ii)]
 The constraint 
(14) regarding the correct next to leading order 
asymptotics in the stiff limit has  not been applied. 
The scaling function (\ref{eq:Ideng}) has the asymptotics 
$I_{\rm Deng}(x) = (c_F/2C_2^{2/3}) x^{-1/3} + {\cal O}(x^{-4/3})$
for $x\gg 1$, which differs from the analytical prediction 
(14). 
\end{itemize}
}

\subsection{Alternative   interpolation function  $I(x)$.}
\label{App_I}

In this appendix, we  use the scaling argument for the 
deflection length $\lambda$ and  $g_c \sim k_BT/\lambda$ to 
motivate a functional form of the  interpolation 
function $I(x)$. The argument is based on a result for the 
thermal displacement $\langle z^2 \rangle(L)
 \equiv \langle (z(L)-z(0))^2 \rangle$ of a free worm-like chain
 in the direction perpendicular to the average preferred 
orientation in $x$-direction.

\begin{figure}
      \includegraphics[width = 0.45\textwidth]{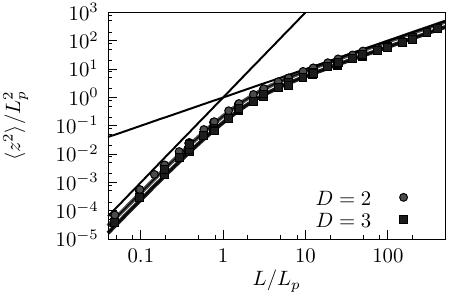}
      \caption{
    MC data for ${\langle z^2\rangle(L)}/{L_p^2}$ as a function of 
    $L/L_p$ for a free worm-like chain in $D=3$ and $D=2$ dimensions confirms
  the scaling form  ${\langle z^2\rangle(L)}/{L_p^2}=c  f_{d_1}(L/L_p)$, 
see eq.\  (\ref{eq:z2}), with $d_1 = 0.76 {\pm} 0.08$ and a 
prefactor $d_2= 0.61 {\pm} 0.01$ in three and 
 $d_2= 0.94 {\pm} 0.01$, $d_1 = 0.84 {\pm} 0.09$ in two dimensions (for  $L/b_0=
        50,100,200,300,400$ and $L_p/b_0 = 2,4,8,\ldots,1024$).
   We fix the first tangent  to have a well-defined  $z$-direction.
}
 \label{fig:zquad}
\end{figure}

For a free worm-like chain in two dimensions ($D=2$), the 
thermal displacement $\langle z^2 \rangle(L)$
can be calculated analytically. 
In $D=2$,  we can parametrize the configuration 
by a single angle $\theta(s)$ by 
$\bt(s) = \partial_s \br(s) = (\cos \theta(s), \sin\theta(s))$.
The angular correlations are 
\begin{align*}
  \langle (\theta(s)-\theta(s'))^2 \rangle &= 
    \frac{k_BT}{\kappa}|s-s'| = \frac{2|s-s'|}{L_{p,D}}
\end{align*}
and 
\begin{align*}
  \langle \theta(s)^2 \rangle &= 
    d_1 L/L_{p,D}
\end{align*}
with a numerical constant $d_1$. 
\begin{widetext}
The angular correlations can be used to calculate 
\begin{align*}
  & \langle (z(L){-}z(0))^2 \rangle = 
    \int_0^L \!\!\!\!ds_1\int_0^L \!\!\!\!ds_2 
   \langle \sin\theta(s_1) \sin\theta(s_2) \rangle 
   \\
  &~= \int_0^L ds_1\int_0^L ds_2 \frac{1}{2}\left(
     e^{-\frac{1}{2} 
            \langle (\theta(s_1) - \theta(s_2))^2\rangle} -
    e^{-\frac{1}{2} 
          \langle (\theta(s_1) + \theta(s_2))^2\rangle} 
  \right)
\\
  &~= \int_0^L ds_1\int_0^L ds_2 \frac{1}{2}\left(
     e^{-\frac{1}{2} 
            \langle (\theta(s_1) - \theta(s_2))^2\rangle} -
    e^{-\langle \theta^2(s_1)\rangle - \langle \theta^2(s_2)\rangle
       +\frac{1}{2} 
          \langle (\theta(s_1) - \theta(s_2))^2\rangle } 
  \right)
\\
 &~= \int_0^L ds_1\int_0^{s_1} ds_2 \left(
     e^{- (s_1-s_2)/L_{p,D}} -
    e^{-2d_1L/L_{p,D}}     e^{ (s_1-s_2)/L_{p,D}}
  \right)
\\
 &~= L_{p,D}^2 \left( \frac{L}{L_{p,D}} -1 +e^{-L/L_{p,D}}\right) 
- e^{-2d_1 L/L_{p,D}} L_{p,D}^2 \left( -\frac{L}{L_{p,D}} -1 +e^{L/L_{p,D}}\right)
  \\
   &~\equiv  L_{p,D}^2 f_{d_1}(L/L_{p,D})
\end{align*}
\end{widetext}
Although this  calculation is difficult to adapt to
 three spatial dimensions, we expect a
similar behavior for $D=3$ with an eventually different numerical 
prefactor $d_2$:
\begin{equation}
 \frac{\langle z^2\rangle(L)}{L_{p,D}^2}=d_2  f_{d_1}(L/L_{p,D}).
 \label{eq:z2}
\end{equation} 
Because of ${\langle z^2\rangle(L)} 
= \langle (\br(L)-\br(0))^2 \rangle /D$  for a free worm-like chain
in the flexible limit $L\gg L_{p,D}$,
we expect $d_2 = 2/D$. 
Simulation results for a free
SHC in $D=3$, which we present in Fig.\ \ref{fig:zquad},
confirm the scaling form (\ref{eq:z2})  with $d_1 = 0.76 {\pm} 0.08$ and a 
prefactor $d_2= 0.61 {\pm} 0.01$ close to the expectation $d_2 = 2/3$.
For the results in $D=2$ we get $d_2= 0.94 {\pm} 0.01$, 
which is close to $d_2=1$, and $d_1 = 0.84 {\pm} 0.09$. 
Because the scaling form is very insensitive 
to variation of  $d_2$ for $L/L_{p,D} \gg 1$, we determine 
the parameters $d_1$ and $d_2$ only with values $L/L_{p,D} \leq 10$.


Therefore, the  inverse function $f^{-1}_{d_1}(x)$ can be used 
 to solve the condition  ${\langle z^2\rangle(\lambda)}/{L_{p,D}^2}=
d_2 f_{d_1}(\lambda/L_{p,D}) = \ell^2/L_{p,D}^2$ for 
the deflection length $\lambda$. This suggests a critical potential 
strength  
$g_c = (k_BT/\ell) I\left(L_{p,D}/\ell\right)$ with 
a scaling function 
\begin{equation}
  I(x)   =     \frac{1}{xf^{-1}_{d_1}(x^{-2}/d_2)}
\label{eq:fit2}
\end{equation}
with only two free parameters $d_1$ and $d_2$. 
In the flexible limit $x=\lambda/L_{p,D} \gg 1$, we use  $f_{d_1}(x) \approx x$,
in the stiff limit $x=\lambda/L_{p,D} \ll 1$, we have 
$f_{d_1}(x) \approx (d_1-1/3)x^3$. 
The choices 
$d_2=c_F$ and  
$d_1  =  (c_{SF}/c_F)^3 + 1/3$
 will reproduce the known flexible and stiff 
limits. In contrast to the interpolation function 
 from eq.\ (15), the  function $I(x)$ in eq.\ 
(\ref{eq:fit2}) contains only two free parameters. 
Therefore, the maximum of interpolation function $I(x)$ 
is already determined by $d_1$ and $d_2$. 

We have determined the fit parameters $d_1$ and $d_2$ from 
the MC simulation results for  the critical potential strength
both by the cumulant method and finite size scaling and 
both in $D=2$ and $D=3$. The theoretical expectation
$d_2/c_F=1$ and  
$\frac{d_2}{c^3_{SF}}(d_1{-}\frac{1}{3})=1$ for the parameters 
 $d_1$ and $d_2$ agrees reasonably well with the simulation 
results. 

\begin{table}
{\small
    \begin{tabular}{c||ccc}
\hline
data set  & 
  $\frac{d_2}{c^3_{SF}}(d_1{-}\frac{1}{3})$  & $d_2/ c_{F}$ & $\text{max}(I)$\\ 
\hline
theory(D=3) &  $1$ & $1$&$1.84$\\
cumulant      &   $1.21 {\pm} 0.04$ &  $0.82 {\pm} 0.01$ & $2.47$ \\
finite size   &  $1.0{\pm}0.1$  & $0.54{\pm}0.02$  & $3.36$ \\
theory(D=2) &   $1$  & $1$& $1.26$\\
cumulant  &    $1.5{\pm}0.1$  &   $0.60 {\pm} 0.02$
    & $1.24$  \\
finite  size   &  $1.19 {\pm} 0.07$ &  $0.35{\pm}0.01$      & $1.92$  \\
\hline
\end{tabular}
}
\caption{Simulation results for the fit  parameters
   $d_1$ and $d_2$  for the  interpolation
  function $I(x)$ from eq.\ (\ref{eq:fit2}) in comparison 
  with  theoretical expectations. 
  The maximum value of the resulting interpolation function
  is given for comparison. 
  All fits for the cumulant method are performed for 
MC data from simulations with 
 $N=200$, $\ell=2b_0$, $k=1000\,k_BT/b_0^2$.
  For the analysis of simulation data we use the cumulant method 
  or finite size scaling as explained in the text. 
}
 \label{tab:fits_d}
\end{table}

In addition, the scaling argument leading to eq.\  (\ref{eq:fit2})
strongly suggests a constraint on the  functional form 
of the scaling function $I(x)$ for the critical potential strength $g_c$:
The asymptotics for the stiff limit shows that the scaling function $f_{d_1}(x)$
has a series expansion  $f_{d_1}(x) =  x^3 g(x)$ with some analytical 
function $g(x)$ with $g(0) \neq 0$. Therefore, the inverse function 
should have a functional  form 
$f_{d_1}^{-1}(y)  = y^{1/3} \tilde{g}(y^{1/3})$ with another analytical 
function $\tilde{g}(x)$ with $\tilde{g}(0)\neq 0$. 
It follows from  eq.\  (\ref{eq:fit2}) that 
 the scaling function $I(x)$  should have an 
asymptotic form  
\begin{equation}
  I(x) \sim \frac{x^{-1/3}}{\tilde{g}({\rm const}\, x^{-2/3})}~~\mbox{for}~x\gg 1.
\label{eq:constraint_app}
\end{equation}
Our first choice  $I(x) \sim c_1 x^{-1/3}(c_2 + c_3x^{-2/3} +
x^{-4/3})^{-1}$, see eq.\ (15), 
 also fulfills  this constraint, whereas the scaling function used in 
Ref.\ \citenum{sdeng10} does not meet this constraint.

\subsection{Analytical transfer matrix calculation 
  in the semiflexible limit}
\label{App_II}

Using the transfer matrix method in the weakly bent or stiff limit 
$L_p \gg \ell$
we will give an analytical derivation  of the 
 critical adsorption strength for weakly bent semiflexible polymers
for a planar surface and a short-range adsorption potential,
i.e., determine the numerical prefactor $c_{SF}$ in (8)
analytically.

In the following we measure all length scales in Kuhn 
lengths $2L_p= 2\kappa/k_BT$
 and  all energies in $k_BT$, i.e., we replace
\begin{equation}
z \to \frac{z}{2L_p},~ \ell \to \frac{ \ell}{2L_p},~
L \to \frac{L}{2L_p},~ g \to \frac{2L_p}{k_BT}g.
\label{eq:rescaling}
\end{equation}
In the stiff limit we have $\ell \ll 1$  in rescaled units. 
We consider  the  restricted partition sum 
$Z(z,v,z_0,v_0,L)$ of a semiflexible polymer of length $L$ with 
initial point $z(0) = z_0$, initial tangent $\partial_xz(0)=v_0$,
end point $z(L)=z$, and end tangent $\partial_xz(L)=v$ in the
Monge representation (4) appropriate 
for a weakly bent polymer.
The restricted partition sum $Z(z,v,z_0,v_0,L)$ fulfills a 
transfer matrix equation of the Klein-Kramers type \cite{sfreed}
\begin{equation}
    \partial_L Z = \left( - v\partial_z  
       + \partial^2_v-V(z)\right) Z
\end{equation}
with boundary condition $Z(z,v,z_0,v_0,0) = \delta(z-z_0)\delta(v-v_0)$
at $L=0$. 

For an adsorbed polymer we make the Ansatz
$Z(x,v,z_0,v_0,L) \sim  Z_E(z,v) e^{-EL}$
where $E = \Delta f <0$ is the adsorption free energy per length 
of the polymer, i.e., the free 
energy difference of the adsorbed state as compared 
to the free state ($V=0$).
We approach the desorption transition 
for $E\nearrow 0$. 
 The ``stationary'' restricted
partition function $Z_E(z,v)$ (which we normalize according to 
$\int dz\int dv Z_E(z,v)Z_E(z,-v)=1$)  fulfills\cite{sMaggs1989,sGompper1989}
\begin{equation}
      -EZ_E =\left( - v\partial_z  
       + \partial^2_v-V(z)\right)Z_E. 
  \label{eq:sDGL} 
\end{equation}
In general we obtain a complete spectrum of solutions for energy 
eigenvalues $E_n$ with a ground state energy $E_0$. The 
solution $Z(z,v,z_0,v_0,L)$ satisfying the boundary conditions 
at $L=0$ is obtained by summing over all solutions,
$Z(z,v,z_0,v_0,L) = \sum_n Z_{E_n}(z,v)Z_{E_n}(z_0,-v_0) e^{-E_nL}$.
On length scales $L\gg \xi = 1/|E_0|$ exceeding the 
correlation length $\xi$ of the adsorption transition, 
the ground state dominates and 
\begin{equation}
Z(z,v,z_0,v_0,L) \approx Z_{E_0}(z,v)Z_{E_0}(z_0,-v_0) e^{-E_0L}.
\end{equation}
The ground state  partition function $Z_{E_0}(z,v)$ contains 
the information about the segment distribution 
$c(z,v) \sim Z_{E_0}(z,v)Z_{E_0}(z,-v)$ of a polymer segment 
in the adsorbed state. 
The ground 
state energy $E_0$ determines the free energy of adsorption 
$\Delta f= E_0<0$ and the correlation length of the 
adsorption transition via $\xi = 1/|E_0|$. 
The condition $E_0=0$ determines the 
critical potential strength $g_c$ for adsorption. 
The partition function 
 $Z_0(z,v)$ at $E_0=0$ gives the critical segment distribution.  
Our main aim will be to determine $g_c$ from the 
condition $E_0=0$ in the following.   
Scaling properties of $Z_{E_0}(z,v)$ and $Z(z,v,z_0,v_0,L)$ have already 
been discussed in Refs.\ \cite{skierfeld03,sKierfeld2005a}.

In order to calculate the ground state energy $E$ (we leave out to 
subscript ``0'' in the following) and the corresponding 
``stationary'' restricted partition function $Z_E(z,v)$, 
we first consider the region $z>\ell$
{\em outside} the potential range, where 
 $V(z)=0$ and we can 
separate the $z$-dependence for the adsorbed state using 
$Z_E = e^{-\alpha z}\Psi_{\alpha,E}(v)$, because the operators $\partial_z$ and
$v\alpha + \partial^2_v$ commute. 
The function $\Psi_{\alpha,E}(v)$ satisfies 
\begin{equation*}
 (\alpha v +\partial^2_v) \Psi_{\alpha,E} = -E \Psi_{\alpha,E},
\end{equation*}
(analogous to the Schr{\"o}dinger equation of a 
 quantum particle in an electric field $\alpha$), 
 which gives
\begin{equation*}
      \Psi_{\alpha,E} (v) = \alpha^{-1/6} 
   \Airy{- \left( v\alpha^{1/3} + E \alpha^{-2/3} \right) },
\end{equation*}
where $\Airy{x}$ is the Airy function \cite{sabramowitz1972}. 
The ground state solution for $z>\ell$ 
has to be a linear combination of the 
eigenfunctions of $\partial_z$ and $\alpha v +\partial^2_v$
\begin{equation}
      Z_E(z,v) =  \int_0^\infty \D{\alpha} A_E(\alpha) e^{-\alpha z}
      \Psi_{\alpha,E}(v). 
\label{eq:loesung_sDGL}
\end{equation}
with a coefficient function $A_E(\alpha)$.

The coefficient function has to be determined by a family of 
matching and boundary conditions  {\em at} the potential 
well and the wall, that is at $z=\ell$ and $z=0$. 
  In the limit of a small potential depth we  approximate the
 square-well adsorbing   potential $V_a(z)$ by a delta-function in the
    middle of the square-well, $V_a/z) = -g\ell\delta(z-\ell/2)$,
with the same integrated potential strength $\int_0^\ell V_a(z) = -g\ell$
 (shaded  area in Fig.\ 2). 
This approximation is valid in the stiff limit $\ell \to 0$.
 Integrating the stationary transfer matrix equation \eqref{eq:sDGL}
    over $z$ from $0$ to $\ell$ and neglecting the terms of 
   higher order in $\ell$
    we get matching conditions 
 \begin{equation}
   v \left(Z_E(\ell,v)- Z_E(0,v)\right) 
     = g\ell  Z_E(\ell/2,v)
   \label{eq:anschluss} 
 \end{equation}
for each  $v$. 
We also have to obey   boundary conditions $Z_E(0,v) =0$ for all $v>0$: It is not possible that the last tangent is
 starting at the wall at $z=0$ and pointing away ($v > 0$)  because
 continuity of tangents would lead to configurations crossing the 
wall.

\subsubsection{Critical potential strength}
\label{App_IIgc}

In order to determine the coefficient function $A_E(\tilde \alpha)$
we make use of a set\cite{sburkhardt} of functions $\Phi_{\alpha,E}$, which are 
  biorthogonal to $\Psi_{\alpha,E}(z,v)$
 \begin{align}
      \int_0^\infty \D{v} v \Psi_{\alpha,E} (v) \Phi_{\tilde \alpha,E} (v) =
      \delta(\alpha - \tilde \alpha).
  \label{eq:biortho}
 \end{align}
 on the half-space $v>0$. 
We use the representation \eqref{eq:loesung_sDGL} in the 
matching condition  \eqref{eq:anschluss} and apply
$\int_0^\infty \D{v} ... \Phi_{\tilde \alpha,E} (v)$ on both 
sides of the matching condition to make use of the biorthogonality 
(\ref{eq:biortho}). 
 Assuming a small potential width we  approximate
 $\exp(\tilde \alpha \ell)\approx 1$ and obtain a
self-consistent integral equation for the coefficient
function $A_E(\tilde \alpha)$
\begin{align*}
      A_E(\tilde \alpha) &= g\ell  \int_0^\infty \D{v} \int_0^\infty
      \D{\alpha} 
      A_E(\alpha) e^{-\ell\alpha/2} \times \\
  & ~~~~~~~~~~~~~~~~~~  \Psi_{\alpha,E} (v)  \Phi_{\tilde \alpha,E} (v).
\end{align*}
Investigating this integral equation for $E\approx 0$ in the 
vicinity of the adsorption transition 
will allow us to (i) determine $A_E(\alpha)$ and thus
the polymer segment distribution and (ii) to find the critical 
potential strength $g_c$ at the transition  $E=0$.

We substitute $\alpha = {2\beta}/{\ell}$ and 
    $v = w (\ell/2)^{1/3}$, which implies
  $\Psi_{\alpha,E}(v) = (\ell/2)^{1/6}
    \Psi_{\beta,E(\ell/2)^{2/3}}(w)$, and 
obtain 
 \begin{align}
   A_E(\tilde \alpha) &= g\ell \left(\frac{\ell}{2}\right)^{-\frac{1}{2}}
   \int_0^\infty \! \! \D{w} \int_0^\beta \!\! \D{\beta}
e^{-\beta}\times
   \nonumber\\
   &\Psi_{\beta,E(\ell/2)^{2/3}}(w)\Phi_{\tilde \alpha,E} \left(w (\ell/2)^{1/3}\right)
A_E(2\beta/\ell). 
 \label{eq:koefffkt}
\end{align}
In principle, this integral equation can  be solved by iteration. 
We start  with  $A_E(\alpha) \approx  A_E$, which is 
 assumed to be {\em independent} of $\alpha$. 
Iterating this equation once for  $E\approx 0$ close to the 
transition and in the stiff limit $\ell \ll 1$, we  find that 
the resulting first iteration  for $A_E(\alpha)$ is 
indeed only weakly dependent on $\alpha$ for $E\approx 0$ and 
 remains {\em  independent}
of $\alpha$ exactly at the transition $E=0$. 
Therefore, the first iteration 
already gives  the correct scaling behavior of $A_E(\alpha)$ and allows 
to determine the critical potential strength $g_c$ exactly for $E=0$. 
For constant $A_E$ and $E(\ell/2)^{2/3}\approx 0$, 
 we can perform the $\beta$-integration to obtain 
\begin{align} 
    \Beta(w) &\equiv \int_0^\beta \D{\beta} e^{-\beta}
                 \Psi_{\beta,0}(w)
  \nonumber\\
       &= 2^{1/3}3^{-2/3}\dfrac{\Gamma\left(
          \frac{1}{2} \right)}{\Gamma\left( \frac{1}{3} \right)}
      \text{M}_{\frac{5}{6},\frac{2}{3}}\left(-\frac{w^3}{9}\right)
    +\nonumber\\
   &~~~~ 2^{-4/3}3^{-5/6}\dfrac{\Gamma\left(
          \frac{1}{3} \right)}{\Gamma\left( \frac{1}{2} \right)}
       w \text{M}_{\frac{7}{6},\frac{4}{3}}\left(-\frac{w^3}{9}\right),
\label{eq:Betaw}
 \end{align}
 where $\text{M}_{a,b}(x)$ is Kummer's confluent hypergeometric function of
 the first kind \cite{sabramowitz1972}, and 
eq.\  (\ref{eq:koefffkt}) becomes
 \begin{align}
   A_E(\tilde \alpha) &= g\ell \left(\frac{\ell}{2}\right)^{-1/2}
   \int_0^\infty \D{w} \Beta(w)
   \nonumber\\
   &~~~\times\Phi_{\tilde \alpha,E} \left(w (\ell/2)^{1/3}\right)A_E 
 \label{eq:koefffkt2}
\end{align} 
 The $\alpha$-dependence of the coefficient function 
$A_E$ in eq.\ (\ref{eq:koefffkt2}) 
stems from the biorthogonal function, and we find 
\begin{align}
A_E(\alpha) = \tilde{\cal N}_{E,\ell} \int_0^\infty \D{w} \Beta(w)
     \Phi_{\alpha,E} \left(w (\ell/2)^{1/3}\right)
\label{eq:AEalpha1}
\end{align}
with a normalization 
factor $\tilde{\cal N}_{E,\ell}$, which is  independent of $\alpha$
but can depend on $E$ and  $\ell$ in general.  

    \begin{figure}
	\includegraphics[width = 0.45\textwidth]{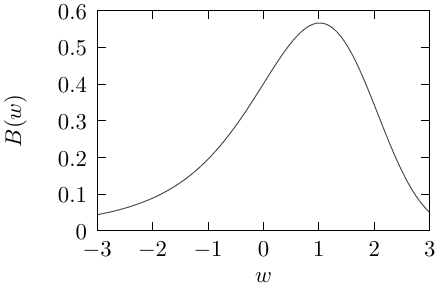}
	\caption{Function $\Beta(w)$, see eq.\ (\ref{eq:Betaw})
         with a maximum at $w\simeq 1.01$. }
	\label{fig:saddle}
    \end{figure}

Because $\Beta(w)$ decreases
  exponentially for $w \gg 1$, see  Fig.\ \ref{fig:saddle}, 
 and  $\ell \ll 1$ 
 in the stiff limit, we can  use
 an approximation for small arguments, 
\begin{equation}
\Phi_{\alpha,E} \left(w(\ell/2)^{1/3}\right) 
\approx \frac{\sqrt{3w}}{\pi}
 \left(\frac{\ell}{2}\right)^{1/6} e^{-\frac{2}{3}{(-E)^{3/2}}/{\alpha}}.
\label{eq:Phismall}
\end{equation}
This leads to
\begin{equation}
A_E(\alpha) \approx {\cal N}_{\ell} \exp\left(-\frac{2}{3}
   \frac{(-E)^{3/2}}{\alpha} \right)
\label{eq:AEalpha}
\end{equation}
in the stiff limit $\ell \ll 1$
with a modified normalization factor ${\cal N}_{\ell}$.
Using this result for $A_E(\alpha)$, 
we find that the  normalization factor  ${\cal N}_{\ell} \propto 
\ell^{-1/3}$ is independent on $E$. 
The coefficient function   $A_E(\alpha)$ 
becomes indeed independent of $\alpha$ for $E\approx 0$,
i.e., close to  the transition. 
This justifies our initial assumption of 
 a {\em constant}  coefficient function 
 $A_E(\alpha) \approx  A_E(0)$.  Therefore, the first 
 iteration already   provides 
 a  self-consistent solution of equation
    \eqref{eq:koefffkt} in the limits of interest.

In order to obtain the critical potential strength $g_c$ we set $E=0$ in 
(\ref{eq:koefffkt2}) and use $\int_0^\infty \D{w} B(w) w^{1/2} 
  =2^{1/3} 3^{-1/6}\pi/ \Gamma\left(1/3\right)$ to  find  
 \begin{align}
    \frac{1}{g_c\ell} &= 
    \frac{\sqrt{3}}{\pi}\left(\frac{\ell}{2}\right)^{-1/3}
   \left(\int_0^\infty \D{w} B(w) w^{1/2}\right)   \nonumber\\
     &=  \frac{2^{2/3}3^{1/3} }{\Gamma\left(1/3\right)}\ell^{-1/3} 
\label{eq:gc1_app}
  \end{align}
In original unrescaled units, see eq.\ (\ref{eq:rescaling}),  
this corresponds to a critical potential 
strength
\begin{equation}
    g_c =2^{-1}3^{-1/3} 
       \Gamma\left(1/3\right)
      \frac{k_BT}{\ell}  \left(\frac{L_p}{\ell}\right)^{-1/3}
\label{eq:gc_app}
\end{equation}
  The scaling behavior of $g_c$ 
agrees with the result (8)
   from the scaling argument and    we quantify the numerical 
prefactor in  (8)
 to be  $c_{SF} = 2^{-1}3^{-1/3}\Gamma(1/3) \simeq 0.929$.

\subsubsection{Critical  exponent $\nu$}
\label{App_IInu}

The exponent $\nu$ characterizes  the critical behavior of the ground state 
energy $|E|\sim |g-g_c|^{\nu}$
as a function of $g-g_c$ close to the adsorption transition
at  $E= 0$. Because of the relations $|\Delta f| = |E| = 1/\xi$,
the exponent $\nu$ characterizes both the  critical behavior of 
 the  correlation length $\xi$ and of  the free energy of adsorption 
$\Delta f$ (i.e., hyperscaling holds). 
\Cr{
For $g>g_c$, 
the correlation length $\xi$ of the adsorption transition is defined
by the distribution of loops lengths, which decays exponentially 
for large loop lengths with a characteristic decay length given by 
the correlation length $\xi$. 
} 

Because the exponent $\nu$ characterizes the critical free 
energy behavior, it also 
 determines the order of the adsorption 
transition: For $\nu >1$ the transition is continuous, whereas it is 
a first order transition for $\nu<1$.
It is a remarkable feature of the polymer adsorption transition that
a  correlation length
 $\xi = 1/|E|$, which describes the typical length scale 
of loops, can always be defined  and diverges at the transition, 
even if the transition is of first order.

Using the result (\ref{eq:AEalpha1}) for $A_E(\alpha)$  in 
the self-consistent equation (\ref{eq:koefffkt}), we obtain 
the relation $g=g(E)$ in the form 
  \begin{align}
   \frac{1}{g\ell}  &\approx \frac{1}{ \tilde{\cal N}_\ell}
                 \int dv Z_E(\ell,v)
\end{align} 
Expanding about $E\approx 0$ gives the exponent $\nu$. A leading 
$|E|$-dependence 
$Z_E(\ell,v)- Z_0(\ell,v)\sim |E|^{3/2}$ 
has been obtained in Ref.\ \citenum{sburkhardt}
 and suggests $\int dv (Z_E(\ell,v) - Z_0(\ell,v))\sim |E|$, 
 corresponding to  $\nu=1$.
 Thus, the transfer matrix 
approach in  the approximation 
of a weakly bent semiflexible polymer  gives\cite{skierfeld03} $\nu=\nu_{SF}=1$
for  purely position-dependent 
adsorption potentials as we use here.
This suggests that the   adsorption transition is 
first order or second order with a weak logarithmic correction
\cite{skierfeld03,sKierfeld2005a}.

\subsubsection{Corrections from
Crossover to an effective flexible polymer model}
\label{App_IIcorr}

The transfer matrix calculation in  the approximation 
of a weakly bent polymer is, strictly speaking, only valid in the stiff limit 
 $L_p \to \infty$. Corrections  start to  arise if the unrescaled 
correlation length 
$\xi$ exceeds the persistence length\footnote{Throughout 
this section we use scaling arguments. 
The distinction between $L_p$ and $L_{p,D}$ is therefore 
unnecessary.} $L_p$:
Because $\xi$ specifies the typical length of an unbound 
desorbed loop of the polymer,  loops start to  loose orientation 
and to develop overhangs if $\xi > L_p$
 or  $\xi  = 1/|E| >2$
in rescaled units (\ref{eq:rescaling}).
This happens 
for potential strengths close to the critical value  
$g_c= g_{c,SF}= c_{SF} ({k_BT}/{\ell}) ({L_p}/{\ell})^{-1/3}$
in the semiflexible limit as given by 
 eq.\ (\ref{eq:gc_app}) or eq.\ (8) in the main text
(in unrescaled units), 
where the correlations length $\xi$ starts to  increase
and the  transfer matrix ground state energy $E$ becomes 
small   according to $|E|\sim |g-g_c|^{\nu}$ with $\nu=\nu_{SF}=1$. 
The condition $|E| <1/2$ corresponds to $|g-g_c|< 1/2$
(or  $|g-g_{c}| < k_BT/L_p$ in unrescaled units). 


Because $g_c \sim \ell^{-2/3}$
(or $g_c \sim (k_BT/L_p) (L_p/\ell)^{2/3}$ in unrescaled units,  
see eq.\ (\ref{eq:gc_app})), 
corrections will always dominate if $\ell \gg 1$ 
such that  $|g-g_{c}| <g_c \ll  1/2$ 
for all $g<g_c$. 
In this regime the  weak  bending approximation breaks down completely. 
The regime $\ell \gg 1$ corresponds to
$\ell \gg L_p$ in unrescaled units, which 
 is the flexible limit.

Corrections to the weak bending results also arise 
in the stiff limit $\ell \ll 1$ or  $\ell \ll L_p$ in unrescaled units.
In the stiff limit corrections arise only
in a small 
interval  $|g-g_{c,SF}| < k_BT/L_p \ll g_{c,SF}$ around $g_{c,SF}$. 
If   $|g-g_{c,SF}| < k_BT/L_p$  we have to use an
{\rm effective flexible} polymer model  with a Kuhn length $b_K=2L_{p}$
 and an effective adsorption potential 
 per length $g_{\rm eff} \sim |\Delta f|= |E|\sim g-g_{c,SF}$,
which derives from  the free energy exponent $\nu=\nu_{SF}=1$ in the 
weak bending approximation,
and 
an effective potential range $\ell_{\rm eff} \sim \langle z^2 \rangle^{1/2}(\xi)
\sim \xi^3/L_p \sim L_p$.

This effective flexible model determines the actual 
free energy exponent \cite{sKierfeld2005a} $\nu_F=2$. 
Close to the transition, where $\xi >L_p$ or 
$|g-g_{c,SF}| < k_BT/L_p$, we  expect a 
 crossover from an apparent exponent $\nu_{SF}=1$ 
to the actual exponent $\nu = \nu_F=2$  for a flexible 
polymer, and the adsorption transition becomes 
 continuous.
However, in a system of
finite size $L$, this crossover should only become apparent 
 if $L>L_p$ such that a hierarchy of length scales $L>\xi>L_p$
is possible. Otherwise, $\xi>L_p$ also implies $\xi >L$, and 
finite size effects mask the crossover.

The crossover to an effective flexible behavior also 
 leads to a shift of the critical potential 
strength. 
For  the  effective 
 flexible polymer the critical potential strength for adsorption is 
given by $g_{c, \rm eff} = c_F \frac{k_BTL_p}{\ell_{\rm eff}^2}$, cf.\ 
 eq.\ (10) in the main text. 
The actual critical potential strength
$g_c$ is given by the condition $g_c-g_{c,SF} = g_{\rm c, eff}$
or 
\begin{equation}
    g_{c} =  g_{c,SF} + c_F \frac{k_BT}{L_p},
\label{eq:gc_corr}
\end{equation}
which is slightly higher than the stiff limit result $g_c = g_{c,SF}$.
Equation (\ref{eq:gc_corr}) 
 shows that the leading corrections to the critical 
potential strength  (11)  in the stiff limit are 
of the form $I(x) \approx c_{SF} x^{-1/3} +  {\cal O}(x^{-1})$, which 
is exactly the third constraint (14).

\section{Determination 
of critical potential strength in simulations}
\label{App_III}

In order to determine the critical potential strength 
in simulations we use two methods --  a cumulant method and 
finite size scaling --  both of which are 
explained in detail in this section.

\subsection{Third cumulant method}
\label{App_IIIc}

\begin{figure}
      \includegraphics[width=0.45\textwidth]{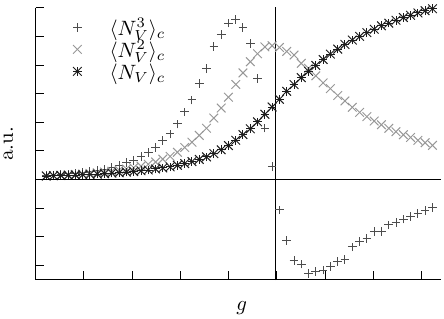}
  \caption{
    Typical shape of the first three cumulants of the order parameter $N_V$, 
the number of beads of the  SHC within the potential range. 
     We locate the adsorption transition by the criterion 
   $\langle N_V^3\rangle_c=0$ in simulations. 
}
      \label{fig:cumulants}
\end{figure}

 An effective method to determine the critical potential strength
 uses the fact that the derivative of the free energy density
 with respect to the potential strength $g$ gives the 
mean  fraction of polymer length  in the square-well potential, 
which provides an  order parameter for the adsorption transition. 
For the discrete SHC, we have 
 $\partial_g f = \langle N_V\rangle /N$, where $N_V$ is 
the number of beads of the  SHC within the potential range 
$0<z<\ell$. 
Because of the crossover to an effective flexible behavior close 
to the adsorption transition, we expect a 
 continuous adsorption transition and 
the second cumulant of the order parameter, i.e., 
the second derivative $\partial^2_g f = \langle (N_V-\langle N_V
\rangle)^2\rangle = \langle N_V^2 \rangle_c$ of the free energy should diverge. 
Because of finite size effects we find a maximum rather than a 
divergence in the simulations, 
which results in a vanishing third cumulant 
$\partial^3_g f =\langle N_V^3 \rangle_c=0$
at the transition, as shown in   Fig.\ \ref{fig:cumulants}.
Therefore, we can use  the vanishing third cumulant to locate the 
adsorption transition in simulations. 
In order to find the zero of the third cumulant we interpolate between the
 first negative and last positive value to determine the critical potential
 strength.
\Cr{We use this criterion both for the planar and for curved 
geometries to locate the adsorption transition.}

\subsection{Finite size scaling procedure}
\label{App_IIIfs}

Finite size scaling of the specific heat 
allows to determine  the critical 
potential strength $g_c$ and to calculate the 
 critical exponent $\nu$ for the correlation length and the 
free energy. 
\Cr{
We apply this method to analyze the simulation data for the planar
 substrate.}

\begin{figure}[ht]
    \includegraphics[width = 0.45\textwidth]{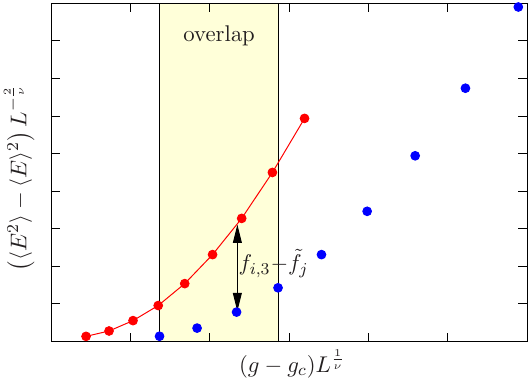}     
    \caption{
    Example for the overlap region of two data sets for  $L_p = 2b_0$ and
 $\ell = b_0$ for two different lengths  $L= 800\,b_0$
      and $L= 400\,b_0$.
      In this example we have  $N_{\mathrm{over}} = 4$ and use
      $g_c\ell/k_BT =0.3$ and  $\nu = 1.4$.
    \label{fig:overlap}
}
  \end{figure}

To systematically find  the best parameter set
$(\nu,\,g_c)$, we calculate a quantity $S(\nu,g_c)$ which measures the
squared differences from one specific heat data set $f$ to the
interpolated curves $\tilde f$ of another set for  different length\cite{sbhattacharjee2001measure} $L$
,  as
shown in  Fig.\ \ref{fig:overlap}. As contour lengths we use $L/b_0= 50$, $100$, $150$, $200$, $300$, $400$, $600$ and $800$. To be able to compare different parameter
  sets $(\nu,\,g_c)$ we take only the relative differences.  

\begin{figure}[ht]
    \includegraphics[width=0.45\textwidth]{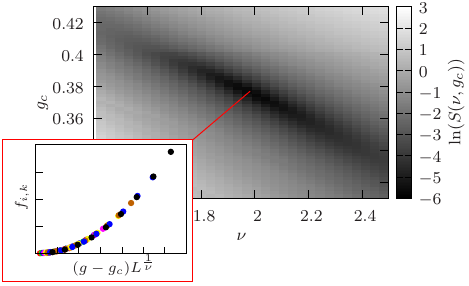}
    \caption{ 
   Logarithmic overall error $\ln\left(S(\nu,g_c)\right)$  as a function of parameters 
$g_c\ell/k_BT$ and $\nu$. Inset:  Scaling function $f_{i,k}$ for the optimal 
  choices $\nu_{\mathrm{min}} = 1.99$ 
  and $g_{c,{\mathrm{min}}}\ell/k_BT = 0.37$. 
   Analysis for MC data for parameters 
   $L_p = 2b_0$, $k=100k_BT/b_0^2$ and  $\ell=b_0$. }
    \label{fig:fss1}
\end{figure}

For this analysis  only a limited number  $N_{\mathrm{over}}$ of 
data points in the overlapping region can be used.
 Our best estimate for $(\nu,\,g_c)$ is
    the parameter set that minimizes the overall error
\begin{align*}
   S(\nu,g_c) =\frac{1}{N_{\mathrm{over}}} \sum_{i} \sum_{j \neq i} 
     \sum_{k=1}^{N_{\mathrm{over}}}  \left(1- \frac{\tilde f_{j}}{f_{i,k}}\right)^2  
\end{align*}
The determination of $\nu$ is quite difficult because 
the minimum is often rather shallow as shown  in the example in 
Fig.\ \ref{fig:fss1}.

A simple approach
to estimate the error in this procedure is given by
\cite{sbhattacharjee2001measure} 
  \begin{align*}
    \Delta g_c &= \eta g_{c,\,\mathrm{min}} 
   \left( 2 \ln{ \frac{S(\nu_\mathrm{min},g_{c,\,\mathrm{min}}(1{\pm} \eta) )}
          {S(\nu_\mathrm{min},g_{c,\,\mathrm{min}} )}} \right)^{{1}/{2}} \\
    \Delta \nu &= \eta \nu_{\,\mathrm{min}} \left( 2
      \ln{\frac{S(\nu_\mathrm{min}(1{\pm} \eta),g_{c,\,\mathrm{min}}
          )}{S(\nu_\mathrm{min},g_{c,\,\mathrm{min}} )}}
    \right)^{{1}/{2}} \Kma
\end{align*}
where $\eta \equiv 0.01$ gives the relative distance to the minimum.  We note
that this method of error estimation might be flawed, because the
determination of $g_c$ is much more precise than the determination of $\nu$.
While changes in $g_c$ mainly shift data points  in Fig.\ \ref{fig:overlap}
horizontally and influence the overlap region, variation of $\nu$
affects mostly the rescaled specific heat values $f_{i,k}$ themselves and,
thus,  shift data points vertically in Fig.\ \ref{fig:overlap}. 
If the overlap region becomes smaller the
overall error $S(\nu,g_c)$ increases fast. This explains why the variation of
$g_c$ influences $S(\nu,g_c)$ much more than $\nu$ such that  the
determination of $\nu$ is more difficult. To take this into account, we
compute the minimal and maximal value of $g_c$ and $\nu$, where $S(\nu,g_c) <
(1+\eta_2) S(\nu_\mathrm{min, max},g_{c,\mathrm{min, max}})$, where $\eta_2$
is an arbitrary parameter. These minimal and maximal values for $g_c$ and
$\nu$ should be a valid estimation of the error.


\section{Simulation results for critical exponent $\nu$
 for planar substrate}
\label{App_IV}

\begin{figure}[ht]
\begin{center}
    \includegraphics[width =0.45\textwidth]{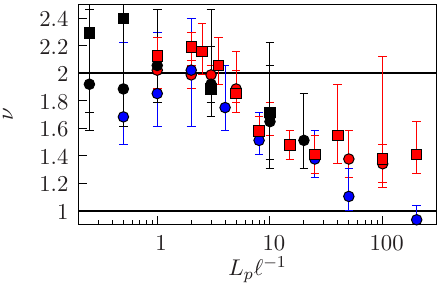} 
    \caption{  
 Finite size scaling results for
  the critical exponent $\nu$ in $D=3$  dimensions (circles)
and  $D=2$
dimensions (squares) as a function of the dimensionless stiffness parameter
$L_p/\ell$ for $\ell = 1$ (red), $\ell = 2$ (blue) and $\ell = 4$ (black).
 Simulation parameters are as in Fig.\ 4.
}
    \label{fig:fss2}
\end{center}
\end{figure}

The finite size scaling procedure  also allows us to 
determine the critical exponent $\nu$ for adsorption 
to a planar substrate.
Results for the exponent $\nu$ 
are shown in Fig.\ \ref{fig:fss2} 
as a function of the stiffness parameter $L_p/\ell$.

The exponent
$\nu$ is around $\nu=2$ for
small bending rigidity and lowers towards $\nu=1$ 
 with increasing stiffness. This is in agreement with the theoretical 
expectation that adsorption of flexible polymers is a continuous
 transition with $\nu_F=2$. A semiflexible polymer  should 
exhibit a critical behavior corresponding to  $\nu_{SF}=1$
with a  crossover to a flexible behavior with $\nu=\nu_F=2$ 
in the small regime $|g-g_{c,SF}| < k_BT/L_p$ around the transition,
where the correlation length $\xi$ exceeds  $L_p$ as explained 
in section \ref{App_IInu}. 
This crossover might be the reason that we obtain values 
$\nu\approx 1.4$ significantly larger than $\nu=1$ for 
stiff polymers using  the finite size scaling. 

\section{Additional simulation snapshots}
\label{App_V}

\begin{figure*}[ht]
\begin{center}
    \includegraphics[width =0.8\textwidth]{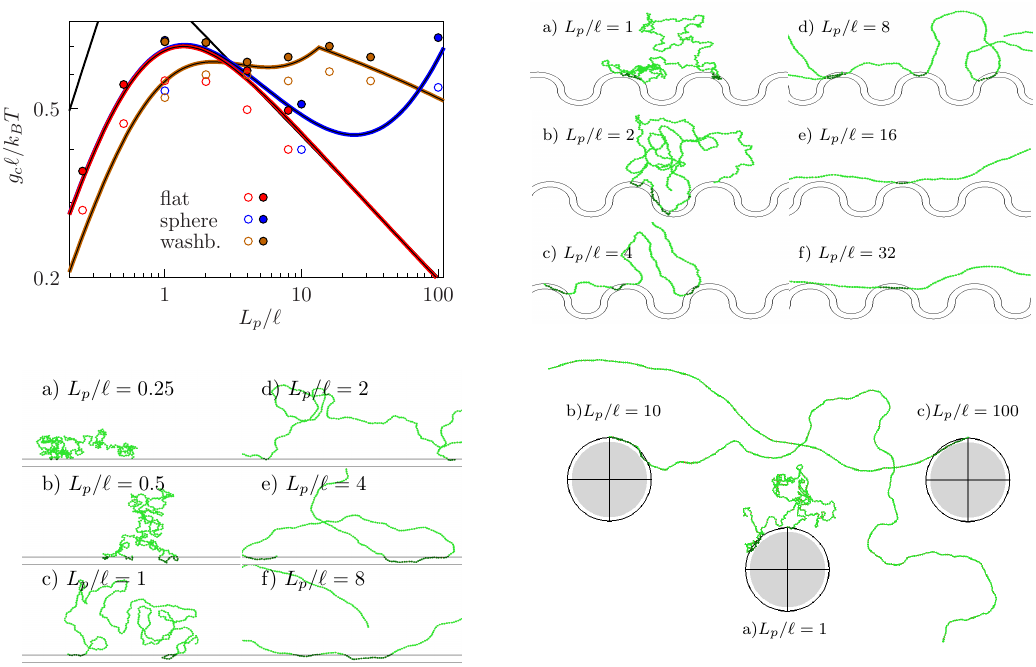} 
    \caption{  
  Phase diagrams for a planar substrate (red line), an adsorbing sphere 
  (blue line) and an adsorbing washboard in $D=2$ (brown line) and 
  simulation snapshots in the desorbed phases corresponding to the 
  open circles in the phase diagram. Solid circles correspond 
to the simulation snapshots in the adsorbed phase presented 
in the main text. 
}
    \label{fig:desorbed}
\end{center}
\end{figure*}

In Fig.\ \ref{fig:desorbed} we present additional simulation snapshots 
in the desorbed phase for all three adsorption geometries. 

\begin{acknowledgments}

We acknowledge financial support by  the Deutsche Forschungsgemeinschaft
(KI 662/2-1).
 T.A.K. thanks the NRW 
Forschungsschule ``Forschung mit Synchrotronstrahlung in 
den Nano- und Biowissenschaften'' for financial support.
\end{acknowledgments}


\begin{thebibliography}{62}%
\makeatletter
\providecommand \@ifxundefined [1]{%
 \@ifx{#1\undefined}
}%
\providecommand \@ifnum [1]{%
 \ifnum #1\expandafter \@firstoftwo
 \else \expandafter \@secondoftwo
 \fi
}%
\providecommand \@ifx [1]{%
 \ifx #1\expandafter \@firstoftwo
 \else \expandafter \@secondoftwo
 \fi
}%
\providecommand \natexlab [1]{#1}%
\providecommand \enquote  [1]{``#1''}%
\providecommand \bibnamefont  [1]{#1}%
\providecommand \bibfnamefont [1]{#1}%
\providecommand \citenamefont [1]{#1}%
\providecommand \href@noop [0]{\@secondoftwo}%
\providecommand \href [0]{\begingroup \@sanitize@url \@href}%
\providecommand \@href[1]{\@@startlink{#1}\@@href}%
\providecommand \@@href[1]{\endgroup#1\@@endlink}%
\providecommand \@sanitize@url [0]{\catcode `\\12\catcode `\$12\catcode
  `\&12\catcode `\#12\catcode `\^12\catcode `\_12\catcode `\%12\relax}%
\providecommand \@@startlink[1]{}%
\providecommand \@@endlink[0]{}%
\providecommand \url  [0]{\begingroup\@sanitize@url \@url }%
\providecommand \@url [1]{\endgroup\@href {#1}{\urlprefix }}%
\providecommand \urlprefix  [0]{URL }%
\providecommand \Eprint [0]{\href }%
\providecommand \doibase [0]{http://dx.doi.org/}%
\providecommand \selectlanguage [0]{\@gobble}%
\providecommand \bibinfo  [0]{\@secondoftwo}%
\providecommand \bibfield  [0]{\@secondoftwo}%
\providecommand \translation [1]{[#1]}%
\providecommand \BibitemOpen [0]{}%
\providecommand \bibitemStop [0]{}%
\providecommand \bibitemNoStop [0]{.\EOS\space}%
\providecommand \EOS [0]{\spacefactor3000\relax}%
\providecommand \BibitemShut  [1]{\csname bibitem#1\endcsname}%
\let\auto@bib@innerbib\@empty
\bibitem [{\citenamefont {Skolnick}\ and\ \citenamefont
  {Fixman}(1977)}]{Skolnick1977}%
  \BibitemOpen
  \bibfield  {author} {\bibinfo {author} {\bibfnamefont {J.}~\bibnamefont
  {Skolnick}}\ and\ \bibinfo {author} {\bibfnamefont {M.}~\bibnamefont
  {Fixman}},\ }\href {\doibase 10.1021/ma60059a011} {\bibfield  {journal}
  {\bibinfo  {journal} {Macromolecules}\ }\textbf {\bibinfo {volume} {10}},\
  \bibinfo {pages} {944} (\bibinfo {year} {1977})}\BibitemShut {NoStop}%
\bibitem [{\citenamefont {F{\"o}rster}\ \emph {et~al.}(1999)\citenamefont
  {F{\"o}rster}, \citenamefont {Neubert}, \citenamefont {Schl{\"u}ter},\ and\
  \citenamefont {Lindner}}]{foerster1999}%
  \BibitemOpen
  \bibfield  {author} {\bibinfo {author} {\bibfnamefont {S.}~\bibnamefont
  {F{\"o}rster}}, \bibinfo {author} {\bibfnamefont {I.}~\bibnamefont
  {Neubert}}, \bibinfo {author} {\bibfnamefont {A.~D.}\ \bibnamefont
  {Schl{\"u}ter}}, \ and\ \bibinfo {author} {\bibfnamefont {P.}~\bibnamefont
  {Lindner}},\ }\href@noop {} {\bibfield  {journal} {\bibinfo  {journal}
  {Macromolecules}\ }\textbf {\bibinfo {volume} {32}},\ \bibinfo {pages} {4043}
  (\bibinfo {year} {1999})}\BibitemShut {NoStop}%
\bibitem [{\citenamefont {Gutjahr}, \citenamefont {Lipowsky},\ and\
  \citenamefont {Kierfeld}(2006)}]{Gutjahr2006}%
  \BibitemOpen
  \bibfield  {author} {\bibinfo {author} {\bibfnamefont {P.}~\bibnamefont
  {Gutjahr}}, \bibinfo {author} {\bibfnamefont {R.}~\bibnamefont {Lipowsky}}, \
  and\ \bibinfo {author} {\bibfnamefont {J.}~\bibnamefont {Kierfeld}},\
  }\href@noop {} {\bibfield  {journal} {\bibinfo  {journal} {Europhys. Lett.}\
  }\textbf {\bibinfo {volume} {76}},\ \bibinfo {pages} {994} (\bibinfo {year}
  {2006})}\BibitemShut {NoStop}%
\bibitem [{\citenamefont {Bednar}\ \emph {et~al.}(1995)\citenamefont {Bednar},
  \citenamefont {Furrer}, \citenamefont {Katritch}, \citenamefont {Stasiak},
  \citenamefont {Dubochet},\ and\ \citenamefont {Stasiak}}]{bednar1995}%
  \BibitemOpen
  \bibfield  {author} {\bibinfo {author} {\bibfnamefont {J.}~\bibnamefont
  {Bednar}}, \bibinfo {author} {\bibfnamefont {P.}~\bibnamefont {Furrer}},
  \bibinfo {author} {\bibfnamefont {V.}~\bibnamefont {Katritch}}, \bibinfo
  {author} {\bibfnamefont {A.}~\bibnamefont {Stasiak}}, \bibinfo {author}
  {\bibfnamefont {J.}~\bibnamefont {Dubochet}}, \ and\ \bibinfo {author}
  {\bibfnamefont {A.}~\bibnamefont {Stasiak}},\ }\href@noop {} {\bibfield
  {journal} {\bibinfo  {journal} {J. Mol. Biol.}\ }\textbf {\bibinfo {volume}
  {254}},\ \bibinfo {pages} {579} (\bibinfo {year} {1995})}\BibitemShut
  {NoStop}%
\bibitem [{\citenamefont {Ott}\ \emph {et~al.}(1993)\citenamefont {Ott},
  \citenamefont {Magnasco}, \citenamefont {Simon},\ and\ \citenamefont
  {Libchaber}}]{ott1993}%
  \BibitemOpen
  \bibfield  {author} {\bibinfo {author} {\bibfnamefont {A.}~\bibnamefont
  {Ott}}, \bibinfo {author} {\bibfnamefont {M.}~\bibnamefont {Magnasco}},
  \bibinfo {author} {\bibfnamefont {A.}~\bibnamefont {Simon}}, \ and\ \bibinfo
  {author} {\bibfnamefont {A.}~\bibnamefont {Libchaber}},\ }\href@noop {}
  {\bibfield  {journal} {\bibinfo  {journal} {Phys. Rev. E}\ }\textbf {\bibinfo
  {volume} {48}},\ \bibinfo {pages} {R1642} (\bibinfo {year}
  {1993})}\BibitemShut {NoStop}%
\bibitem [{\citenamefont {Venier}\ \emph {et~al.}(1994)\citenamefont {Venier},
  \citenamefont {Maggs}, \citenamefont {Carlier},\ and\ \citenamefont
  {Pantaloni}}]{venier1994}%
  \BibitemOpen
  \bibfield  {author} {\bibinfo {author} {\bibfnamefont {P.}~\bibnamefont
  {Venier}}, \bibinfo {author} {\bibfnamefont {A.}~\bibnamefont {Maggs}},
  \bibinfo {author} {\bibfnamefont {M.}~\bibnamefont {Carlier}}, \ and\
  \bibinfo {author} {\bibfnamefont {D.}~\bibnamefont {Pantaloni}},\ }\href@noop
  {} {\bibfield  {journal} {\bibinfo  {journal} {J. Biol. Chem.}\ }\textbf
  {\bibinfo {volume} {269}},\ \bibinfo {pages} {13353} (\bibinfo {year}
  {1994})}\BibitemShut {NoStop}%
\bibitem [{\citenamefont {de~Gennes}(1979)}]{degennes}%
  \BibitemOpen
  \bibfield  {author} {\bibinfo {author} {\bibfnamefont {P.}~\bibnamefont
  {de~Gennes}},\ }\href@noop {} {\emph {\bibinfo {title} {Scaling Concepts in
  Polymer Physics}}}\ (\bibinfo  {publisher} {Cornell University Press},\
  \bibinfo {address} {Ithaca and London},\ \bibinfo {year} {1979})\BibitemShut
  {NoStop}%
\bibitem [{\citenamefont {Eisenriegler}(1993)}]{eisenriegler}%
  \BibitemOpen
  \bibfield  {author} {\bibinfo {author} {\bibfnamefont {E.}~\bibnamefont
  {Eisenriegler}},\ }\href@noop {} {\emph {\bibinfo {title} {Polymers near
  Surfaces}}}\ (\bibinfo  {publisher} {World Scientific},\ \bibinfo {address}
  {London},\ \bibinfo {year} {1993})\BibitemShut {NoStop}%
\bibitem [{\citenamefont {Netz}\ and\ \citenamefont
  {Andelman}(2003)}]{Netz2003}%
  \BibitemOpen
  \bibfield  {author} {\bibinfo {author} {\bibfnamefont {R.~R.}\ \bibnamefont
  {Netz}}\ and\ \bibinfo {author} {\bibfnamefont {D.}~\bibnamefont
  {Andelman}},\ }\href@noop {} {\bibfield  {journal} {\bibinfo  {journal}
  {Phys. Rep.}\ }\textbf {\bibinfo {volume} {380}},\ \bibinfo {pages} {1}
  (\bibinfo {year} {2003})}\BibitemShut {NoStop}%
\bibitem [{\citenamefont {Angelini}\ \emph {et~al.}(2003)\citenamefont
  {Angelini}, \citenamefont {Liang}, \citenamefont {Wriggers},\ and\
  \citenamefont {Wong}}]{angelini2003}%
  \BibitemOpen
  \bibfield  {author} {\bibinfo {author} {\bibfnamefont {T.}~\bibnamefont
  {Angelini}}, \bibinfo {author} {\bibfnamefont {H.}~\bibnamefont {Liang}},
  \bibinfo {author} {\bibfnamefont {W.}~\bibnamefont {Wriggers}}, \ and\
  \bibinfo {author} {\bibfnamefont {G.}~\bibnamefont {Wong}},\ }\href@noop {}
  {\bibfield  {journal} {\bibinfo  {journal} {Proc. Nat. Acad. Sci. USA}\
  }\textbf {\bibinfo {volume} {100}},\ \bibinfo {pages} {8634} (\bibinfo {year}
  {2003})}\BibitemShut {NoStop}%
\bibitem [{\citenamefont {Dos~Remedios}\ \emph {et~al.}(2003)\citenamefont
  {Dos~Remedios}, \citenamefont {Chhabra}, \citenamefont {Kekic}, \citenamefont
  {Dedova}, \citenamefont {Tsubakihara}, \citenamefont {Berry},\ and\
  \citenamefont {Nosworthy}}]{dos2003}%
  \BibitemOpen
  \bibfield  {author} {\bibinfo {author} {\bibfnamefont {C.}~\bibnamefont
  {Dos~Remedios}}, \bibinfo {author} {\bibfnamefont {D.}~\bibnamefont
  {Chhabra}}, \bibinfo {author} {\bibfnamefont {M.}~\bibnamefont {Kekic}},
  \bibinfo {author} {\bibfnamefont {I.}~\bibnamefont {Dedova}}, \bibinfo
  {author} {\bibfnamefont {M.}~\bibnamefont {Tsubakihara}}, \bibinfo {author}
  {\bibfnamefont {D.}~\bibnamefont {Berry}}, \ and\ \bibinfo {author}
  {\bibfnamefont {N.}~\bibnamefont {Nosworthy}},\ }\href@noop {} {\bibfield
  {journal} {\bibinfo  {journal} {Physiol. Rev.}\ }\textbf {\bibinfo {volume}
  {83}},\ \bibinfo {pages} {433} (\bibinfo {year} {2003})}\BibitemShut
  {NoStop}%
\bibitem [{\citenamefont {Winder}\ and\ \citenamefont
  {Ayscough}(2005)}]{winder2005}%
  \BibitemOpen
  \bibfield  {author} {\bibinfo {author} {\bibfnamefont {S.~J.}\ \bibnamefont
  {Winder}}\ and\ \bibinfo {author} {\bibfnamefont {K.~R.}\ \bibnamefont
  {Ayscough}},\ }\href@noop {} {\bibfield  {journal} {\bibinfo  {journal} {J.
  Cell Sci.}\ }\textbf {\bibinfo {volume} {118}},\ \bibinfo {pages} {651}
  (\bibinfo {year} {2005})}\BibitemShut {NoStop}%
\bibitem [{\citenamefont {Fletcher}\ and\ \citenamefont
  {Mullins}(2010)}]{Fletcher2010}%
  \BibitemOpen
  \bibfield  {author} {\bibinfo {author} {\bibfnamefont {D.~A.}\ \bibnamefont
  {Fletcher}}\ and\ \bibinfo {author} {\bibfnamefont {R.~D.}\ \bibnamefont
  {Mullins}},\ }\href@noop {} {\bibfield  {journal} {\bibinfo  {journal}
  {Nature}\ }\textbf {\bibinfo {volume} {463}},\ \bibinfo {pages} {485}
  (\bibinfo {year} {2010})}\BibitemShut {NoStop}%
\bibitem [{\citenamefont {Maggs}, \citenamefont {Huse},\ and\ \citenamefont
  {Leibler}(1989)}]{Maggs1989}%
  \BibitemOpen
  \bibfield  {author} {\bibinfo {author} {\bibfnamefont {A.~C.}\ \bibnamefont
  {Maggs}}, \bibinfo {author} {\bibfnamefont {D.~A.}\ \bibnamefont {Huse}}, \
  and\ \bibinfo {author} {\bibfnamefont {S.}~\bibnamefont {Leibler}},\
  }\href@noop {} {\bibfield  {journal} {\bibinfo  {journal} {Europhys. Lett.}\
  }\textbf {\bibinfo {volume} {8}},\ \bibinfo {pages} {615} (\bibinfo {year}
  {1989})}\BibitemShut {NoStop}%
\bibitem [{\citenamefont {Kierfeld}(2006)}]{Kierfeld2006}%
  \BibitemOpen
  \bibfield  {author} {\bibinfo {author} {\bibfnamefont {J.}~\bibnamefont
  {Kierfeld}},\ }\href@noop {} {\bibfield  {journal} {\bibinfo  {journal}
  {Phys. Rev. Lett.}\ }\textbf {\bibinfo {volume} {97}},\ \bibinfo {pages}
  {058302} (\bibinfo {year} {2006})}\BibitemShut {NoStop}%
\bibitem [{\citenamefont {Birshtein}, \citenamefont {Zhulina},\ and\
  \citenamefont {Skvortsov}(1979)}]{Birshtein1979}%
  \BibitemOpen
  \bibfield  {author} {\bibinfo {author} {\bibfnamefont {T.~M.}\ \bibnamefont
  {Birshtein}}, \bibinfo {author} {\bibfnamefont {E.~B.}\ \bibnamefont
  {Zhulina}}, \ and\ \bibinfo {author} {\bibfnamefont {A.~M.}\ \bibnamefont
  {Skvortsov}},\ }\href@noop {} {\bibfield  {journal} {\bibinfo  {journal}
  {Biopolymers}\ }\textbf {\bibinfo {volume} {18}},\ \bibinfo {pages} {1171}
  (\bibinfo {year} {1979})}\BibitemShut {NoStop}%
\bibitem [{\citenamefont {van~der Linden}, \citenamefont {Leermakers},\ and\
  \citenamefont {Fleer}(1996)}]{VanderLinden1996}%
  \BibitemOpen
  \bibfield  {author} {\bibinfo {author} {\bibfnamefont {C.~C.}\ \bibnamefont
  {van~der Linden}}, \bibinfo {author} {\bibfnamefont {F.~A.~M.}\ \bibnamefont
  {Leermakers}}, \ and\ \bibinfo {author} {\bibfnamefont {G.~J.}\ \bibnamefont
  {Fleer}},\ }\href@noop {} {\bibfield  {journal} {\bibinfo  {journal}
  {Macromolecules}\ }\textbf {\bibinfo {volume} {29}},\ \bibinfo {pages} {1172}
  (\bibinfo {year} {1996})}\BibitemShut {NoStop}%
\bibitem [{\citenamefont {Kramarenko}\ \emph {et~al.}(1996)\citenamefont
  {Kramarenko}, \citenamefont {Winkler}, \citenamefont {Khalatur},
  \citenamefont {Khokhlov},\ and\ \citenamefont {Reineker}}]{kramarenko96}%
  \BibitemOpen
  \bibfield  {author} {\bibinfo {author} {\bibfnamefont {E.}~\bibnamefont
  {Kramarenko}}, \bibinfo {author} {\bibfnamefont {R.}~\bibnamefont {Winkler}},
  \bibinfo {author} {\bibfnamefont {P.}~\bibnamefont {Khalatur}}, \bibinfo
  {author} {\bibfnamefont {A.}~\bibnamefont {Khokhlov}}, \ and\ \bibinfo
  {author} {\bibfnamefont {P.}~\bibnamefont {Reineker}},\ }\href@noop {}
  {\bibfield  {journal} {\bibinfo  {journal} {J. Chem. Phys.}\ }\textbf
  {\bibinfo {volume} {104}},\ \bibinfo {pages} {4806} (\bibinfo {year}
  {1996})}\BibitemShut {NoStop}%
\bibitem [{\citenamefont {Sintes}, \citenamefont {Sumithra},\ and\
  \citenamefont {Straube}(2001)}]{sintes01}%
  \BibitemOpen
  \bibfield  {author} {\bibinfo {author} {\bibfnamefont {T.}~\bibnamefont
  {Sintes}}, \bibinfo {author} {\bibfnamefont {K.}~\bibnamefont {Sumithra}}, \
  and\ \bibinfo {author} {\bibfnamefont {E.}~\bibnamefont {Straube}},\ }\href
  {\doibase 10.1021/ma000493s} {\bibfield  {journal} {\bibinfo  {journal}
  {Macromolecules}\ }\textbf {\bibinfo {volume} {34}},\ \bibinfo {pages} {1352}
  (\bibinfo {year} {2001})}\BibitemShut {NoStop}%
\bibitem [{\citenamefont {Netz}\ and\ \citenamefont
  {Joanny}(1999{\natexlab{a}})}]{Netz1999}%
  \BibitemOpen
  \bibfield  {author} {\bibinfo {author} {\bibfnamefont {R.~R.}\ \bibnamefont
  {Netz}}\ and\ \bibinfo {author} {\bibfnamefont {J.-F.}\ \bibnamefont
  {Joanny}},\ }\href@noop {} {\bibfield  {journal} {\bibinfo  {journal}
  {Macromolecules}\ }\textbf {\bibinfo {volume} {32}},\ \bibinfo {pages} {9013}
  (\bibinfo {year} {1999}{\natexlab{a}})}\BibitemShut {NoStop}%
\bibitem [{\citenamefont {Kong}\ and\ \citenamefont
  {Muthukumar}(1998)}]{Kong1998}%
  \BibitemOpen
  \bibfield  {author} {\bibinfo {author} {\bibfnamefont {C.~Y.}\ \bibnamefont
  {Kong}}\ and\ \bibinfo {author} {\bibfnamefont {M.}~\bibnamefont
  {Muthukumar}},\ }\href@noop {} {\bibfield  {journal} {\bibinfo  {journal} {J.
  Chem. Phys.}\ }\textbf {\bibinfo {volume} {109}},\ \bibinfo {pages} {1522}
  (\bibinfo {year} {1998})}\BibitemShut {NoStop}%
\bibitem [{\citenamefont {Caravenna}\ and\ \citenamefont
  {Deuschel}(2008)}]{Caravenna2008}%
  \BibitemOpen
  \bibfield  {author} {\bibinfo {author} {\bibfnamefont {F.}~\bibnamefont
  {Caravenna}}\ and\ \bibinfo {author} {\bibfnamefont {J.-D.}\ \bibnamefont
  {Deuschel}},\ }\href@noop {} {\bibfield  {journal} {\bibinfo  {journal} {Ann.
  Probab.}\ }\textbf {\bibinfo {volume} {36}},\ \bibinfo {pages} {2388}
  (\bibinfo {year} {2008})}\BibitemShut {NoStop}%
\bibitem [{\citenamefont {Hsu}\ and\ \citenamefont {Binder}(2013)}]{Hsu2013}%
  \BibitemOpen
  \bibfield  {author} {\bibinfo {author} {\bibfnamefont {H.-P.}\ \bibnamefont
  {Hsu}}\ and\ \bibinfo {author} {\bibfnamefont {K.}~\bibnamefont {Binder}},\
  }\href@noop {} {\bibfield  {journal} {\bibinfo  {journal} {Macromolecules}\
  }\textbf {\bibinfo {volume} {46}},\ \bibinfo {pages} {2496} (\bibinfo {year}
  {2013})}\BibitemShut {NoStop}%
\bibitem [{\citenamefont {Freed}(1972)}]{freed}%
  \BibitemOpen
  \bibfield  {author} {\bibinfo {author} {\bibfnamefont {K.~F.}\ \bibnamefont
  {Freed}},\ }\href {\doibase 10.1002/9780470143728} {\bibfield  {journal}
  {\bibinfo  {journal} {Adv. in Chem. Phys.}\ }\textbf {\bibinfo {volume}
  {22}},\ \bibinfo {pages} {1} (\bibinfo {year} {1972})}\BibitemShut {NoStop}%
\bibitem [{\citenamefont {Gompper}\ and\ \citenamefont
  {Burkhardt}(1989)}]{Gompper1989}%
  \BibitemOpen
  \bibfield  {author} {\bibinfo {author} {\bibfnamefont {G.}~\bibnamefont
  {Gompper}}\ and\ \bibinfo {author} {\bibfnamefont {T.}~\bibnamefont
  {Burkhardt}},\ }\href@noop {} {\bibfield  {journal} {\bibinfo  {journal}
  {Phys. Rev. A}\ }\textbf {\bibinfo {volume} {40}},\ \bibinfo {pages} {6124}
  (\bibinfo {year} {1989})}\BibitemShut {NoStop}%
\bibitem [{\citenamefont {Gompper}\ and\ \citenamefont
  {Seifert}(1990)}]{Gompper1990}%
  \BibitemOpen
  \bibfield  {author} {\bibinfo {author} {\bibfnamefont {G.}~\bibnamefont
  {Gompper}}\ and\ \bibinfo {author} {\bibfnamefont {U.}~\bibnamefont
  {Seifert}},\ }\href@noop {} {\bibfield  {journal} {\bibinfo  {journal} {J.
  Phys. A: Math. Gen.}\ }\textbf {\bibinfo {volume} {23}},\ \bibinfo {pages}
  {L1161} (\bibinfo {year} {1990})}\BibitemShut {NoStop}%
\bibitem [{\citenamefont {Kuznetsov}\ and\ \citenamefont
  {Sung}(1997)}]{Kuznetsov1997}%
  \BibitemOpen
  \bibfield  {author} {\bibinfo {author} {\bibfnamefont {D.~V.}\ \bibnamefont
  {Kuznetsov}}\ and\ \bibinfo {author} {\bibfnamefont {W.}~\bibnamefont
  {Sung}},\ }\href@noop {} {\bibfield  {journal} {\bibinfo  {journal} {J. Chem.
  Phys.}\ }\textbf {\bibinfo {volume} {107}},\ \bibinfo {pages} {4729}
  (\bibinfo {year} {1997})}\BibitemShut {NoStop}%
\bibitem [{\citenamefont {Bundschuh}, \citenamefont {L{\"a}ssig},\ and\
  \citenamefont {Lipowsky}(2000)}]{bundschuh00}%
  \BibitemOpen
  \bibfield  {author} {\bibinfo {author} {\bibfnamefont {R.}~\bibnamefont
  {Bundschuh}}, \bibinfo {author} {\bibfnamefont {M.}~\bibnamefont
  {L{\"a}ssig}}, \ and\ \bibinfo {author} {\bibfnamefont {R.}~\bibnamefont
  {Lipowsky}},\ }\href@noop {} {\bibfield  {journal} {\bibinfo  {journal} {Eur.
  Phys. J. E}\ }\textbf {\bibinfo {volume} {3}},\ \bibinfo {pages} {295}
  (\bibinfo {year} {2000})}\BibitemShut {NoStop}%
\bibitem [{\citenamefont {Stepanow}(2001)}]{Stepanow2001}%
  \BibitemOpen
  \bibfield  {author} {\bibinfo {author} {\bibfnamefont {S.}~\bibnamefont
  {Stepanow}},\ }\href@noop {} {\bibfield  {journal} {\bibinfo  {journal} {J.
  Chem. Phys.}\ }\textbf {\bibinfo {volume} {115}},\ \bibinfo {pages} {1565}
  (\bibinfo {year} {2001})}\BibitemShut {NoStop}%
\bibitem [{\citenamefont {Semenov}(2002)}]{Semenov2002}%
  \BibitemOpen
  \bibfield  {author} {\bibinfo {author} {\bibfnamefont {A.~N.}\ \bibnamefont
  {Semenov}},\ }\href@noop {} {\bibfield  {journal} {\bibinfo  {journal} {Eur.
  Phys. J. E}\ }\textbf {\bibinfo {volume} {9}},\ \bibinfo {pages} {353}
  (\bibinfo {year} {2002})}\BibitemShut {NoStop}%
\bibitem [{\citenamefont {Kierfeld}\ and\ \citenamefont
  {Lipowsky}(2003)}]{kierfeld03}%
  \BibitemOpen
  \bibfield  {author} {\bibinfo {author} {\bibfnamefont {J.}~\bibnamefont
  {Kierfeld}}\ and\ \bibinfo {author} {\bibfnamefont {R.}~\bibnamefont
  {Lipowsky}},\ }\href {\doibase 10.1209/epl/i2003-00139-0} {\bibfield
  {journal} {\bibinfo  {journal} {Europhys. Lett.}\ }\textbf {\bibinfo {volume}
  {62}},\ \bibinfo {pages} {285} (\bibinfo {year} {2003})}\BibitemShut
  {NoStop}%
\bibitem [{\citenamefont {Benetatos}\ and\ \citenamefont
  {Frey}(2003)}]{Benetatos2003}%
  \BibitemOpen
  \bibfield  {author} {\bibinfo {author} {\bibfnamefont {P.}~\bibnamefont
  {Benetatos}}\ and\ \bibinfo {author} {\bibfnamefont {E.}~\bibnamefont
  {Frey}},\ }\href@noop {} {\bibfield  {journal} {\bibinfo  {journal} {Phys.
  Rev. E}\ }\textbf {\bibinfo {volume} {67}},\ \bibinfo {pages} {051108}
  (\bibinfo {year} {2003})}\BibitemShut {NoStop}%
\bibitem [{\citenamefont {Kierfeld}\ and\ \citenamefont
  {Lipowsky}(2005)}]{Kierfeld2005a}%
  \BibitemOpen
  \bibfield  {author} {\bibinfo {author} {\bibfnamefont {J.}~\bibnamefont
  {Kierfeld}}\ and\ \bibinfo {author} {\bibfnamefont {R.}~\bibnamefont
  {Lipowsky}},\ }\href@noop {} {\bibfield  {journal} {\bibinfo  {journal} {J.
  Phys. A: Math. Gen.}\ }\textbf {\bibinfo {volume} {38}},\ \bibinfo {pages}
  {L155} (\bibinfo {year} {2005})}\BibitemShut {NoStop}%
\bibitem [{\citenamefont {Deng}\ \emph {et~al.}(2010)\citenamefont {Deng},
  \citenamefont {Jiang}, \citenamefont {Liang},\ and\ \citenamefont
  {Chen}}]{deng10}%
  \BibitemOpen
  \bibfield  {author} {\bibinfo {author} {\bibfnamefont {M.}~\bibnamefont
  {Deng}}, \bibinfo {author} {\bibfnamefont {Y.}~\bibnamefont {Jiang}},
  \bibinfo {author} {\bibfnamefont {H.}~\bibnamefont {Liang}}, \ and\ \bibinfo
  {author} {\bibfnamefont {J.}~\bibnamefont {Chen}},\ }\href@noop {} {\bibfield
   {journal} {\bibinfo  {journal} {J. Chem. Phys.}\ }\textbf {\bibinfo {volume}
  {133}},\ \bibinfo {pages} {034902} (\bibinfo {year} {2010})}\BibitemShut
  {NoStop}%
\bibitem [{\citenamefont {Wallin}\ and\ \citenamefont
  {Linse}(1996)}]{Wallin1996}%
  \BibitemOpen
  \bibfield  {author} {\bibinfo {author} {\bibfnamefont {T.}~\bibnamefont
  {Wallin}}\ and\ \bibinfo {author} {\bibfnamefont {P.}~\bibnamefont {Linse}},\
  }\href@noop {} {\bibfield  {journal} {\bibinfo  {journal} {Langmuir}\
  }\textbf {\bibinfo {volume} {12}},\ \bibinfo {pages} {305} (\bibinfo {year}
  {1996})}\BibitemShut {NoStop}%
\bibitem [{\citenamefont {Netz}\ and\ \citenamefont
  {Joanny}(1999{\natexlab{b}})}]{Netz1999b}%
  \BibitemOpen
  \bibfield  {author} {\bibinfo {author} {\bibfnamefont {R.~R.}\ \bibnamefont
  {Netz}}\ and\ \bibinfo {author} {\bibfnamefont {J.-F.}\ \bibnamefont
  {Joanny}},\ }\href@noop {} {\bibfield  {journal} {\bibinfo  {journal}
  {Macromolecules}\ }\textbf {\bibinfo {volume} {32}},\ \bibinfo {pages} {9026}
  (\bibinfo {year} {1999}{\natexlab{b}})}\BibitemShut {NoStop}%
\bibitem [{\citenamefont {Schiessel}\ \emph {et~al.}(2000)\citenamefont
  {Schiessel}, \citenamefont {Rudnick}, \citenamefont {Bruinsma},\ and\
  \citenamefont {Gelbart}}]{Schiessel2000}%
  \BibitemOpen
  \bibfield  {author} {\bibinfo {author} {\bibfnamefont {H.}~\bibnamefont
  {Schiessel}}, \bibinfo {author} {\bibfnamefont {J.}~\bibnamefont {Rudnick}},
  \bibinfo {author} {\bibfnamefont {R.}~\bibnamefont {Bruinsma}}, \ and\
  \bibinfo {author} {\bibfnamefont {W.~M.}\ \bibnamefont {Gelbart}},\
  }\href@noop {} {\bibfield  {journal} {\bibinfo  {journal} {Europhys. Lett.}\
  }\textbf {\bibinfo {volume} {51}},\ \bibinfo {pages} {237} (\bibinfo {year}
  {2000})}\BibitemShut {NoStop}%
\bibitem [{\citenamefont {Kunze}\ and\ \citenamefont {Netz}(2000)}]{netz00}%
  \BibitemOpen
  \bibfield  {author} {\bibinfo {author} {\bibfnamefont {K.-K.}\ \bibnamefont
  {Kunze}}\ and\ \bibinfo {author} {\bibfnamefont {R.~R.}\ \bibnamefont
  {Netz}},\ }\href@noop {} {\bibfield  {journal} {\bibinfo  {journal} {Phys.
  Rev. Lett.}\ }\textbf {\bibinfo {volume} {85}},\ \bibinfo {pages} {4389}
  (\bibinfo {year} {2000})}\BibitemShut {NoStop}%
\bibitem [{\citenamefont {Cherstvy}\ and\ \citenamefont
  {Winkler}(2011)}]{Cherstvy2011}%
  \BibitemOpen
  \bibfield  {author} {\bibinfo {author} {\bibfnamefont {A.~G.}\ \bibnamefont
  {Cherstvy}}\ and\ \bibinfo {author} {\bibfnamefont {R.~G.}\ \bibnamefont
  {Winkler}},\ }\href@noop {} {\bibfield  {journal} {\bibinfo  {journal} {Phys.
  Chem. Chem. Phys.}\ }\textbf {\bibinfo {volume} {13}},\ \bibinfo {pages}
  {11686} (\bibinfo {year} {2011})}\BibitemShut {NoStop}%
\bibitem [{\citenamefont {Cherstvy}(2012)}]{Cherstvy2012}%
  \BibitemOpen
  \bibfield  {author} {\bibinfo {author} {\bibfnamefont {A.~G.}\ \bibnamefont
  {Cherstvy}},\ }\href@noop {} {\bibfield  {journal} {\bibinfo  {journal}
  {Biopolymers}\ }\textbf {\bibinfo {volume} {97}},\ \bibinfo {pages} {311}
  (\bibinfo {year} {2012})}\BibitemShut {NoStop}%
\bibitem [{\citenamefont {Hochrein}\ \emph {et~al.}(2007)\citenamefont
  {Hochrein}, \citenamefont {Leierseder}, \citenamefont {Golubovi\'{c}},\ and\
  \citenamefont {R\"{a}dler}}]{Hochrein2007}%
  \BibitemOpen
  \bibfield  {author} {\bibinfo {author} {\bibfnamefont {M.}~\bibnamefont
  {Hochrein}}, \bibinfo {author} {\bibfnamefont {J.}~\bibnamefont
  {Leierseder}}, \bibinfo {author} {\bibfnamefont {L.}~\bibnamefont
  {Golubovi\'{c}}}, \ and\ \bibinfo {author} {\bibfnamefont {J.}~\bibnamefont
  {R\"{a}dler}},\ }\href@noop {} {\bibfield  {journal} {\bibinfo  {journal}
  {Phys. Rev. E}\ }\textbf {\bibinfo {volume} {75}},\ \bibinfo {pages} {021901}
  (\bibinfo {year} {2007})}\BibitemShut {NoStop}%
\bibitem [{\citenamefont {Gutjahr}, \citenamefont {Lipowsky},\ and\
  \citenamefont {Kierfeld}(2010)}]{Gutjahr2010}%
  \BibitemOpen
  \bibfield  {author} {\bibinfo {author} {\bibfnamefont {P.}~\bibnamefont
  {Gutjahr}}, \bibinfo {author} {\bibfnamefont {R.}~\bibnamefont {Lipowsky}}, \
  and\ \bibinfo {author} {\bibfnamefont {J.}~\bibnamefont {Kierfeld}},\
  }\href@noop {} {\bibfield  {journal} {\bibinfo  {journal} {Soft Matter}\ ,\
  \bibinfo {pages} {5461}} (\bibinfo {year} {2010})}\BibitemShut {NoStop}%
\bibitem [{\citenamefont {Kraikivski}, \citenamefont {Lipowsky},\ and\
  \citenamefont {Kierfeld}(2004)}]{Kraikivski2004}%
  \BibitemOpen
  \bibfield  {author} {\bibinfo {author} {\bibfnamefont {P.}~\bibnamefont
  {Kraikivski}}, \bibinfo {author} {\bibfnamefont {R.}~\bibnamefont
  {Lipowsky}}, \ and\ \bibinfo {author} {\bibfnamefont {J.}~\bibnamefont
  {Kierfeld}},\ }\href@noop {} {\bibfield  {journal} {\bibinfo  {journal}
  {Europhys. Lett.}\ }\textbf {\bibinfo {volume} {66}},\ \bibinfo {pages} {763}
  (\bibinfo {year} {2004})}\BibitemShut {NoStop}%
\bibitem [{\citenamefont {Kraikivski}, \citenamefont {Lipowsky},\ and\
  \citenamefont {Kierfeld}(2005)}]{Kraikivski2005a}%
  \BibitemOpen
  \bibfield  {author} {\bibinfo {author} {\bibfnamefont {P.}~\bibnamefont
  {Kraikivski}}, \bibinfo {author} {\bibfnamefont {R.}~\bibnamefont
  {Lipowsky}}, \ and\ \bibinfo {author} {\bibfnamefont {J.}~\bibnamefont
  {Kierfeld}},\ }\href@noop {} {\bibfield  {journal} {\bibinfo  {journal} {Eur.
  Phys. J. E}\ }\textbf {\bibinfo {volume} {16}},\ \bibinfo {pages} {319}
  (\bibinfo {year} {2005})}\BibitemShut {NoStop}%
\bibitem [{\citenamefont {Pierre-Louis}(2011)}]{pierrelouis2011}%
  \BibitemOpen
  \bibfield  {author} {\bibinfo {author} {\bibfnamefont {O.}~\bibnamefont
  {Pierre-Louis}},\ }\href@noop {} {\bibfield  {journal} {\bibinfo  {journal}
  {Phys. Rev. E}\ }\textbf {\bibinfo {volume} {83}},\ \bibinfo {pages} {011801}
  (\bibinfo {year} {2011})}\BibitemShut {NoStop}%
\bibitem [{\citenamefont {Kratky}\ and\ \citenamefont
  {Porod}(1949)}]{kratkyporod}%
  \BibitemOpen
  \bibfield  {author} {\bibinfo {author} {\bibfnamefont {O.}~\bibnamefont
  {Kratky}}\ and\ \bibinfo {author} {\bibfnamefont {G.}~\bibnamefont {Porod}},\
  }\href {\doibase 10.1002/recl.19490681203} {\bibfield  {journal} {\bibinfo
  {journal} {Recueil des Travaux Chimiques des Pays-Bas}\ }\textbf {\bibinfo
  {volume} {68}},\ \bibinfo {pages} {1106} (\bibinfo {year}
  {1949})}\BibitemShut {NoStop}%
\bibitem [{\citenamefont {Harris}\ and\ \citenamefont
  {Hearst}(1966)}]{harris66}%
  \BibitemOpen
  \bibfield  {author} {\bibinfo {author} {\bibfnamefont {R.}~\bibnamefont
  {Harris}}\ and\ \bibinfo {author} {\bibfnamefont {J.}~\bibnamefont
  {Hearst}},\ }\href@noop {} {\bibfield  {journal} {\bibinfo  {journal} {J.
  Chem. Phys.}\ }\textbf {\bibinfo {volume} {44}},\ \bibinfo {pages} {2595}
  (\bibinfo {year} {1966})}\BibitemShut {NoStop}%
\bibitem [{\citenamefont {Kleinert}(2006)}]{kleinert}%
  \BibitemOpen
  \bibfield  {author} {\bibinfo {author} {\bibfnamefont {H.}~\bibnamefont
  {Kleinert}},\ }\href@noop {} {\emph {\bibinfo {title} {Path Integrals in
  Quantum Mechanics, Statistics, Polymer Physics, and Financial Markets}}}\
  (\bibinfo  {publisher} {World Scientific},\ \bibinfo {address} {Singapore},\
  \bibinfo {year} {2006})\BibitemShut {NoStop}%
\bibitem [{\citenamefont {Odijk}(1977)}]{Odijk1977}%
  \BibitemOpen
  \bibfield  {author} {\bibinfo {author} {\bibfnamefont {T.}~\bibnamefont
  {Odijk}},\ }\href@noop {} {\bibfield  {journal} {\bibinfo  {journal} {J.
  Polym. Sci.}\ }\textbf {\bibinfo {volume} {15}},\ \bibinfo {pages} {477}
  (\bibinfo {year} {1977})}\BibitemShut {NoStop}%
\bibitem [{\citenamefont {Lipowsky}(1989)}]{lipowsky89}%
  \BibitemOpen
  \bibfield  {author} {\bibinfo {author} {\bibfnamefont {R.}~\bibnamefont
  {Lipowsky}},\ }\href@noop {} {\bibfield  {journal} {\bibinfo  {journal}
  {Phys. Rev. Lett.}\ }\textbf {\bibinfo {volume} {62}},\ \bibinfo {pages}
  {704} (\bibinfo {year} {1989})}\BibitemShut {NoStop}%
\bibitem [{\citenamefont {Kierfeld}\ \emph {et~al.}(2004)\citenamefont
  {Kierfeld}, \citenamefont {Niamploy}, \citenamefont {Sa-Yakanit},\ and\
  \citenamefont {Lipowsky}}]{kierfeld04}%
  \BibitemOpen
  \bibfield  {author} {\bibinfo {author} {\bibfnamefont {J.}~\bibnamefont
  {Kierfeld}}, \bibinfo {author} {\bibfnamefont {O.}~\bibnamefont {Niamploy}},
  \bibinfo {author} {\bibfnamefont {V.}~\bibnamefont {Sa-Yakanit}}, \ and\
  \bibinfo {author} {\bibfnamefont {R.}~\bibnamefont {Lipowsky}},\ }\href
  {\doibase 10.1140/epje/i2003-10089-3} {\bibfield  {journal} {\bibinfo
  {journal} {Eur. Phys. J. E}\ }\textbf {\bibinfo {volume} {14}},\ \bibinfo
  {pages} {17} (\bibinfo {year} {2004})}\BibitemShut {NoStop}%
\bibitem [{\citenamefont {MacKintosh}, \citenamefont {K{\"a}s},\ and\
  \citenamefont {Janmey}(1995)}]{MacKintosh1995}%
  \BibitemOpen
  \bibfield  {author} {\bibinfo {author} {\bibfnamefont {F.}~\bibnamefont
  {MacKintosh}}, \bibinfo {author} {\bibfnamefont {J.}~\bibnamefont {K{\"a}s}},
  \ and\ \bibinfo {author} {\bibfnamefont {P.}~\bibnamefont {Janmey}},\
  }\href@noop {} {\bibfield  {journal} {\bibinfo  {journal} {Phys. Rev. Lett.}\
  }\textbf {\bibinfo {volume} {75}},\ \bibinfo {pages} {4425} (\bibinfo {year}
  {1995})}\BibitemShut {NoStop}%
\bibitem [{\citenamefont {Kojima}, \citenamefont {Ishijima},\ and\
  \citenamefont {Yanagida}(1994)}]{Kojima1994}%
  \BibitemOpen
  \bibfield  {author} {\bibinfo {author} {\bibfnamefont {H.}~\bibnamefont
  {Kojima}}, \bibinfo {author} {\bibfnamefont {A.}~\bibnamefont {Ishijima}}, \
  and\ \bibinfo {author} {\bibfnamefont {T.}~\bibnamefont {Yanagida}},\
  }\href@noop {} {\bibfield  {journal} {\bibinfo  {journal} {Proc. Nat. Acad.
  Sci. USA}\ }\textbf {\bibinfo {volume} {91}},\ \bibinfo {pages} {12962}
  (\bibinfo {year} {1994})}\BibitemShut {NoStop}%
\bibitem [{\citenamefont {Odijk}(1983)}]{Odijk1983}%
  \BibitemOpen
  \bibfield  {author} {\bibinfo {author} {\bibfnamefont {T.}~\bibnamefont
  {Odijk}},\ }\href@noop {} {\bibfield  {journal} {\bibinfo  {journal}
  {Macromolecules}\ }\textbf {\bibinfo {volume} {16}},\ \bibinfo {pages} {1340}
  (\bibinfo {year} {1983})}\BibitemShut {NoStop}%
\bibitem [{\citenamefont {K\"{o}ster}\ \emph {et~al.}(2007)\citenamefont
  {K\"{o}ster}, \citenamefont {Stark}, \citenamefont {Pfohl},\ and\
  \citenamefont {Kierfeld}}]{Koster2007}%
  \BibitemOpen
  \bibfield  {author} {\bibinfo {author} {\bibfnamefont {S.}~\bibnamefont
  {K\"{o}ster}}, \bibinfo {author} {\bibfnamefont {H.}~\bibnamefont {Stark}},
  \bibinfo {author} {\bibfnamefont {T.}~\bibnamefont {Pfohl}}, \ and\ \bibinfo
  {author} {\bibfnamefont {J.}~\bibnamefont {Kierfeld}},\ }\href@noop {}
  {\bibfield  {journal} {\bibinfo  {journal} {Biophys. Rev. Lett.}\ }\textbf
  {\bibinfo {volume} {2}},\ \bibinfo {pages} {155} (\bibinfo {year}
  {2007})}\BibitemShut {NoStop}%
\bibitem [{\citenamefont {K\"{o}ster}, \citenamefont {Kierfeld},\ and\
  \citenamefont {Pfohl}(2008)}]{Koster2008}%
  \BibitemOpen
  \bibfield  {author} {\bibinfo {author} {\bibfnamefont {S.}~\bibnamefont
  {K\"{o}ster}}, \bibinfo {author} {\bibfnamefont {J.}~\bibnamefont
  {Kierfeld}}, \ and\ \bibinfo {author} {\bibfnamefont {T.}~\bibnamefont
  {Pfohl}},\ }\href@noop {} {\bibfield  {journal} {\bibinfo  {journal} {Eur.
  Phys J. E}\ }\textbf {\bibinfo {volume} {25}},\ \bibinfo {pages} {439}
  (\bibinfo {year} {2008})}\BibitemShut {NoStop}%
\bibitem [{sup()}]{suppl}%
 \BibitemOpen
{See supplementary material (appended) for
additional information}\BibitemShut {NoStop}%
\bibitem [{\citenamefont {Binder}(1987)}]{Binder1987}%
  \BibitemOpen
  \bibfield  {author} {\bibinfo {author} {\bibfnamefont {K.}~\bibnamefont
  {Binder}},\ }\href@noop {} {\bibfield  {journal} {\bibinfo  {journal} {Rep.
  Prog. Phys.}\ }\textbf {\bibinfo {volume} {50}},\ \bibinfo {pages} {783}
  (\bibinfo {year} {1987})}\BibitemShut {NoStop}%
\bibitem [{\citenamefont {Pierre-Louis}(2008)}]{pierrelouis2008}%
  \BibitemOpen
  \bibfield  {author} {\bibinfo {author} {\bibfnamefont {O.}~\bibnamefont
  {Pierre-Louis}},\ }\href@noop {} {\bibfield  {journal} {\bibinfo  {journal}
  {Phys. Rev. E}\ }\textbf {\bibinfo {volume} {78}},\ \bibinfo {pages} {021603}
  (\bibinfo {year} {2008})}\BibitemShut {NoStop}%
\bibitem [{\citenamefont {Li}\ and\ \citenamefont {Zhang}(2010)}]{Li2010}%
  \BibitemOpen
  \bibfield  {author} {\bibinfo {author} {\bibfnamefont {T.}~\bibnamefont
  {Li}}\ and\ \bibinfo {author} {\bibfnamefont {Z.}~\bibnamefont {Zhang}},\
  }\href@noop {} {\bibfield  {journal} {\bibinfo  {journal} {J. Phys. D: Appl.
  Phys.}\ }\textbf {\bibinfo {volume} {43}},\ \bibinfo {pages} {075303}
  (\bibinfo {year} {2010})}\BibitemShut {NoStop}%
\bibitem [{\citenamefont {Zhang}, \citenamefont {Ohbu},\ and\ \citenamefont
  {Dubin}(2000)}]{Zhang2000}%
  \BibitemOpen
  \bibfield  {author} {\bibinfo {author} {\bibfnamefont {H.}~\bibnamefont
  {Zhang}}, \bibinfo {author} {\bibfnamefont {K.}~\bibnamefont {Ohbu}}, \ and\
  \bibinfo {author} {\bibfnamefont {P.~L.}\ \bibnamefont {Dubin}},\ }\href@noop
  {} {\bibfield  {journal} {\bibinfo  {journal} {Langmuir}\ }\textbf {\bibinfo
  {volume} {16}},\ \bibinfo {pages} {9082} (\bibinfo {year}
  {2000})}\BibitemShut {NoStop}%
\bibitem [{\citenamefont {Cooper}\ \emph {et~al.}(2006)\citenamefont {Cooper},
  \citenamefont {Goulding}, \citenamefont {Kayitmazer}, \citenamefont {Ulrich},
  \citenamefont {Stoll}, \citenamefont {Turksen}, \citenamefont {Yusa},
  \citenamefont {Kumar},\ and\ \citenamefont {Dubin}}]{Cooper2006}%
  \BibitemOpen
  \bibfield  {author} {\bibinfo {author} {\bibfnamefont {C.~L.}\ \bibnamefont
  {Cooper}}, \bibinfo {author} {\bibfnamefont {A.}~\bibnamefont {Goulding}},
  \bibinfo {author} {\bibfnamefont {a.~B.}\ \bibnamefont {Kayitmazer}},
  \bibinfo {author} {\bibfnamefont {S.}~\bibnamefont {Ulrich}}, \bibinfo
  {author} {\bibfnamefont {S.}~\bibnamefont {Stoll}}, \bibinfo {author}
  {\bibfnamefont {S.}~\bibnamefont {Turksen}}, \bibinfo {author} {\bibfnamefont
  {S.-i.}\ \bibnamefont {Yusa}}, \bibinfo {author} {\bibfnamefont
  {A.}~\bibnamefont {Kumar}}, \ and\ \bibinfo {author} {\bibfnamefont {P.~L.}\
  \bibnamefont {Dubin}},\ }\href@noop {} {\bibfield  {journal} {\bibinfo
  {journal} {Biomacromolecules}\ }\textbf {\bibinfo {volume} {7}},\ \bibinfo
  {pages} {1025} (\bibinfo {year} {2006})}\BibitemShut {NoStop}%
\end{thebibliography}

\begin{thebibliography}{10}%
\makeatletter
\providecommand \@ifxundefined [1]{%
 \@ifx{#1\undefined}
}%
\providecommand \@ifnum [1]{%
 \ifnum #1\expandafter \@firstoftwo
 \else \expandafter \@secondoftwo
 \fi
}%
\providecommand \@ifx [1]{%
 \ifx #1\expandafter \@firstoftwo
 \else \expandafter \@secondoftwo
 \fi
}%
\providecommand \natexlab [1]{#1}%
\providecommand \enquote  [1]{``#1''}%
\providecommand \bibnamefont  [1]{#1}%
\providecommand \bibfnamefont [1]{#1}%
\providecommand \citenamefont [1]{#1}%
\providecommand \href@noop [0]{\@secondoftwo}%
\providecommand \href [0]{\begingroup \@sanitize@url \@href}%
\providecommand \@href[1]{\@@startlink{#1}\@@href}%
\providecommand \@@href[1]{\endgroup#1\@@endlink}%
\providecommand \@sanitize@url [0]{\catcode `\\12\catcode `\$12\catcode
  `\&12\catcode `\#12\catcode `\^12\catcode `\_12\catcode `\%12\relax}%
\providecommand \@@startlink[1]{}%
\providecommand \@@endlink[0]{}%
\providecommand \url  [0]{\begingroup\@sanitize@url \@url }%
\providecommand \@url [1]{\endgroup\@href {#1}{\urlprefix }}%
\providecommand \urlprefix  [0]{URL }%
\providecommand \Eprint [0]{\href }%
\providecommand \doibase [0]{http://dx.doi.org/}%
\providecommand \selectlanguage [0]{\@gobble}%
\providecommand \bibinfo  [0]{\@secondoftwo}%
\providecommand \bibfield  [0]{\@secondoftwo}%
\providecommand \translation [1]{[#1]}%
\providecommand \BibitemOpen [0]{}%
\providecommand \bibitemStop [0]{}%
\providecommand \bibitemNoStop [0]{.\EOS\space}%
\providecommand \EOS [0]{\spacefactor3000\relax}%
\providecommand \BibitemShut  [1]{\csname bibitem#1\endcsname}%
\let\auto@bib@innerbib\@empty
\bibitem [{\citenamefont {Deng}\ \emph {et~al.}(2010)\citenamefont {Deng},
  \citenamefont {Jiang}, \citenamefont {Liang},\ and\ \citenamefont
  {Chen}}]{sdeng10}%
  \BibitemOpen
  \bibfield  {author} {\bibinfo {author} {\bibfnamefont {M.}~\bibnamefont
  {Deng}}, \bibinfo {author} {\bibfnamefont {Y.}~\bibnamefont {Jiang}},
  \bibinfo {author} {\bibfnamefont {H.}~\bibnamefont {Liang}}, \ and\ \bibinfo
  {author} {\bibfnamefont {J.}~\bibnamefont {Chen}},\ }\href@noop {} {\bibfield
   {journal} {\bibinfo  {journal} {J. Chem. Phys.}\ }\textbf {\bibinfo {volume}
  {133}},\ \bibinfo {pages} {034902} (\bibinfo {year} {2010})}\BibitemShut
  {NoStop}%
\bibitem [{\citenamefont {Freed}(1972)}]{sfreed}%
  \BibitemOpen
  \bibfield  {author} {\bibinfo {author} {\bibfnamefont {K.~F.}\ \bibnamefont
  {Freed}},\ }\href {\doibase 10.1002/9780470143728} {\bibfield  {journal}
  {\bibinfo  {journal} {Adv. in Chem. Phys.}\ }\textbf {\bibinfo {volume}
  {22}},\ \bibinfo {pages} {1} (\bibinfo {year} {1972})}\BibitemShut {NoStop}%
\bibitem [{\citenamefont {Maggs}, \citenamefont {Huse},\ and\ \citenamefont
  {Leibler}(1989)}]{sMaggs1989}%
  \BibitemOpen
  \bibfield  {author} {\bibinfo {author} {\bibfnamefont {A.~C.}\ \bibnamefont
  {Maggs}}, \bibinfo {author} {\bibfnamefont {D.~A.}\ \bibnamefont {Huse}}, \
  and\ \bibinfo {author} {\bibfnamefont {S.}~\bibnamefont {Leibler}},\
  }\href@noop {} {\bibfield  {journal} {\bibinfo  {journal} {Europhys. Lett.}\
  }\textbf {\bibinfo {volume} {8}},\ \bibinfo {pages} {615} (\bibinfo {year}
  {1989})}\BibitemShut {NoStop}%
\bibitem [{\citenamefont {Gompper}\ and\ \citenamefont
  {Burkhardt}(1989)}]{sGompper1989}%
  \BibitemOpen
  \bibfield  {author} {\bibinfo {author} {\bibfnamefont {G.}~\bibnamefont
  {Gompper}}\ and\ \bibinfo {author} {\bibfnamefont {T.}~\bibnamefont
  {Burkhardt}},\ }\href@noop {} {\bibfield  {journal} {\bibinfo  {journal}
  {Phys. Rev. A}\ }\textbf {\bibinfo {volume} {40}},\ \bibinfo {pages} {6124}
  (\bibinfo {year} {1989})}\BibitemShut {NoStop}%
\bibitem [{\citenamefont {Kierfeld}\ and\ \citenamefont
  {Lipowsky}(2003)}]{skierfeld03}%
  \BibitemOpen
  \bibfield  {author} {\bibinfo {author} {\bibfnamefont {J.}~\bibnamefont
  {Kierfeld}}\ and\ \bibinfo {author} {\bibfnamefont {R.}~\bibnamefont
  {Lipowsky}},\ }\href {\doibase 10.1209/epl/i2003-00139-0} {\bibfield
  {journal} {\bibinfo  {journal} {Europhys. Lett.}\ }\textbf {\bibinfo {volume}
  {62}},\ \bibinfo {pages} {285} (\bibinfo {year} {2003})}\BibitemShut
  {NoStop}%
\bibitem [{\citenamefont {Kierfeld}\ and\ \citenamefont
  {Lipowsky}(2005)}]{sKierfeld2005a}%
  \BibitemOpen
  \bibfield  {author} {\bibinfo {author} {\bibfnamefont {J.}~\bibnamefont
  {Kierfeld}}\ and\ \bibinfo {author} {\bibfnamefont {R.}~\bibnamefont
  {Lipowsky}},\ }\href@noop {} {\bibfield  {journal} {\bibinfo  {journal} {J.
  Phys. A: Math. Gen.}\ }\textbf {\bibinfo {volume} {38}},\ \bibinfo {pages}
  {L155} (\bibinfo {year} {2005})}\BibitemShut {NoStop}%
\bibitem [{\citenamefont {Abramowitz}\ and\ \citenamefont
  {Stegun}(1972)}]{sabramowitz1972}%
  \BibitemOpen
  \bibfield  {author} {\bibinfo {author} {\bibfnamefont {W.}~\bibnamefont
  {Abramowitz}}\ and\ \bibinfo {author} {\bibfnamefont {I.}~\bibnamefont
  {Stegun}},\ }\href@noop {} {\emph {\bibinfo {title} {Handbook of Mathematical
  Functions}}},\ \bibinfo {series} {Applied Mathematics Series}\ No.~\bibinfo
  {number} {55}\ (\bibinfo  {publisher} {National Bureau of Standard},\
  \bibinfo {address} {Washington},\ \bibinfo {year} {1972})\BibitemShut
  {NoStop}%
\bibitem [{\citenamefont {Burkhardt}(1993)}]{sburkhardt}%
  \BibitemOpen
  \bibfield  {author} {\bibinfo {author} {\bibfnamefont {T.}~\bibnamefont
  {Burkhardt}},\ }\href@noop {} {\bibfield  {journal} {\bibinfo  {journal} {J.
  Phys. A: Math. Gen.}\ }\textbf {\bibinfo {volume} {26}},\ \bibinfo {pages}
  {L1157} (\bibinfo {year} {1993})}\BibitemShut {NoStop}%
\bibitem [{Note1()}]{Note1}%
  \BibitemOpen
  \bibinfo {note} {Throughout this section we use scaling arguments. The
  distinction between $L_p$ and $L_{p,D}$ is therefore
  unnecessary.}\BibitemShut {Stop}%
\bibitem [{\citenamefont {Bhattacharjee}\ and\ \citenamefont
  {Seno}(2001)}]{sbhattacharjee2001measure}%
  \BibitemOpen
  \bibfield  {author} {\bibinfo {author} {\bibfnamefont {S.}~\bibnamefont
  {Bhattacharjee}}\ and\ \bibinfo {author} {\bibfnamefont {F.}~\bibnamefont
  {Seno}},\ }\href@noop {} {\bibfield  {journal} {\bibinfo  {journal} {J. Phys.
  A: Math. Gen.}\ }\textbf {\bibinfo {volume} {34}},\ \bibinfo {pages} {6375}
  (\bibinfo {year} {2001})}\BibitemShut {NoStop}%
\end{thebibliography}
\end{document}